\documentclass[a4paper,12pt]{article}

\pdfoutput=1
\usepackage{jheppub}

\usepackage{epsf}
\usepackage{epsfig}
\usepackage{subfigure}
\usepackage{mathtools}
\usepackage{hhline}
\usepackage{float}
\usepackage{multirow}
\usepackage{nicefrac}
\usepackage{epstopdf}
\usepackage{slashed}
\usepackage{xcolor}
\usepackage{url}
\usepackage{titlesec}

\usepackage{afterpage}

\newcommand{\fwidth}{0.82\hsize}

\titleformat*{\section}{\Large\bfseries}
\titleformat*{\subsection}{\large\bfseries}

\newcommand{\ka}{\kappa}

\newcommand{\ve}{\mathbf{e}}

\newcommand{\GeV}{\mathrm{GeV}}
\newcommand{\TeV}{\mathrm{TeV}}
\newcommand{\fbi}{\mathrm{fb}^{-1}}
\newcommand{\abi}{\mathrm{ab}^{-1}}

\newcommand{\jets}{\mathrm{jets}} 

\newcommand{\la}{\lambda}
\newcommand{\eps}{\epsilon}

\newcommand{\Lcal}{\mathcal{L}}

\newcommand{\ol}[1]{\overline{#1}}

\newcommand{\vev}[1]{\langle{#1}\rangle}
\newcommand{\abs}[1]{\left|{#1}\right|}
\newcommand{\order}[1]{\mathcal{O}\left({#1}\right)}
\newcommand{\br}[2]{\mathrm{BR} \left({#1}\to{#2}\right)}
\newcommand{\ga}[2]{\Gamma \left({#1}\to{#2}\right)}

\newcommand{\inv}{\mathrm{inv}}
\newcommand{\FB}{\mathrm{FB}}
\newcommand{\gaga}{{\gamma\gamma}}

\newcommand{\met}{E_T^\mathrm{miss}}

\newcommand{\obs}{\mathrm{obs}}
\newcommand{\CL}{\mathrm{CL}}
\newcommand{\disc}{\mathrm{disc}}
\newcommand{\excl}{\mathrm{excl}}

\toccontinuoustrue

\usepackage{amsmath}	
\usepackage{version}	

\begin{document}

\title{
Leptonic cascade decays of a heavy Higgs boson through vectorlike leptons at the LHC
}
\author{Radovan Dermisek$^1$,}
\author{Junichiro Kawamura$^2$,}
\author{Enrico Lunghi$^1$,}
\author{Navin McGinnis$^3$}
\author{and Seodong Shin$^{2,4}$}
\affiliation{
$^1$Physics Department, Indiana University, Bloomington, IN 47405, USA \\
$^2$Center for Theoretical Physics of the Universe, Institute for Basic Science, Daejeon 34126, Korea \\
$^3$TRIUMF, 4004 Westbrook Mall, Vancouver, BC, Canada V6T 2A3 \\
$^4$Department of Physics, Jeounbuk National University, Jeonju, Jeonbuk 54896, Korea \\
}
\emailAdd{dermisek@indiana.edu} 
\emailAdd{jkawa@ibs.re.kr}
\emailAdd{elunghi@indiana.edu} 
\emailAdd{nmcginnis@triumf.ca}
\emailAdd{sshin@jbnu.ac.kr}

\abstract{
We demonstrate the potential of fully leptonic cascade decays of a heavy neutral Higgs boson through vectorlike leptons as a simultaneous probe for extended Higgs sectors and extra matter particles at the LHC.
The processes we explore are unique in that their event topologies lead to di-boson-like leptonic final states with a lepton pair which does not reconstruct the mass of a gauge boson. By recasting existing $2\ell + E_T^{\rm miss}$ and $3/4\ell$ searches channels using run2 data from the LHC we obtain {\it model independent} bounds on the masses of heavy scalars and vectorlike leptons and use these results to explore future prospects at the HL-LHC.
Our results can be directly applied to any kind of new physics scenarios sharing the final states and the event topology.
For concreteness, we apply our results to a benchmark scenario: a two Higgs doublet model type-II augmented with vectorlike leptons.
Remarkably, even with current data the sensitivity of our analysis shows a reach for masses of a heavy neutral Higgs and vectorlike leptons up to 2 TeV and 1.5 TeV, respectively. Even for low $\tan\beta \gtrsim 1,$ the analysis retains sensitivity to heavy Higgs masses slightly above 1 TeV.
The future sensitivities at the HL-LHC extend the reach for heavy Higgses and new leptons to 2.7 TeV and 2 TeV, respectively.
\newpage
}
\preprint{
\begin{minipage}{4cm}
\small
\flushright
CTPU-PTC-22-06
\end{minipage}} 

\maketitle

\clearpage

\section{Introduction}

Since the discovery of the Higgs boson, unveiling the Higgs sector involved in electroweak symmetry breaking (EWSB) has become a strong focus of the LHC program and is essential to understanding possible features of new physics beyond the Standard Model (SM).
In particular, there are no symmetry arguments prohibiting the existence of an additional isodoublet field in the Higgs sector.
Two Higgs doublet models (2HDMs) are generally considered to be among the simplest and theoretically well-motivated frameworks for possible extensions of the Higgs sector, for a review see, e.g.~Refs.~\cite{Djouadi:2005gj,Branco:2011iw}. 
Popular examples of new theories beyond the SM (BSM) sharing this 2HDM-like Higgs sector include the Minimal Supersymmetric SM (MSSM)~\cite{Dimopoulos:1981zb,Nilles:1983ge}, the DFSZ axion model~\cite{Zhitnitsky:1980tq,Dine:1981rt}, and Twin Higgs model~\cite{Chacko:2005pe}. 

Another possibility which is not forbidden by any symmetry arguments is the existence of extra fermions beyond the three generations of the SM.
Since the existence of new chiral fermions is strongly disfavored both by theoretical and experimental reasons~\cite{Bar-Shalom:2012vvt}, extra matter fermions, if they exist, are likely to be vectorlike with masses which can be generated independently of the Higgs mechanism. 
Vectorlike fermions have also been considered in many compelling BSM theories including both non-supersymmetric and supersymmetric models with complete vectorlike families~\cite{Dermisek:2012as,Dermisek:2012ke,Dermisek:2017ihj,Dermisek:2018hxq,Dermisek:2018ujw}, composite/little Higgs model~\cite{Dugan:1984hq,Arkani-Hamed:2001nha}, the KSVZ axion model~\cite{Kim:1979if,Shifman:1979if}, and gauge mediated supersymmetry breaking model~\cite{Dine:1981gu,Dine:1981rt,Nappi:1982hm,Alvarez-Gaume:1981abe,Dine:1993yw,Dine:1994vc,Dine:1995ag}. Furthermore, models that include both an extension of the Higgs sector and vectorlike leptons were recently studied as possible solutions for the discrepancy in the measured value of the muon anomalous magnetic moment~\cite{Frank:2020smf,Chun:2020uzw,Dermisek:2020cod,Dermisek:2021ajd}.

\begin{figure}[th]
    \centering
    \begin{minipage}[c]{0.32\hsize}
    \centering 
    \includegraphics[height=36mm]{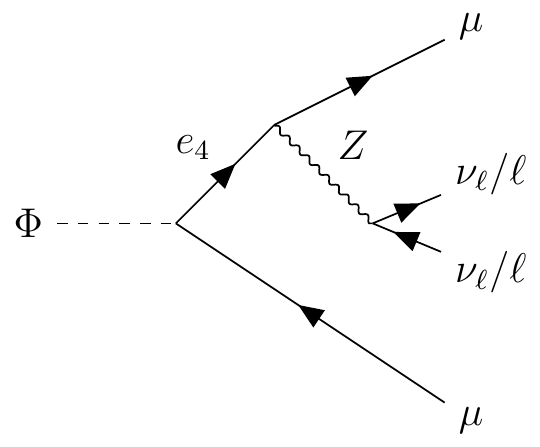}
    \end{minipage}
    \begin{minipage}[c]{0.32\hsize}
    \centering 
    \includegraphics[height=36mm]{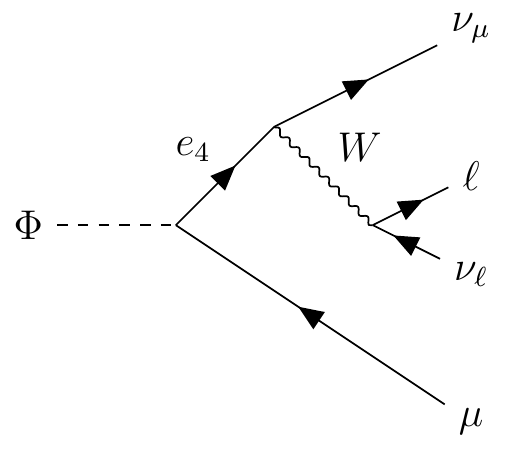}
    \end{minipage}
      \begin{minipage}[c]{0.32\hsize}
    \centering 
    \includegraphics[height=36mm]{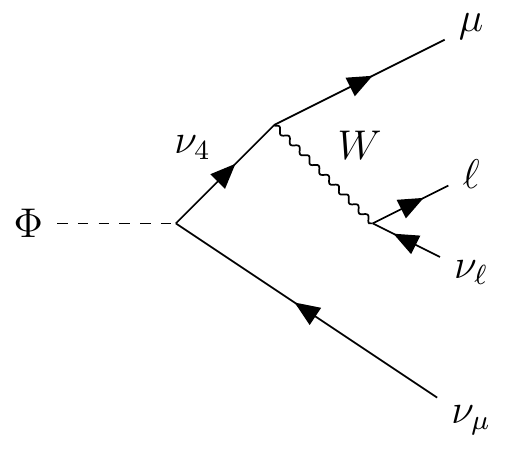}
    \end{minipage}
    \caption{
Leptonic cascade decays of a heavy neutral scalar ($\Phi$) through a charged ($e_4$) and neutral ($\nu_4$) vectorlike lepton. 
The decay in the left panel  contributes to the $3/4\ell$ search if $Z\to \ell\ell$, 
while the other decay modes contribute to the $2\ell+\met$ channel. 
The diagrams are drawn by TikZ-FeynHand~\cite{Ellis:2016jkw,Dohse:2018vqo}.
} 
    \label{fig:Hlcasc}
\end{figure}

Existing searches at the LHC for extra Higgs bosons or vectorlike fermions have been developed independently. 
On the other hand, it has been proposed that signatures involving both types of new particles simultaneously have many advantages over the typical searches~\cite{Dermisek:2015oja,Dermisek:2015vra,Dermisek:2015hue,Dermisek:2016via,CidVidal:2018eel,Dermisek:2019vkc,Dermisek:2019heo,Dermisek:2020gbr,Dermisek:2022kyh}.
In particular, heavy Higgs cascade decays through a vectorlike lepton mixing with the SM fermion, such as those depicted in Fig.~\ref{fig:Hlcasc}, provide relatively clean and distinctive signals due to their unique event topology.
The final states of the processes in Fig.~\ref{fig:Hlcasc} are fully leptonic, possibly with missing energy, and hence we dub these signatures {\it leptonic cascade decays} throughout this paper.
An important feature of our process is that the signature is ``di-boson-like'' but with one lepton pair whose momentum does \textit{not} reconstruct the mass of a gauge boson in any way.

In this paper, we obtain new, model-independent constraints on heavy scalars and vectorlike leptons by recasting recent results of the dilepton with missing energy ($2\ell + \met$) search~\cite{ATLAS:2019lff} and 3 or 4 lepton ($3/4\ell$) search~\cite{ATLAS:2021wob} at the LHC run2. 
We show that, remarkably, even with current data assuming a $50\%$ cascade branching ratio the sensitivity to masses of new scalars and heavy leptons can reach well above 1 TeV, proving the effectiveness of our leptonic cascade processes.
Our analysis results are quite promising compared to the typical consensus that the sensitivities for new leptons are weaker than those for vectorlike quarks considering the conventional production processes through gauge bosons.~\footnote{For example, compare Refs.~\cite{Dermisek:2014qca} and \cite{Sahinsoy:2015clx}.}
For concreteness, we interpret our results in the context of a reference model: a two Higgs doublet model (2HDM) type-II, where $\Phi = H~(A)$ is the heavy CP even (odd) neutral Higgs boson, augmented by vectorlike leptons which mix with the muon with $e_4$ ($\nu_4$) being the lightest charged (neutral) vectorlike lepton. This is exactly the scenario which can explain the muon anomalous magnetic moment, even with multi-TeV Higgs bosons and new leptons~\cite{Dermisek:2020cod,Dermisek:2021ajd}. 
This is the main motivation to pursue signals with muons in the final state. However, results generated from vectorlike leptons mixing with the first generation would be almost identical.
We find that the projected sensitivity of the analysis for the HL-LHC extends the reach for heavy Higgses and new leptons to 2.7 TeV and 2 TeV, respectively.
We emphasize that the search strategies and results in this paper can be readily applied to other new physics scenarios with the same kinematic topology and final states such as models of vectorlike leptons with a $Z^\prime$ boson~\cite{Allanach:2015gkd,Kawamura:2019rth,Kawamura:2019hxp,Kawamura:2021ygg}, singlet-extended 2HDM or Next-to-MSSM type models~\cite{Dermisek:2013cxa}.

This paper is organized as follows. In Sec.~\ref{sec-strategy}, we outline search strategies at the LHC for the leptonic heavy Higgs cascade decays. In Sec.~\ref{sec-results}, we obtain current and projected upper limits on the proposed cross sections and compare them to the corresponding limits from conventional heavy Higgs search channels. As part of the comparison we also obtain sensitivities to masses of heavy Higgses and new leptons in our reference model. Section \ref{sec-cncl} is devoted to our conclusions and outlooks. Details of the 2HDM we consider are explained in Appendix~\ref{appendix-model}.

\section{Search strategies} 
\label{sec-strategy}

In this section, we provide a detailed description of the analysis strategies we propose to obtain new constraints for heavy Higgses and new leptons based on the full LHC run2 data and projected expectations for the HL-LHC. As part of our analysis we recast the conventional searches for heavy Higgses based on decays to $\tau$ lepton pairs, as well as existing searches for leptonic signals in $2\ell + E_T^{\rm miss}$ and $3/4\ell$ final states initially motivated in supersymmetric and seesaw models. Importantly, our recast of the $2\ell + E_T^{\rm miss}$ and $3/4\ell$ searches can be generically applied to other BSM scenarios with the same event topology and final states, whereas our results for the heavy Higgs to di-$\tau$ channel are relevant for our reference model of the two Higgs doublet model type-II augmented by vectorlike leptons in the alignment limit.

For simplicity, events for SM background processes are taken from data available from existing results, giving the relevant references along the way.
Nevertheless, as we will see our approach provides powerful enough experimental sensitivities on our leptonic cascade processes, as will be shown in Sec.~\ref{sec-results}.
Although not implemented here, further investigation for multiple lepton resonances in the $3/4\ell$ channel would increase the sensitivities even more, as already demonstrated in another cascade process in Ref.~\cite{Dermisek:2016via}.

\subsection{Heavy neutral Higgs to $\tau \tau$}
\label{sec-HiggsSM}

\begin{figure}[t]
    \centering
    \includegraphics[height=0.8\textwidth]{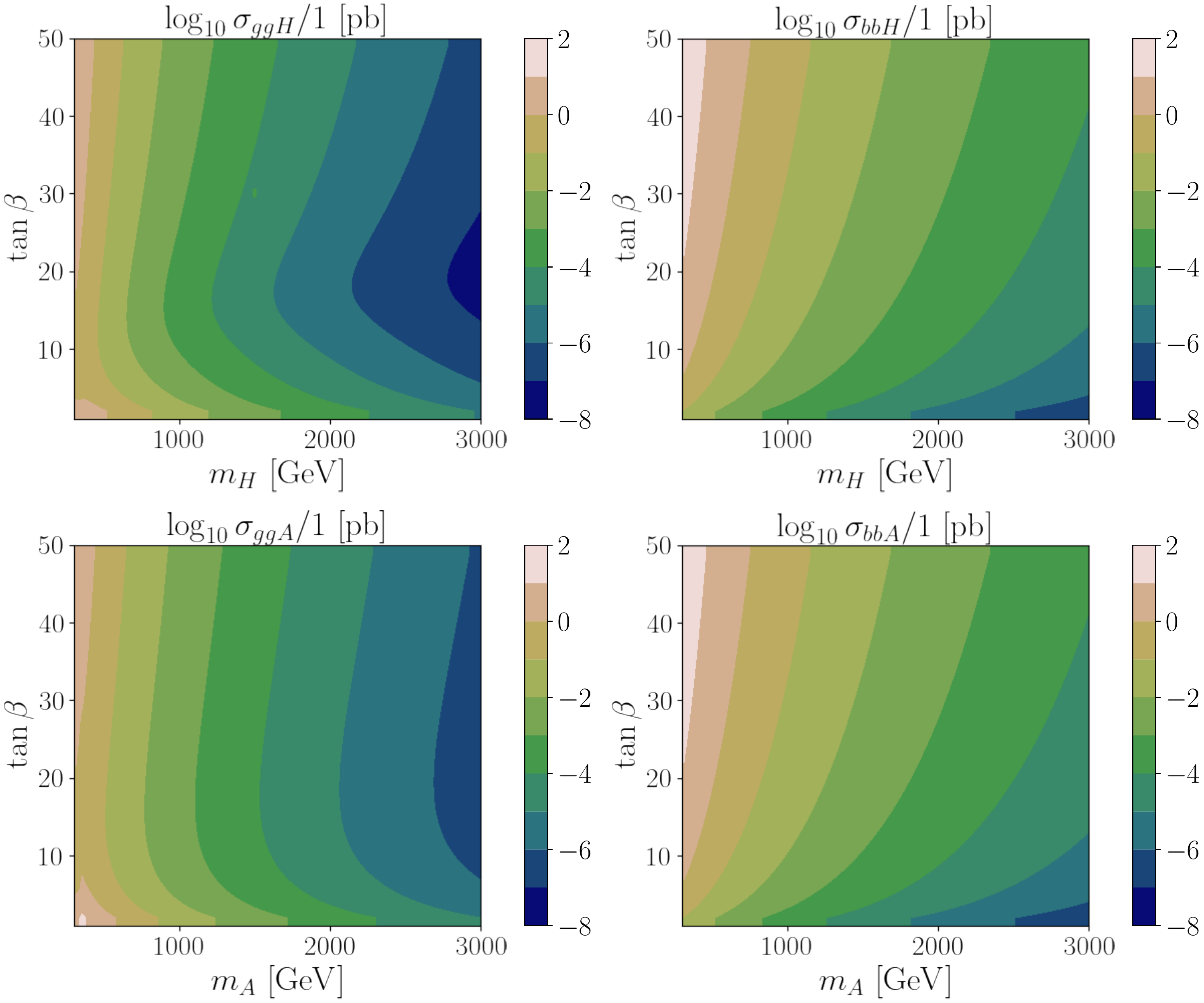} 
    \caption{
Production cross sections of the neutral Higgs bosons $H$ (upper) and $A$ (lower)
via the ggF (left) and bbH (right) processes at $\sqrt{s}=13~\TeV$. 
} 
    \label{fig:mHprod}
\end{figure}

Searches for heavy BSM Higgs bosons have been considered as important tasks in the LHC collaborations.
Among the conventional searches for a heavy neutral Higgs boson decaying to the SM fermions, we impose constraints from the $\Phi \to \tau \tau$ searches~\cite{CMS:2018rmh,ATLAS:2020zms} which provides the most stringent limits in a wide range of parameter space (for example, see Ref.~\cite{ATLAS:2021ayy}). 

The neutral Higgs bosons dominantly decay to the SM fermions in the third generation.  
In our numerical analysis, we calculated $H\to cc$, $bb$, $tt$, $\tau\tau$, $\gamma\gamma$, $gg$, $hh$ and $A\to cc$, $bb$, $tt$, $\tau\tau$, $\gamma\gamma$, $gg$, assuming the MSSM Higgs potential for $H\to hh$.
Although not pursued in this paper, our results could be weakened upon further consideration of the Higgs potential where BR($H \to hh$) is sizable. 
The branching fractions involving weak gauge bosons are vanishing in the alignment limit and we do not consider those decay modes.
Note, however, that our final states nevertheless mimic certain di-boson-like signals and hence can be potentially constrained using the corresponding searches proposed in~\cite{Dermisek:2015hue,Dermisek:2015vra,Dermisek:2015oja}.

We consider the limit on $\sigma \times \mathrm{Br}(\Phi \to \tau\tau)$ 
using the latest ATLAS data~\cite{ATLAS:2020zms} 
for gluon-gluon fusion ($gg\Phi$) and $b$-annihilation ($bb\Phi$).   
In our analysis, the production of the neutral Higgs bosons, $H$ and $A$, 
via these processes are calculated using \texttt{SuShi}~\cite{Harlander:2012pb,Harlander:2016hcx} and we further assume $m_H = m_A$ to avoid bounds from custodial symmetry breaking.
Values of the production cross sections at $\sqrt{s} = 13$ TeV are shown in Fig.~\ref{fig:mHprod}. 
For the constraints discussed later in our reference model, we use these values of production cross sections and consider a parameter set to be excluded if either
\begin{align}
\sum_{\Phi = H,A} \sigma_{gg\Phi} \times \mathrm{Br}\left(\Phi\to \tau\tau\right) 
\quad \mathrm{or} \quad 
\sum_{\Phi = H,A} \sigma_{bb\Phi} \times \mathrm{Br}\left(\Phi\to \tau\tau\right) 
\end{align}
is larger than the experimental bounds on $\sigma \times \br{\Phi}{\tau\tau}$, where $\br{\Phi}{\tau\tau}$ is calculated in context of the model. 
In estimating the future sensitivity at the HL-LHC, we assume the statistical uncertainty dominates and simply rescale the upper bound on the cross section by $\sqrt{R_\Lcal}$, where $R_\Lcal := \Lcal_\mathrm{run2}/\Lcal_\mathrm{HL}$,  
to estimate the future sensitivity at the HL-LHC, 
where the integrated luminosities are $\Lcal_\mathrm{run2} = 139~\fbi$ and $\Lcal_\mathrm{HL}= 3~\abi$.

\subsection{$2\ell+\met$ channel}

\begin{figure}[t]
    \centering
    \includegraphics[width=0.95\hsize]{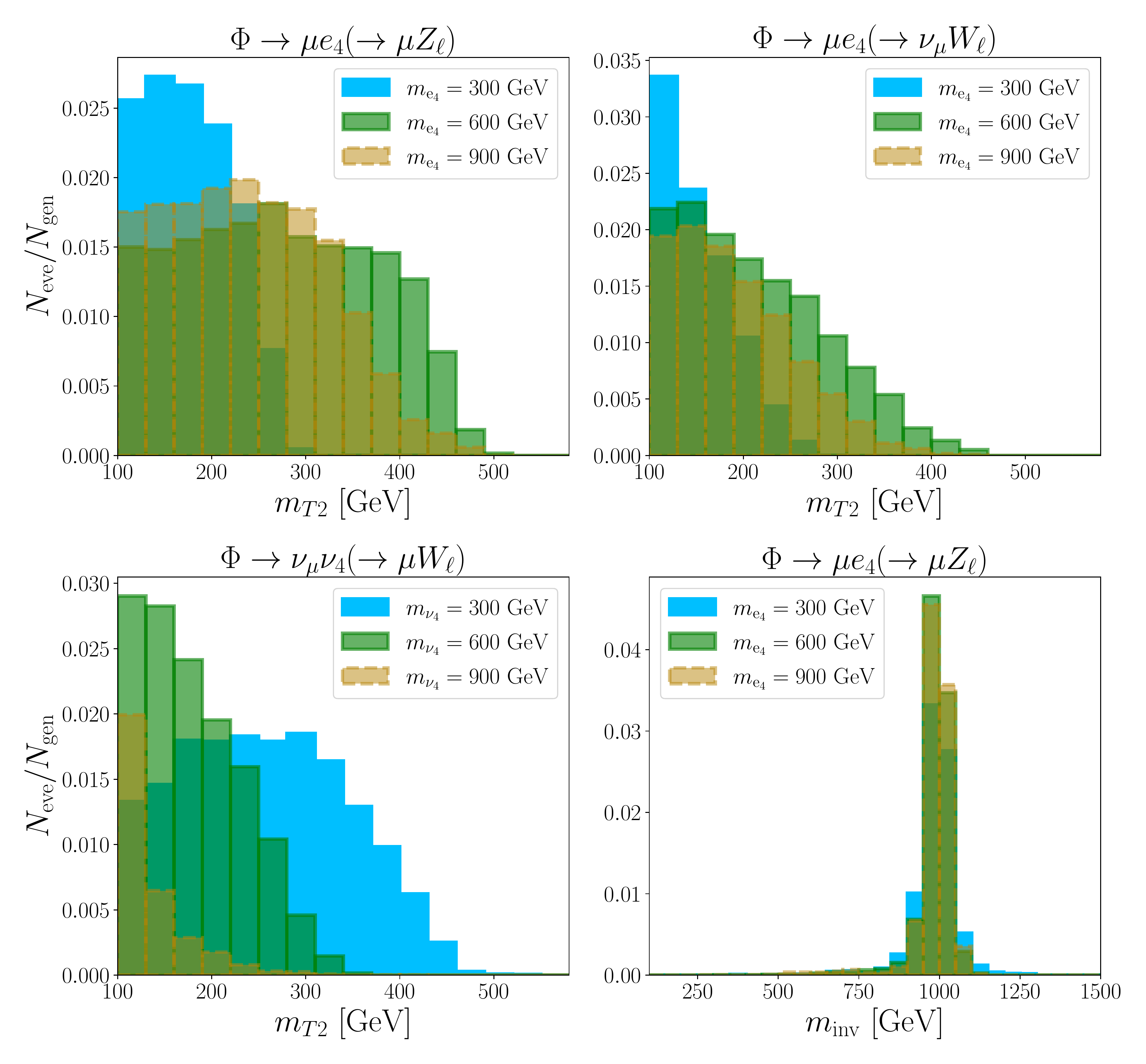}
    \caption{
$m_{T2}$ distributions of the Higgs cascade decays for $m_H = 1000~\GeV$ 
and vectorlike lepton masses 300, 600, and 900 GeV.
The left-lower panel shows the invariant mass of four leptons from the EZ decay.
} 
    \label{fig:mT2hist}
\end{figure}

All three leptonic cascade processes in Fig.~\ref{fig:Hlcasc} contribute to the $2\ell+\met$ channel.
We recast bounds for slepton and chargino masses in SUSY models from Ref.~\cite{ATLAS:2019lff} whose signal region includes a lepton pair which does not reconstruct the mass of a weak gauge boson.

Let us first explain how we obtain the sensitivities. 
We calculate the $95\%$ CL$_{\rm s}$ limits by requiring~\cite{CMS-NOTE-2011-005} 
\begin{align}
 \CL_s(\mu=1) := \frac{\CL_{s+b}}{\CL_{b}} = 
          \frac{1-F\left(\sqrt{q_\mu^\obs}\right)}{F\left(\sqrt{q^A_\mu}-\sqrt{q_\mu^\obs}\right)}  < 0.05, 
\end{align} 
where $F$ is the cummulative distribution function of the normal distribution.

Here, the asymptotic formula for the distribution of test statistic $q_\mu$, defined as
\begin{align}
 q_\mu = -2 \log \frac{L(\mu, \hat{\hat{b}})}{L(\hat{\mu}, \hat{b})}, 
\end{align} 
 is used~\cite{Cowan:2010js}, where the likelihood function is defined as:
\begin{align}
 L(\mu, b) = \prod_{i} \frac{\nu_i^{n_i}}{n_i!} e^{-\nu_i} \times 
                       \frac{1}{\sqrt{2\pi \sigma_i^2}} \mathrm{exp}
                        \left[ -\frac{(b_i-b_i^0)^2}{2\sigma_i^2}\right]\,. 
\end{align}
In the above definition, $\nu_i = \mu s_i + b_i$ with $s_i$, $b^0_i$, $\sigma_i$ being the number of signal events, the number of background events and the error of background in the $i$-th bin, respectively.  
Here, we assume that only background events have uncertainties and obey the Gaussian distribution. $(\hat{\mu}, \hat{b})$ are the values of $\mu$ and $\hat{b} = \{\hat{b}_i\}$ 
which maximize the likelihood, 
and $\hat{\hat{b}} = \{\hat{\hat{b}}_i\}$ is the value of $b_i$ for a given $\mu$.    
Note that $n_i$ corresponds to the observed data (central values of backgrounds) for $q_\mu^{\rm obs}$ ($q_\mu^A$).
For future limits, the significance for exclusion is given by~\cite{Cowan:2010js} 
\begin{align}
 Z_{\mathrm{excl}} \simeq \sqrt{q^{n=b}_{\mu=1}}\,, 
\quad n_i = b_i\,,   
\end{align}
while that for discovery is given by 
\begin{align}
 Z_{\mathrm{disc}} \simeq \sqrt{q^{n=s+b}_{\mu=0}}\,, 
\quad n_i = s_i + b_i\,.
\end{align}
For the future sensitivity with $3~\mathrm{ab}^{-1}$ data,  
the number of events (error) are rescaled by $R_{\mathcal{L}}$ ($\sqrt{R_{\mathcal{L}}}$).

In Ref.~\cite{ATLAS:2019lff}, the number of observed events, fitted SM backgrounds, and their uncertainties are listed in 4 categories of signal regions (SRs): 
\texttt{SR-SF-0J}, \texttt{SR-SF-1J}, \texttt{SR-DF-0J}, and \texttt{SR-DF-1J} , further binned by $m_{T2}$.~\footnote{
We use the values of $m_{T2}$ defined in Ref.~\cite{Lester:2014yga}. 
}
The results in the last bin $m_{T2} > 260~\GeV$ are the most relevant ones constraining our parameter choices of heavy new particles.
Here, SF (DF) means that the two leptons in the final state are the same (different) flavor. 

The number of signal events in each $m_{T2}$ bin is calculated as
\begin{align}
\label{eq-sifull}
 s_{i} = \Lcal \times \sum_{\Phi} 
           \sum_{P=\mathrm{gg},\mathrm{bb}}  
           \sum_{J=\mathrm{EZ},\mathrm{EW},\mathrm{NW}} 
          \sigma_{P\Phi} \times \mathrm{Br}_{\Phi, J}  \times  \eps^{J,P}_{i}, 
\end{align}
where $\Lcal$ is the integrated luminosity. 
Here, 
$\Phi$, $P$ and $J$ run over scalar fields, production process and decay modes, respectively.  
The cut acceptance times the detector efficiency, $\eps^{J,P}_i$, is a function of the scalar and new fermion masses and is the fraction of events that pass the cuts in a given signal region. 
From now on, we simply call the $\eps^{J,P}_i$ values the efficiencies.
These are calculated by means of Monte Carlo simulations. 
In our analysis, we generated events using \texttt{MadGraph5$\_$2$\_$8$\_$2}~\cite{Alwall:2014hca} based on a \texttt{UFO}~\cite{Degrande:2011ua} model file generated with \texttt{FeynRules$\_$2$\_$3$\_$43}~\cite{Alloul:2013bka,Christensen:2008py}. 
In order to boost up the speed of the simulation, decays of the neutral heavy Higgs boson are handled using \texttt{MadSpin}~\cite{Artoisenet:2012st}.  
Showering and hadronization are controlled by \texttt{PYTHIA8}~\cite{Sjostrand:2007gs}, 
and detector simulation by \texttt{Delphes3.4.2}~\cite{deFavereau:2013fsa}. 
We used the default ATLAS card for the fast detector simulation. 
Jets are reconstructed  using the anti-$k_T$ algorithm~\cite{Cacciari:2008gp,Cacciari:2011ma} with $\Delta R = 0.4$. 
For the decay modes $J=$ EZ, EW and NW, which are depicted from the left-to-right panels respectively in Fig.~\ref{fig:Hlcasc}, we simulate the processes 
\begin{align}
  pp \to \Phi + \jets,&\ \quad \Phi \to \mu e_4~(\to \mu Z_\ell ) \,,   \\
  pp \to \Phi + \jets,&\ \quad \Phi \to \mu e_4~(\to \nu_\mu W_\ell ) \,,   \\
  pp \to \Phi + \jets,&\ \quad \Phi \to \nu_\mu \nu_4~(\to \mu W_\ell ) \, ,   
\end{align}
where $V_\ell$ $(V=Z, W)$ indicates leptonic decays including $\tau$ leptons.~\footnote{Note that we include the contributions of the leptonic taus to the final states of our interest.}
Production of $\Phi$ is simulated via the CP-even Higgs boson production from gluon-gluon fusion in the 4 flavor scheme with the 5-dimensional effective interaction and the $b$-annihilation process in the 5 flavor scheme separately.  
In the simulation up to two additional partons are included and these are matched to the showered events by the MLM matching~\cite{Caravaglios:1998yr} with $\texttt{xqcut} = m_H/10$.

The $m_{T2}$ distributions from our leptonic cascade decays are shown in Fig.~\ref{fig:mT2hist}.
We see that $m_{T2}$ has a relatively sharp edge for lighter $e_4$ masses in the EZ and EW decays where all the $E_T^{\rm miss}$ comes wholly from the decay products of $e_4$, while the behavior is the other way around in the NW decay.
Thus, the $m_{T2}$ cuts can also be efficient in a wide range of parameter space in discriminating the leptonic cascade decays from the SM backgrounds, in particular for $WW$ and $t \bar t$ events whose $m_{T2}$ distributions are mostly shifted below $\sim 160$ GeV in all the SRs (See Fig. 5 of Ref.~\cite{ATLAS:2019lff}.).

Among the $m_{T2}$ bins, the highest bins with $m_{T2} > 260~\GeV$ are the most important to discriminate the signals from SM backgrounds. 
From the $gg\Phi$ ($bb\Phi$) production followed by the EZ decay, 
the efficiencies to the $m_{T2} > 260~\GeV$ bins are roughly $1$-$5\%$ ($5$-$10\%$) in SF-0J, 
and $5$-$10\%$ ($5$-$10\%$) in SF-1J, where $n$J ($n=0,1$) is the number of jets in the SRs.

\subsection{$3/4\ell$ channel}

Leptonic cascade decays involving the $Z$ boson, i.e., the left-most panel of Fig.~\ref{fig:Hlcasc} (the EZ mode), can produce fully leptonic final states when including $Z \to \ell \ell$, which yields clear multiple leptonic resonance signals.
The process is hence constrained by the searches for $3/4 \ell$ events depending on the cuts for the softest lepton.
For this analysis, we recast the search results of Ref.~\cite{ATLAS:2021wob}.

Among the SRs defined in Ref.~\cite{ATLAS:2021wob}, the $4\ell$ SRs with one $Z$-pair and small $\met$ have the strongest sensitivities, although there could be sub-dominant contributions from the other SRs due to the mis-reconstruction of leptons.  
There are two bins with $m_\inv < 400~\GeV$ and $m_\inv > 400~\GeV$, where $m_\inv$ is an invariant mass of the four leptons, which corresponds to the mass of the neutral Higgs $\Phi$ in our reference model. 
This is shown in the lower-right panel of Fig.~\ref{fig:mT2hist}, where the events are clearly clustered around $m_\Phi = 1$ TeV. 
Since all SRs in Ref.~\cite{ATLAS:2021wob} are mutually exclusive, we combine them all in the same way as in the $2\ell + \met$ search.
The efficiencies of the EZ decay in the SR requiring four leptons, with one $Z$-like lepton pair with $E_T^\mathrm{miss}<50~\GeV$ and $m_\mathrm{inv} > 400~\GeV$ are roughly $5$-$10\%$ for both $gg\Phi$ and $bb\Phi$ productions.

\section{Current constraints and future prospects} 
\label{sec-results}

In the following subsections, we present the current limits based on the LHC run2 data with $139~\fbi$ integrated luminosity and future prospects at the HL-LHC with luminosity $3~\abi$ based on the analysis strategies explained in the previous section.
We emphasize that our analysis results can be generically applied to other BSM scenarios which share the same kinematic topology and final states; thereby model-independent upper limits and prospects on the total cross sections will be shown first.
For scenarios with a resonant particle production like a Higgs boson, i.e., via gluon-gluon fusion and/or $b$-annihilation, our results on the branching ratios can be readily applied.
Finally, we provide the model-dependent constraints and expected sensitivities for the vectorlike lepton mass and heavy neutral Higgs mass in a 2HDM type-II.

\subsection{Model independent limits on cross sections}

\begin{figure}[th]
  \centering
  \includegraphics[width=\fwidth]{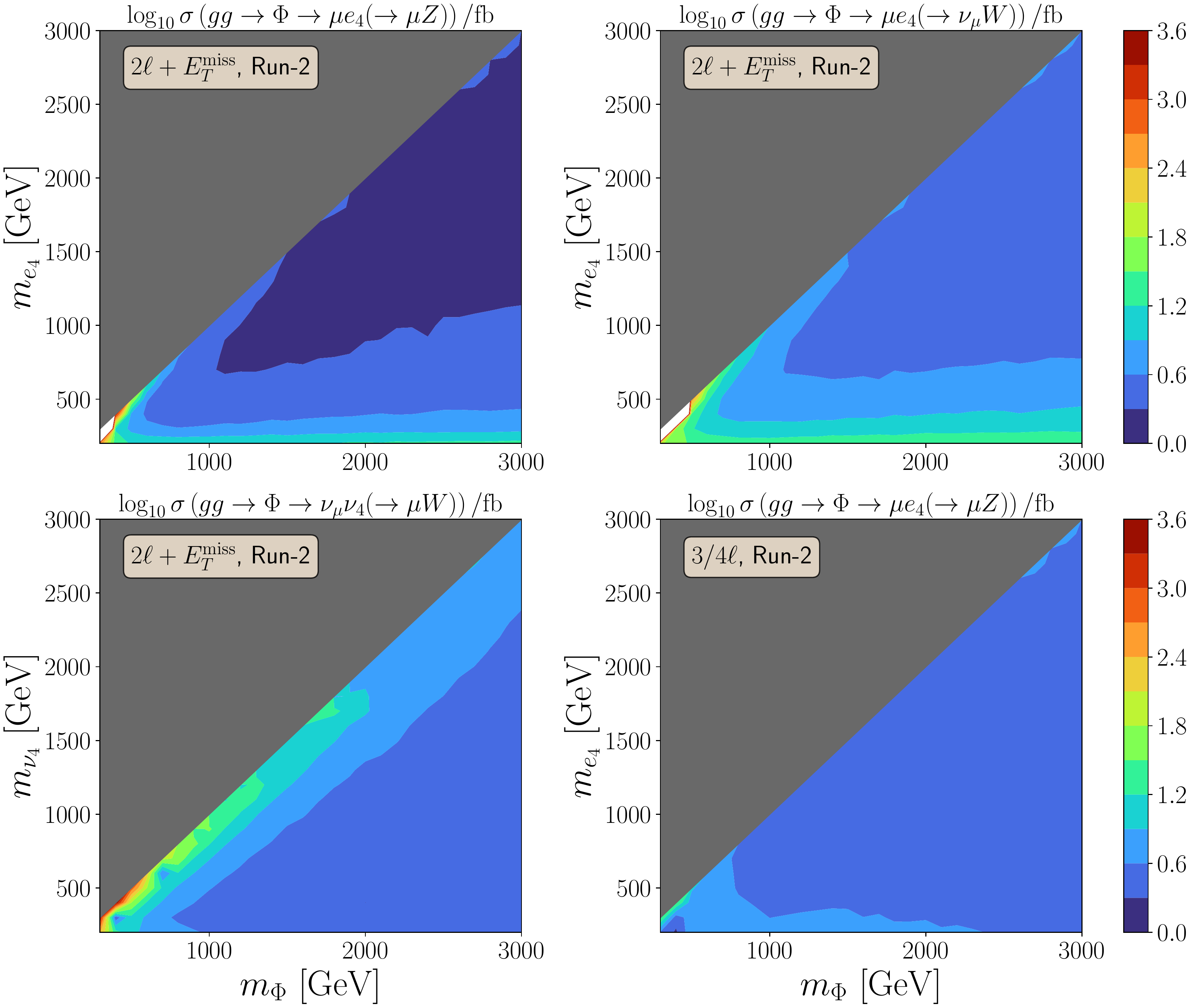}
    \caption{ \label{fig-xlim-ggR2}
Current upper bounds on $\sigma(gg\to\Phi\to\ell \ell_4(\to\ell^\prime V))$ from our recast of the $2\ell + \met$ and $3/4\ell$ searches.  
See the texts for the details.
}
\end{figure}

\begin{figure}[th]
    \centering
    \includegraphics[width=\fwidth]{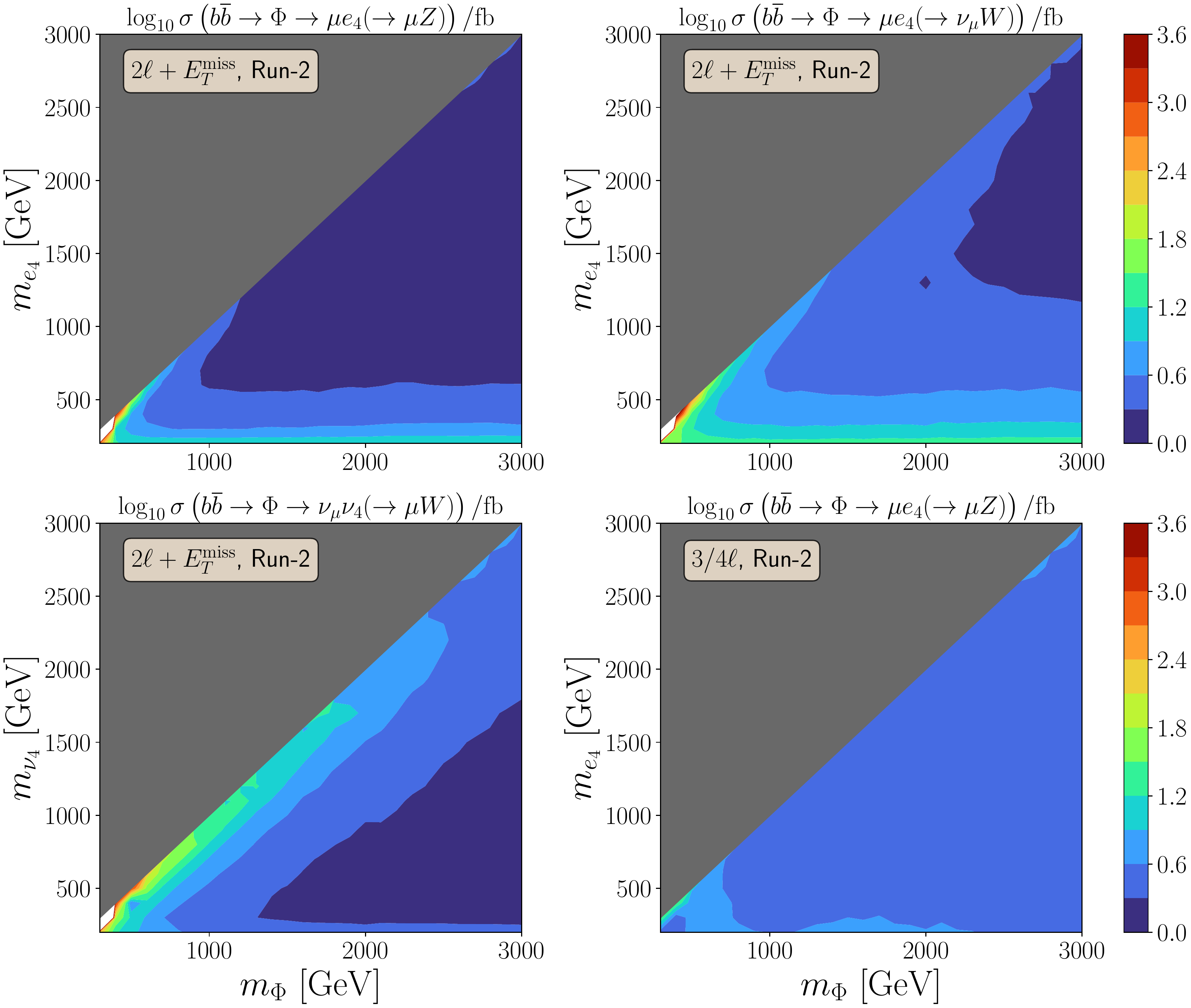}
    \caption{ \label{fig-xlim-bbR2}
Current upper bounds on $\sigma(b\ol{b}\to\Phi\to\ell \ell_4(\to\ell^\prime V))$.  
} 
\end{figure}

\begin{figure}[th]
    \centering
     \includegraphics[width=\fwidth]{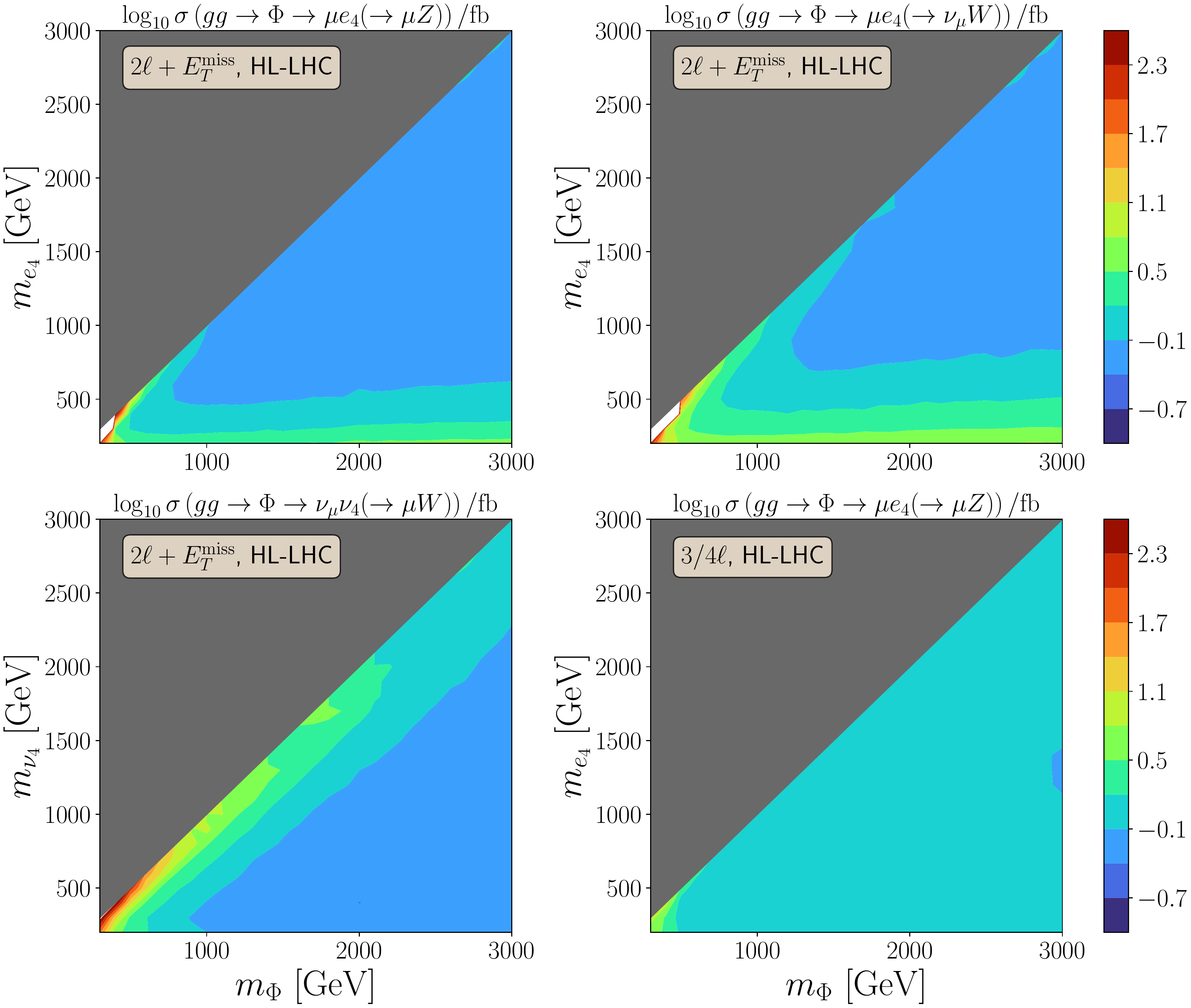}
    \caption{ \label{fig-xlim-ggHL}
Projected upper bounds on $\sigma(gg\to\Phi\to\ell \ell_4(\to\ell^\prime V))$ at the HL-LHC.
See the texts for the details.
} 
\end{figure}

\begin{figure}[th]
    \centering
    \includegraphics[width=\fwidth]{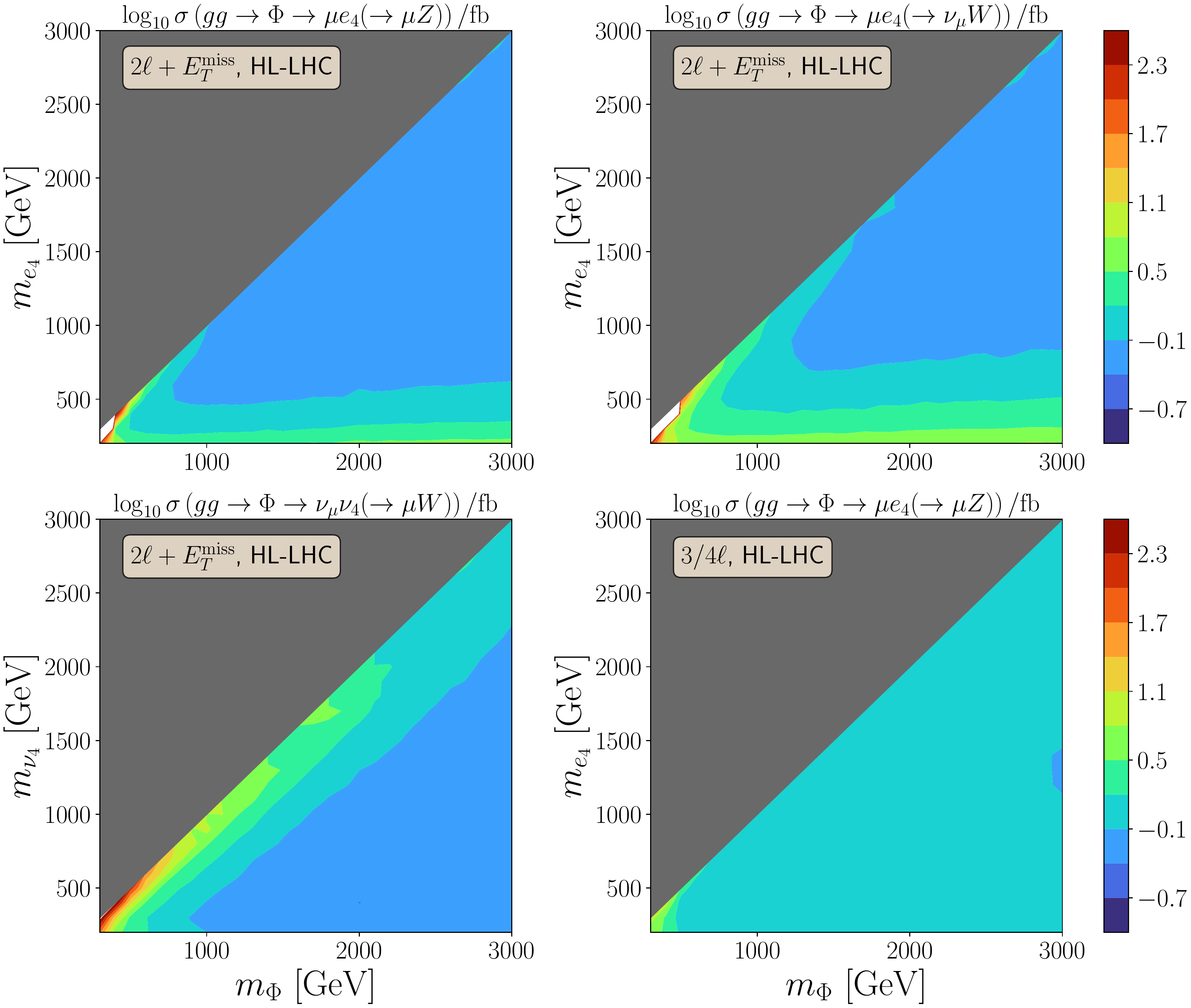}%
    \caption{\label{fig-xlim-bbHL}
Projected upper bounds on $\sigma(b\ol{b}\to\Phi\to\ell \ell_4(\to\ell^\prime V))$ at the HL-LHC. 
} 
\end{figure}

Figures~\ref{fig-xlim-ggR2} (\ref{fig-xlim-ggHL}) 
and~\ref{fig-xlim-bbR2} (\ref{fig-xlim-bbHL}) show the current (future) upper limits on
\begin{align}
  \sigma\left( g g \to \Phi+\jets\right)&\times 
  \mathrm{Br}\left(\Phi \to \ell \ell_4 (\to V \ell^\prime) \right)\,, 
\end{align}
and 
\begin{align}
  \sigma\left( b \ol{b} \to \Phi+\jets \right)&\times 
  \mathrm{Br}\left(\Phi\to \ell \ell_4 (\to V \ell^\prime) \right)\,, 
\end{align}
respectively~\footnote{
In the small white region at $m_H \sim m_{e_4}\lesssim 400~\GeV$, 
the efficiencies in all of the signal regions of the $2\ell+\met$ search 
are zero according to our simulation.
}. 
Here, $(\ell_4, \ell, V, \ell^\prime) = (e_4, \mu, Z, \mu),~(e_4,\mu,W,\nu_\mu),~(\nu_4,\nu_\mu,W,\mu)$ refers to the EZ, EW and NW decays, respectively. 
Tables of the values used in these figures are attached in supplemental material.
Note that the production and decays of $\Phi$ are handled separately by \texttt{MadGraph} and \texttt{MadSpin} respectively to boost up the speed of our event generations; thereby our method does not technically include the cases of an off-shell production of $\Phi$ or the breakdown of the narrow width approximation, which might provide different values of the cut acceptances.
The number of signal events in the $i$-th $m_{T2}$ bin is calculated by
\begin{align}
s_i = \Lcal \times \sigma_{P\Phi} \times \mathrm{Br}_{\Phi,J} \times \eps^{J,P}_i\;,
\end{align}
without summations for each choice of $\Phi, J, P$, and we obtain upper limits on $\sigma_{P\Phi} \times \mathrm{Br}_{\Phi,J}$.

The current limits are $\order{1\mathrm{-}10}~\mathrm{fb}$ for $m_\Phi \gtrsim 1~\TeV$ for the EZ and EW decay modes. 
The search is more sensitive for larger $m_{e_4}$ due to the increasing number of events passing the $m_{T2}$ cut for larger lepton masses, as displayed in Fig.~\ref{fig:mT2hist}. 
The limits on the NW mode is of the same order, but the search is more sensitive to the smaller $m_{\nu_4}$ since the $m_{T2}$ distribution shows the opposite behavior with the vectorlike lepton mass.
In both cases the sensitivities increase with increasing $m_H$ as a larger Higgs mass tends to produce events with larger values of $m_{T2}$. 
The bottom-right panels in Figs.~\ref{fig-xlim-ggR2} - \ref{fig-xlim-bbHL} show limits on the EZ mode, where $Z \to \ell \ell$, which is constrained by the $3/4\ell$ search channel.
The limit is almost independent with respect to the vectorlike lepton mass because $m_\mathrm{inv}$ is determined by $m_\Phi$, so the only requirement is that four leptons should be reconstructed with a sufficient $p_T$. 
At the HL-LHC, we expect the experimental sensitivities to the cross sections would be increased by about $\sqrt{R_\Lcal} \simeq 4.7$ and hence they can be improved to be $\order{0.2-1}~\mathrm{fb}$.

\subsection{Constraints on the branching fractions}

\begin{figure}[hp]
    \centering
    \includegraphics[width=\fwidth]{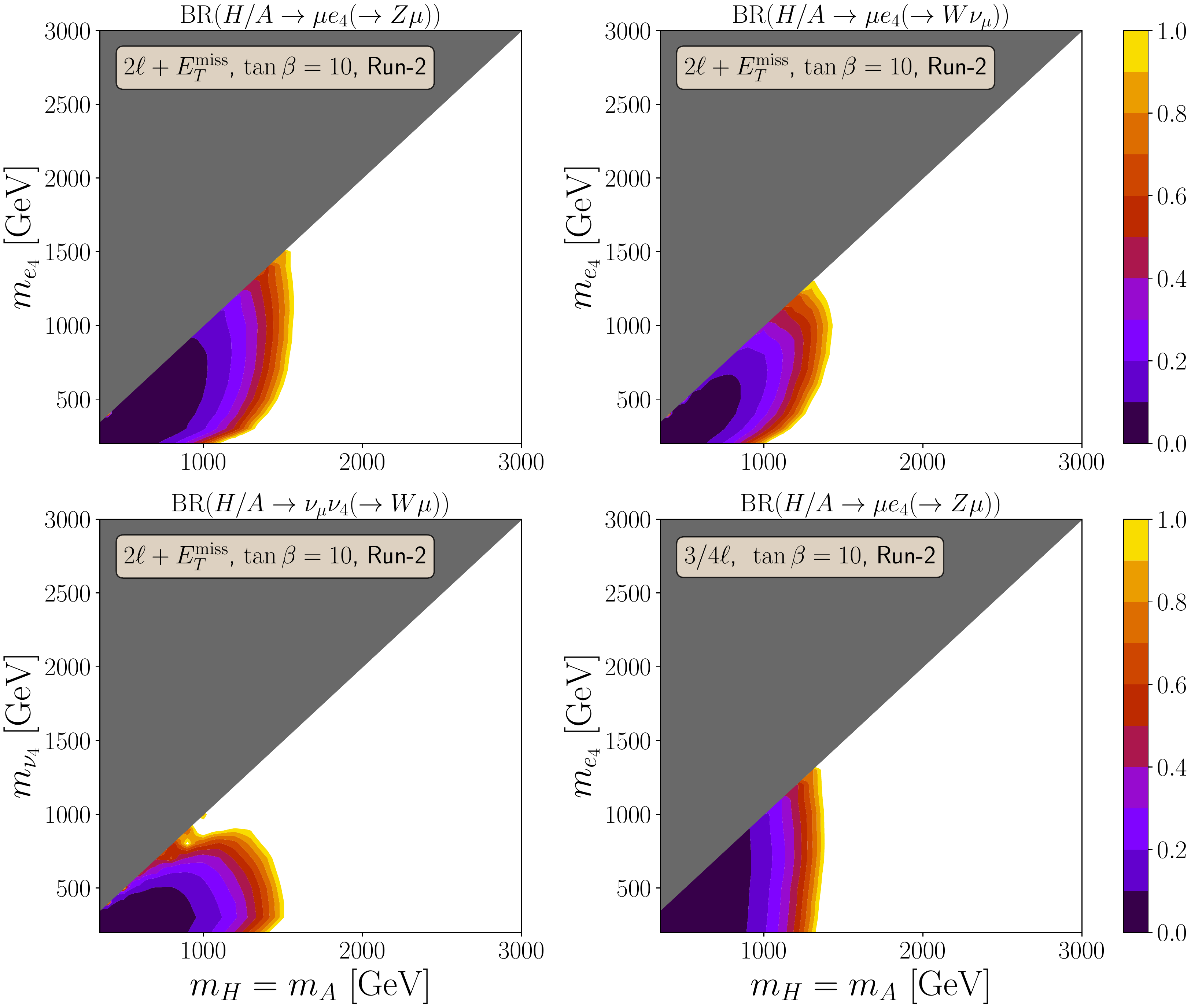}
    \caption{
Current upper bounds on the branching fractions when $\tan\beta = 10$.  
} 
    \label{fig:BRlim-tb10}
\vspace{0.5cm}
    \centering
     \includegraphics[width=\fwidth]{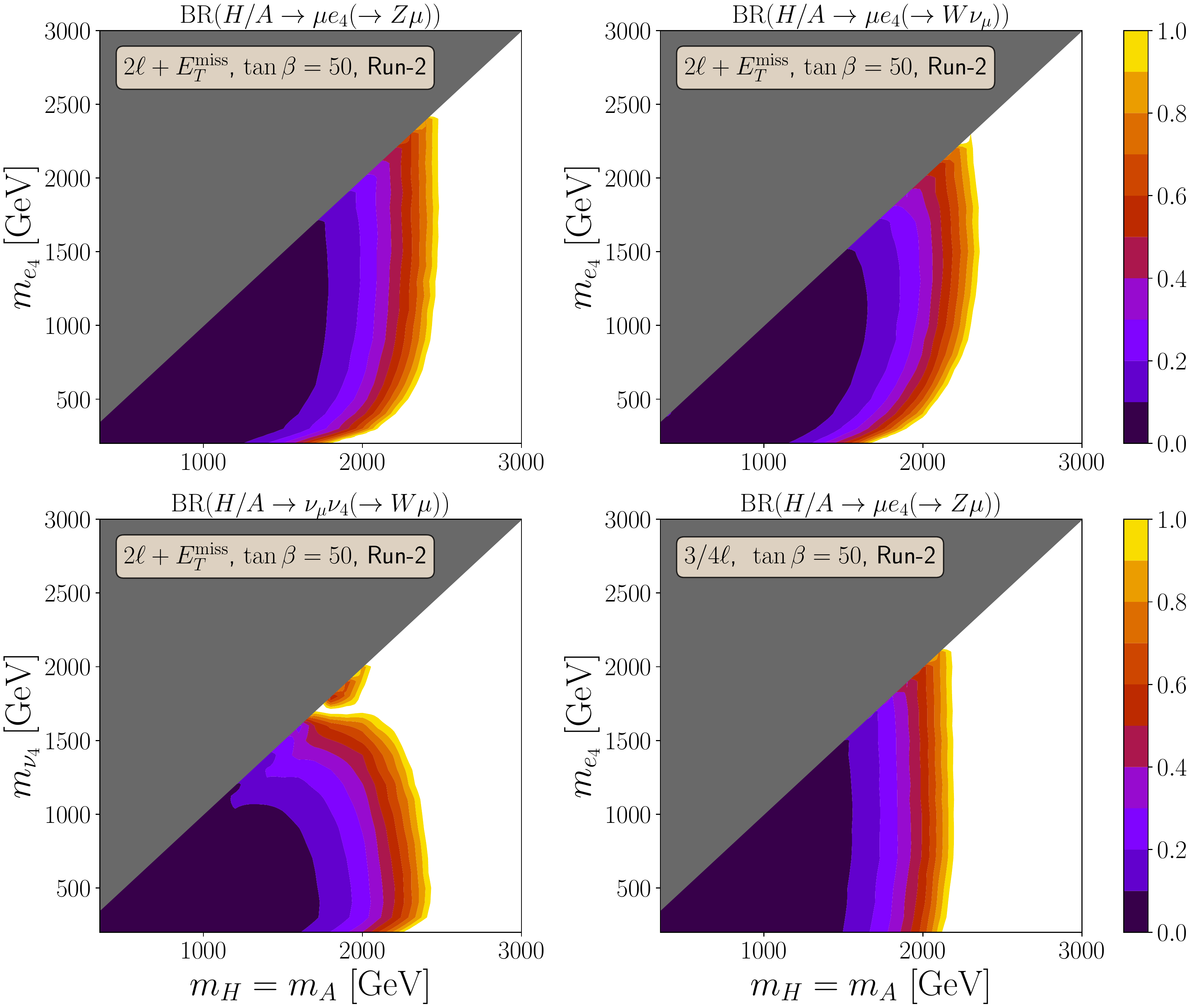}
    \caption{    \label{fig:BRlim-tb50}
Current upper bounds on the branching fractions when $\tan\beta = 50$.  
} 
\end{figure}

\begin{figure}[hp]
    \centering
     \includegraphics[width=\fwidth]{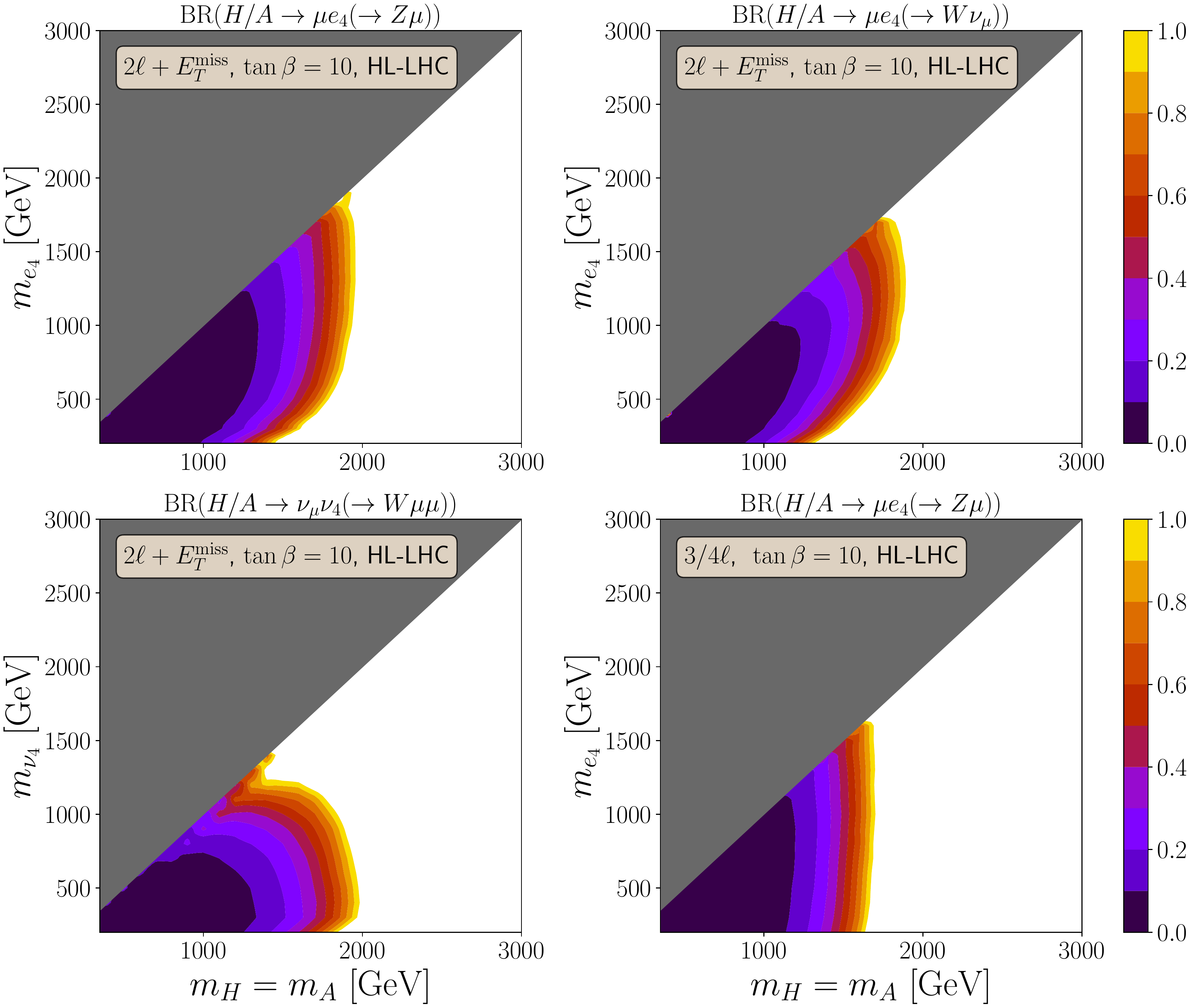}
    \caption{
Future upper bounds on the branching fractions when $\tan\beta = 10$.  
} 
    \label{fig:BRexp-tb10}
\vspace{0.5cm}
    \centering
     \includegraphics[width=\fwidth]{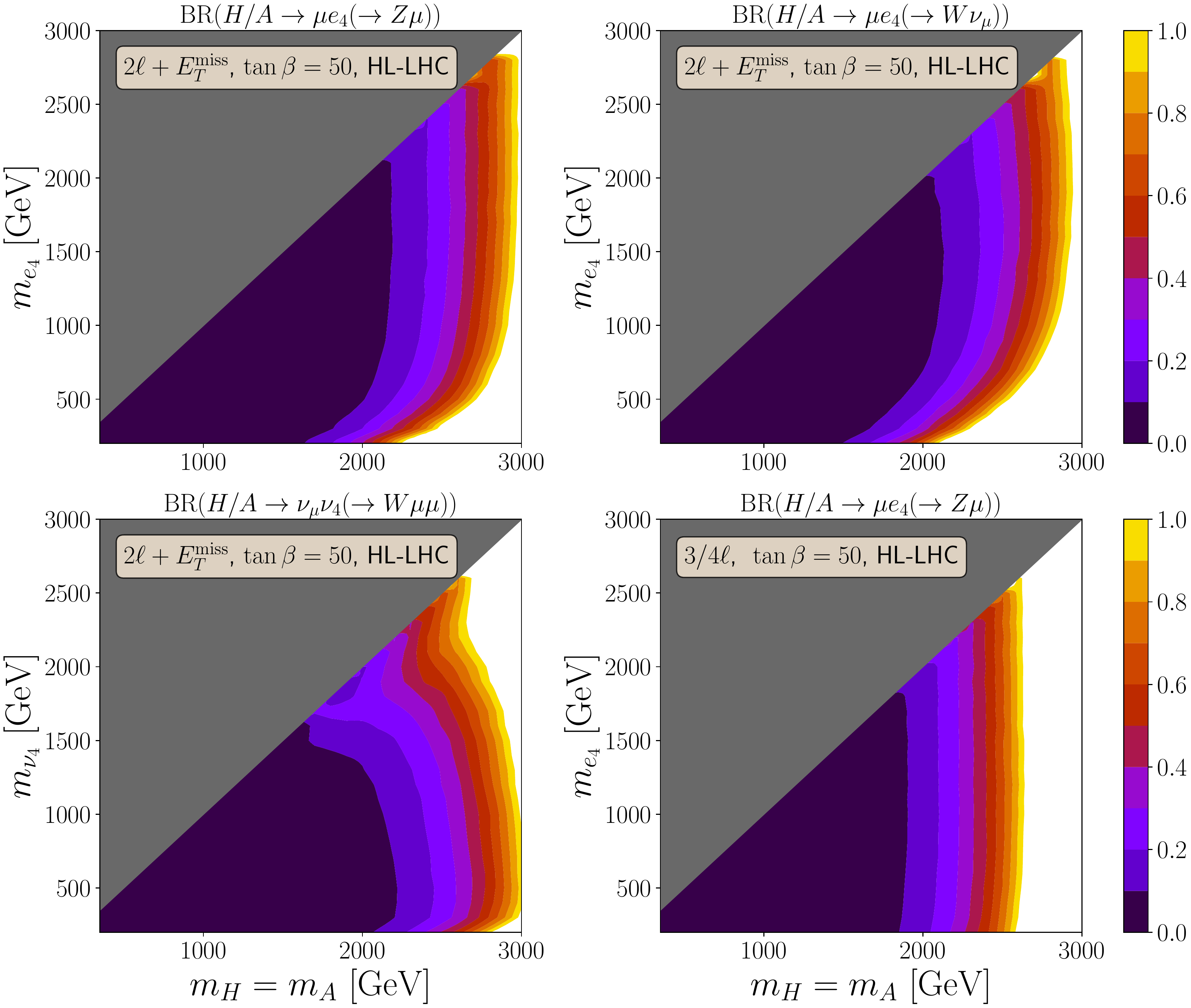}
    \caption{
Future upper bounds on the branching fractions when $\tan\beta = 50$.  
} 
    \label{fig:BRexp-tb50}
\end{figure}

We now discuss the constraints on the branching fractions of the neutral heavy Higgs bosons. 
As stated earlier, our bounds on the branching fractions can be applied to a wide range of other scenarios with a resonant particle production like a Higgs boson.
Figures~\ref{fig:BRlim-tb10} and~\ref{fig:BRlim-tb50} (\ref{fig:BRexp-tb10} and~\ref{fig:BRexp-tb50}) show the upper bounds on the branching fractions from the current (future) data assuming a production cross section for $\Phi$ as in a 2HDM type-II for $\tan\beta = 10$ and $50$, respectively.  
The number of signal events is calculated in a combined way by 
\begin{align}
s_i = \Lcal \times \sum_{P,\Phi} \sigma_{P,\Phi} \times \mathrm{Br}_{\Phi,J} \times \eps^{J,P}_i,     
\end{align}
where $\Phi = H,A$ and $P=gg,bb$, 
but there is no summation over the decay modes $J = \mathrm{EZ},~\mathrm{EW},~\mathrm{NW}$.
The production cross sections from gluon fusion and $b$-annihilation are calculated using 
\texttt{SuShi}~\cite{Harlander:2012pb,Harlander:2016hcx}. 
In these figures, we neglect the difference in the branching fractions between the CP-even and CP-odd Higgs bosons, which is expected to be small for $m_t^2/m_H^2 \ll 1$, i.e., we set $\mathrm{Br}_{H} = \mathrm{Br}_A$. 
We assume that only one of the three decay modes contributes to the signal regions to extract the corresponding limit. 
Here, we also show the limits on the NW decay mode for completeness, 
although this branching fraction for a mostly singlet-like $\nu_4$ is constrained to be below $\sim 5$ \% (0.1 \%) for $m_H \lesssim 340$ GeV (above 340 GeV) in our reference model. This decay is constrained mainly by the $\Phi \to \tau \tau$ search results but also from electroweak precision measurements (see the discussion in Appendix~\ref{appendix-model}.).
The limits obtained by combining all decay modes are presented in the next section. 

The $2\ell + \met$ search excludes the Higgs boson mass up to about $1.3~(2.1)~\TeV$ for $m_{e_4}\gtrsim 500~\GeV$ and $\tan\beta = 10~(50)$ if the branching fraction is 50\%.  
The sensitivity of this search becomes weaker for smaller $m_{e_4}$, because the $m_{T2}$ distribution drops off around $m_{e_4}$ as shown in the top panels of Fig.~\ref{fig:mT2hist}, 
and hence cannot pass the $m_{T2}$ cut.   
The limits for the EZ decay mode are tighter than those for the EW decay mode because of the broader $m_{T2}$ distribution. 
For the NW decay, although the branching fraction cannot be sizable in our reference model, a similar range of Higgs masses can be probed unless the mass difference is small, implying a smaller $m_{T2}$. 
At the HL-LHC with $3~\abi$ data, the limits will be strengthened to about $1.8~(2.8)~\TeV$ 
for $\tan\beta = 10~(50)$ and 50\% branching fraction.

The limits on the branching fraction of the EZ decay mode from the $3/4\ell$ search
are shown in the lower-right panels of Figs.~\ref{fig:BRlim-tb10},~\ref{fig:BRexp-tb10},~\ref{fig:BRlim-tb50} and~\ref{fig:BRexp-tb50}.    
The current (future) sensitivities on $m_H$ with the 50\% total branching fraction are about 1.2 and 1.5~TeV (1.9 and 2.4~TeV) for $\tan\beta = 10$ and $50$, respectively. 
These are nearly independent of the masses of the vectorlike leptons because the key kinematic cut is given by $m_\inv = m_\Phi$.

\afterpage{\clearpage}

\subsection{Model dependent constraints} 
\label{sec-cmodel}

Finally, we apply our results to the reference model: a 2HDM type-II augmented by vectorlike leptons.
The details of our reference model are described in Refs.~\cite{Dermisek:2015oja,Dermisek:2015hue} but we simply show the Lagrangian and explain the field contents briefly to aid with our discussion.
The most general Lagrangian of Yukawa interactions and mass terms relevant for our processes include:
\begin{align}\nonumber
{\cal L} \supset \;  & - y_{\mu} \bar \mu_{L}  \mu_{R} H_d - \lambda_E \bar \mu_{L}  E_{R} H_d  -  \lambda_L \bar L_{L} \mu_{R} H_d -  \lambda \bar L_{L}  E_{R} H_d - \bar \lambda H_d^\dagger \bar E_{L}  L_{R} \nonumber \\
& - \kappa_N  \bar \mu_L  N_R H_u - \kappa  \bar{L}_L  N_R H_u - \bar \kappa H_u^\dagger \bar{N}_L L_R \nonumber \\
&  - m_L \bar L_L L_R - m_E \bar E_L E_R - m_N \bar N_L N_R + {\rm h.c.}\,,
\label{eq:lagrangian}
\end{align}
where the first term is the Yukawa interaction of the down-type Higgs boson $H_d$ with the SM muon, followed by the Yukawa interactions with the vectorlike lepton isodoublet $L_{L,R} = (L^0_{L,R}, L^-_{L,R})$ and the charged isosinglet $E_{L,R}$ which have the same quantum numbers as SM leptons (denoted by various $\lambda$'s).
The second line denotes the Yukawa interactions between the up-type Higgs $H_u$ and the neutral vectorlike leptons including a SM singlet lepton $N_{L,R}$ (denoted by $\kappa$s).
The final line denotes the vectorlike mass terms of the new leptons. In order to avoid strong bounds in the Higgs sector, we take the alignment limit~\cite{Gunion:2002zf,Craig:2013hca,Carena:2013ooa,Haber:2013mia}, i.e., $\alpha = \beta -\pi/2$ where $\alpha$ is the neutral Higgs mixing angle and $\tan\beta = \vev{H_u}/\vev{H_d}$ is the ratio of the vacuum expectation values of the two Higgs doublets.
More details on the model, the field contents, mass mixing, and the interactions among the mass eigenstates are explained in Ref.~\cite{Dermisek:2015oja}.

In our numerical analysis, we scan the parameters to vary in the ranges: 
\begin{align}
  m_H,~m_L,~m_E,~ m_N \in&\ [300,~3000]~\GeV\,, \\ 
 \lambda_L, \lambda_E, \la,~\ol{\la},~\ka,~\ol{\ka} \in&\ [-c_\mathrm{max},c_\mathrm{max}]\,, \\ 
  \tan\beta \in&\ [1,50]\,, 
\end{align}
where we consider $c_{\rm max} = 1$ and $3.5\sim \sqrt{4\pi}$, 
with the latter being  motivated by the upper limit of couplings near the weak scale from perturbativity.
Since there are many parameters in the Lagrangian above, we consider an optimized parameter scan strategy; two representative cases maximizing BR($H \to e_4 \mu$) are picked to focus on emphasizing the current and future sensitivities of our processes:
\begin{enumerate}
 \item {light-L: $m_L < m_H < m_E, m_N$, $\ka_N = 0$,
 $\la_L > 0.5$}\,,
 \item {light-E: $m_E < m_H < m_L, m_N$, $\ka_N = 0$,
 $\la_E > 0.5$\,,
 }
\end{enumerate}
The ``light-L'' denotes the case where the lightest new leptons $e_4, \nu_4$ are almost isodoublet, i.e., $(e_4^-, \nu_4)  \sim (L^-, L^0)$.
The ``light-E'' denotes the case where $e_4$ is almost isosinglet, i.e., $e_4 \sim E$.
In order to focus on our leptonic cascade processes, we require the other vectorlike leptons to be heavier than the neutral Higgses, $H/A$.
Note that these two representative cases correspond to the simple scenarios where the branching ratios of $e_4$ typically follow the pattern expected by the Goldstone boson equivalence theorem, i.e., BR($e_4 \to W \mu$):BR($e_4 \to Z \mu$):BR($e_4 \to h \mu$) = 2:1:1 (for isosinglet) and 0:1:1 (for isodoublet), and the approximation $(\lambda_L, \lambda_E, \lambda, \bar \lambda) v / (m_L, m_E) \ll 1$ is  valid in most of the parameter space; the couplings among the mass eigenstates are easily expressed with the Lagrangian parameters as in Ref.~\cite{Dermisek:2013gta,Dermisek:2015hue}.
As pointed out in Ref.~\cite{Dermisek:2019heo} for the vectorlike quarks, general vectorlike lepton scenarios can include the possibilities of small $\lambda_L$/$\lambda_E$ as well as sizable mixing between the isodoublet and isosinglet, which allow all the values between 0 and 1 for the $e_4$ branching ratios.

In the parameter scan, we do not study the case in which $\nu_4\sim N$ when  
$\Phi \to \nu_4 \nu_\mu \to W \mu \nu_\mu$ can dominate for $\ka_N > 0.5$.
This is because, in the limit where $(\lambda, \bar \lambda, \lambda_E, \lambda_L, \kappa, \bar \kappa,  \kappa_N)v / (m_N, m_L) \ll 1$ is valid, large values of $\kappa_N$ maximizing the decay width $\Phi \to \nu_4 \nu_\mu$ are strongly constrained by the electroweak precision measurements, especially from the Fermi constant $G_F$ at large $\tan\beta$.~\footnote{Although we use the constraints from the Particle Data Group~\cite{Zyla:2020zbs}, it is worth noting that the recent measurement of $M_W$ from the CDF collaboration at Tevatron~\cite{CDF:2022hxs} claims a central value in tension with respect to other experiments as well as an increase in precision, which would give stronger bounds.} 
Moreover, the values of BR($\Phi \to \nu_4 \nu_\mu$) are limited by the competing decay modes, $\Phi \to \tau^+ \tau^-$ and $\Phi \to t \bar t$, and hence we find the total values of BR($\Phi \to \nu_4 \nu_\mu \to W \mu \nu_\mu$) are preferred to be smaller than 4\%.
See Appendix~\ref{appendix-model} for more details.

In the light-L (light-E) case, we scan the absolute value of the Yukawa coupling constant $\la_L$ ($\la_E$) in the $[0.5, c_\mathrm{max}]$ range, so that the branching fraction BR($\Phi \to e_4 \mu$) covers the full range of possible values. 
For each point, we calculate the contributions to the electroweak precision observables (EWPOs), cross sections to the di-tau channels,  
$\CL_s$, $Z_{\excl}$, and $Z_{\disc}$. 

Note that the heavy Higgs $\Phi$ can decay into the vectorlike lepton pair, e.g., $\Phi \to e_4 e_4$, but the decay width is suppressed by $(\lambda_L, \lambda_E) v_u / (m_L, m_E)$ compared to that of $\Phi \to e_4 \mu$ in our simple scenarios ``light-L" and ``light-E".
Nevertheless, in general scenarios, such a double $e_4$ production can be sizable and provide another interesting signature, which is beyond the scope of this paper. Recall also that the vectorlike leptons can be pair produced through gauge boson interactions which can be subjected to robust bounds, as in Ref.~\cite{Dermisek:2014qca}.
The most recent analysis by ATLAS in Ref.~\cite{ATLAS:2020wop} shows that a nominal bound of $m_{e_4, \nu_4} \gtrsim 800$ GeV can be obtained assuming our total cross section of $p p \to e_4 \nu_4 \to W W \nu_\mu \mu$ is similar to what is expected in the type-III seesaw model.
We do not include this bound (after rescaling the production cross section) in our parameter scan but we emphasize that some light $e_4$ region can be constrained further by this complementary search.

\begin{figure}[p]
    \centering
    \begin{minipage}[c]{0.48\hsize}
    \centering 
    \includegraphics[height=70mm]{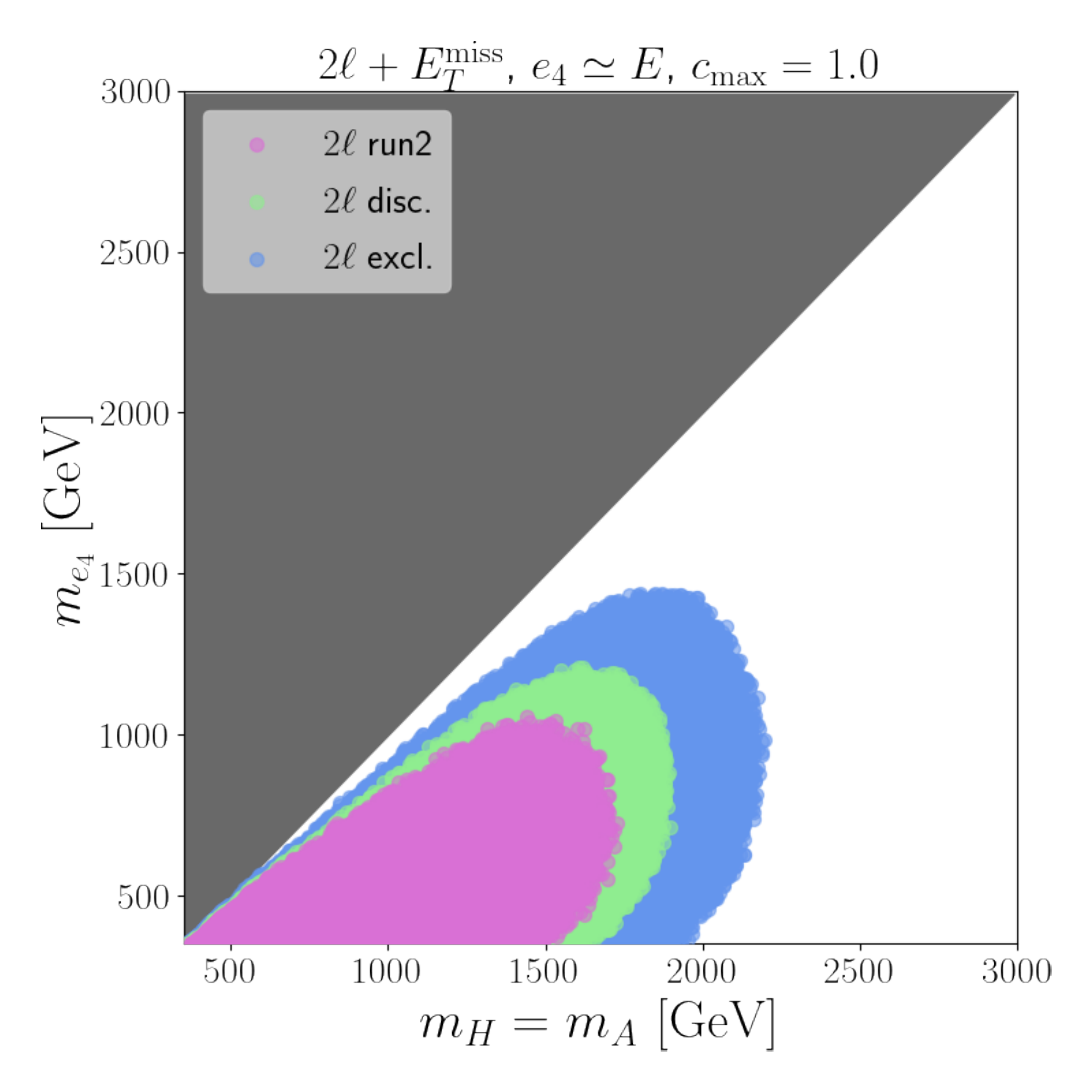}
    \end{minipage}
    \begin{minipage}[c]{0.48\hsize}
    \centering 
    \includegraphics[height=70mm]{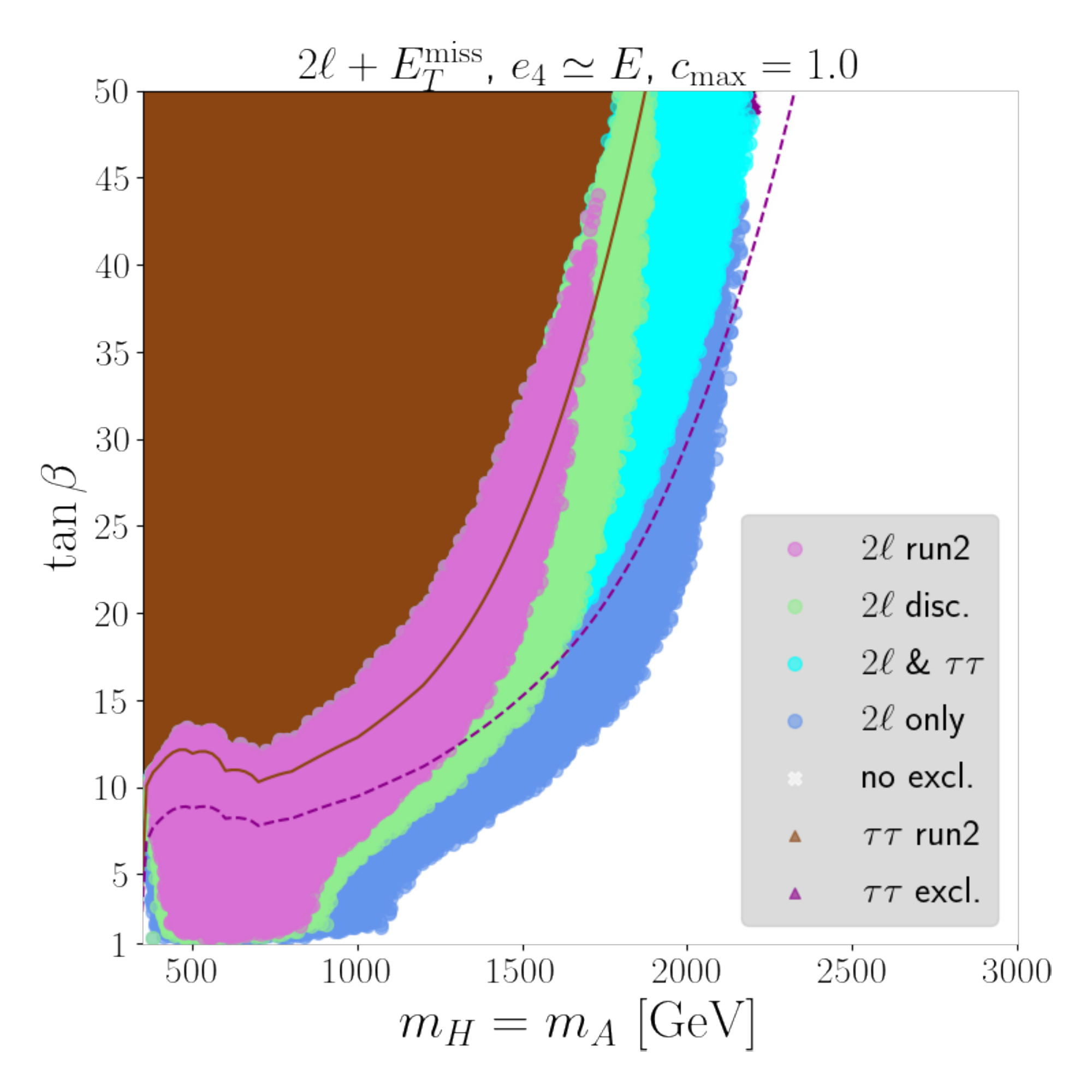}
    \end{minipage}
    \begin{minipage}[c]{0.48\hsize}
    \centering 
    \includegraphics[height=70mm]{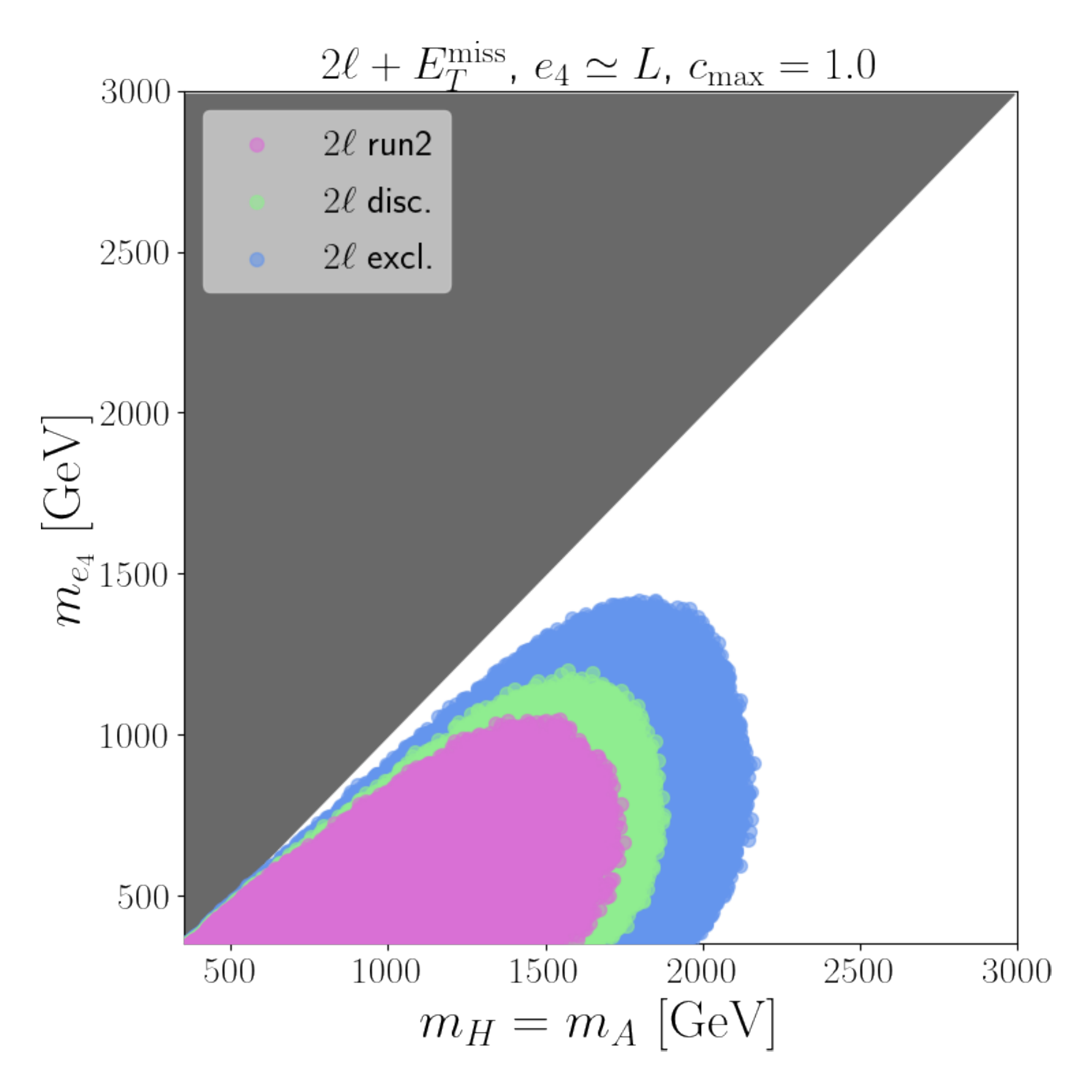}
    \end{minipage}
    \begin{minipage}[c]{0.48\hsize}
    \centering 
    \includegraphics[height=70mm]{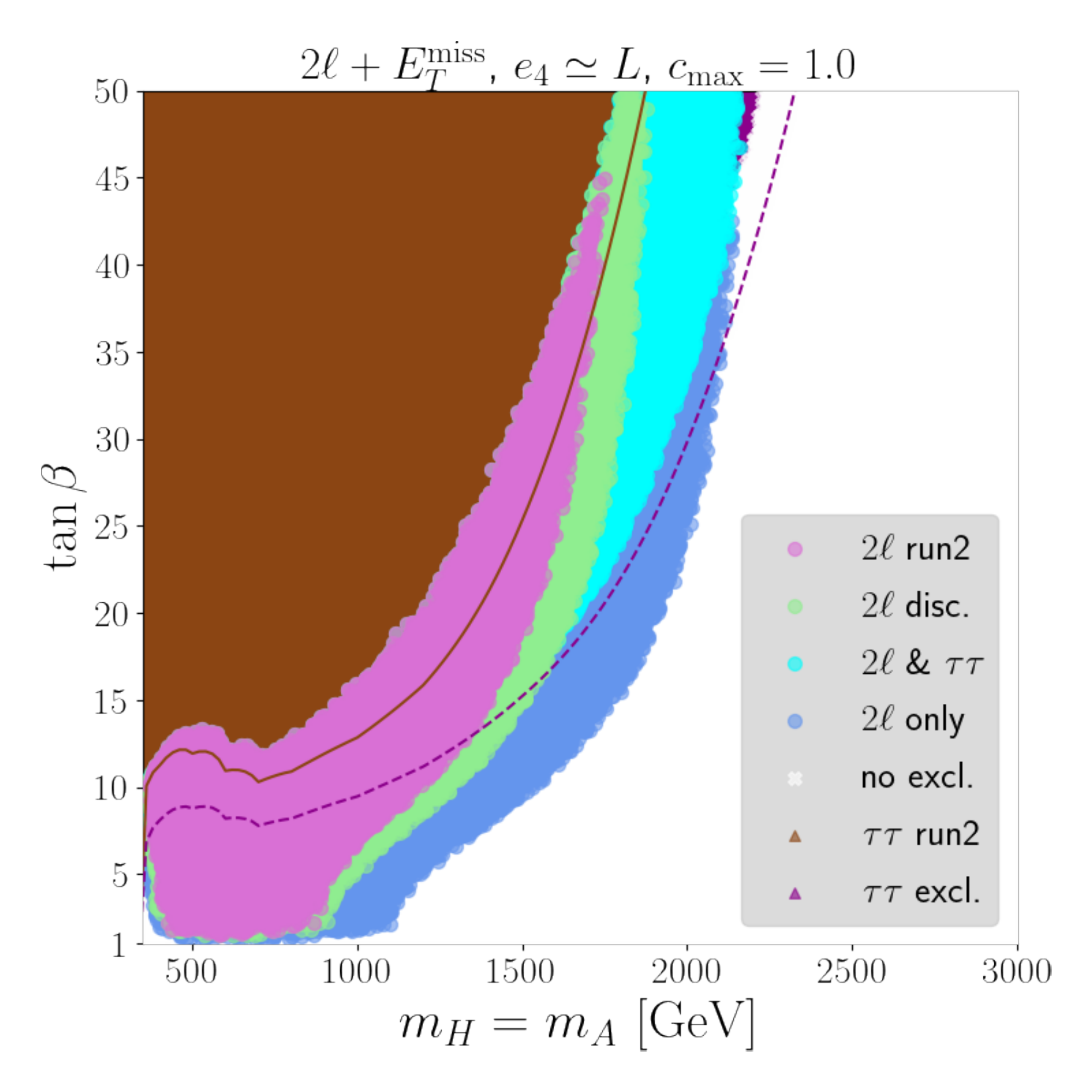}
    \end{minipage}

   \caption{    \label{fig:lim2lMET}
Sensitivities of the $2\ell+\met$ search 
to the light-E (light-L) scenario with $c_\mathrm{max} = 1$ 
in the upper (lower) panels. 
In the labels, ``$2\ell$'' is a short for $2\ell+\met$ search and ``$\tau\tau$'' is the di-tau search. 
In addition, ``run2'', ``disc.'' and ``excl.'' indicate that the points are constrained by the current run-2 data, expected to be discovered and constrained by the future HL-LHC data, respectively. 
See the texts for more details.
}
\end{figure}

\begin{figure}[p]
    \centering
    \begin{minipage}[c]{0.48\hsize}
    \centering 
    \includegraphics[height=70mm]{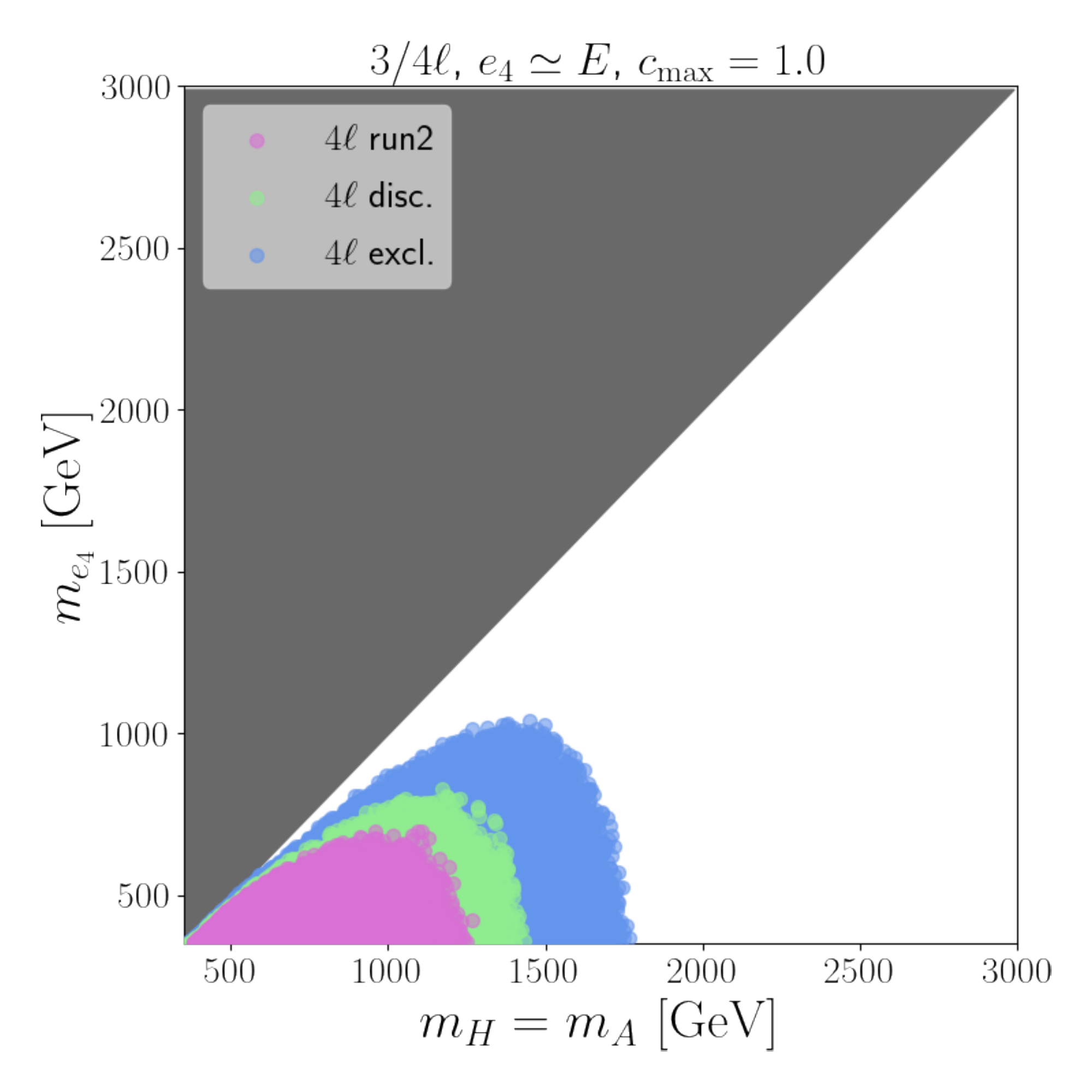}
    \end{minipage}
    \begin{minipage}[c]{0.48\hsize}
    \centering 
    \includegraphics[height=70mm]{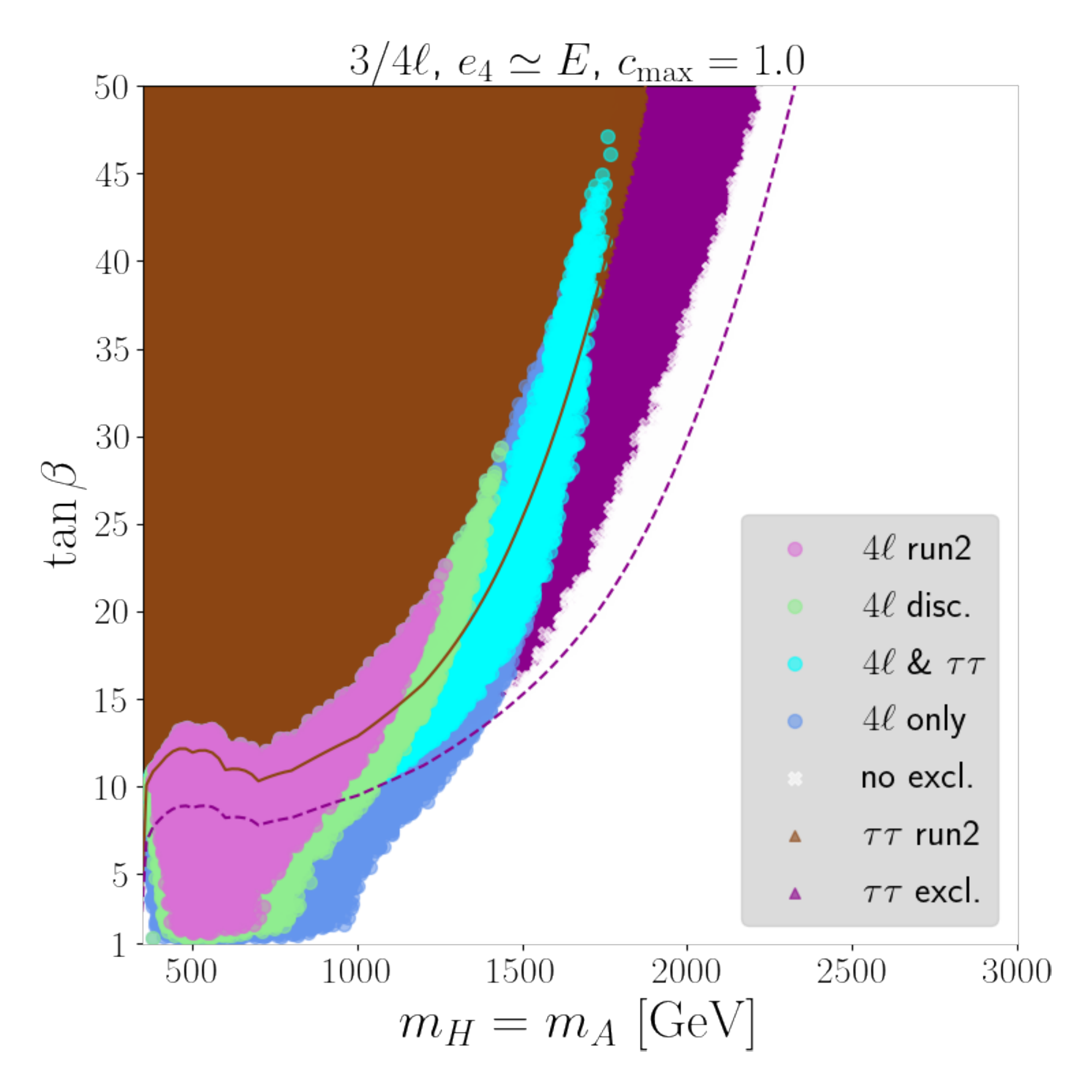}
    \end{minipage}
    \begin{minipage}[c]{0.48\hsize}
    \centering 
    \includegraphics[height=70mm]{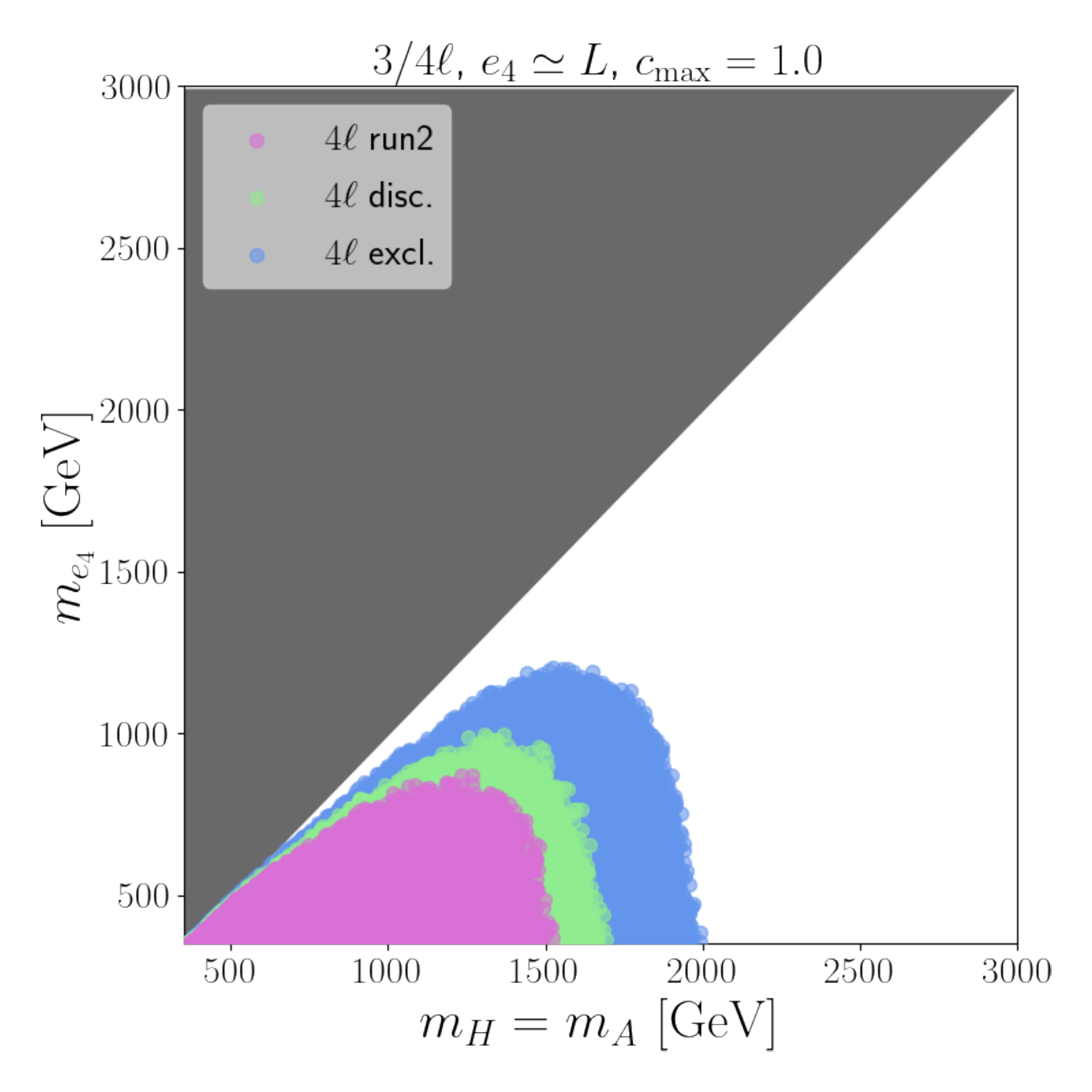}
    \end{minipage}
    \begin{minipage}[c]{0.48\hsize}
    \centering 
    \includegraphics[height=70mm]{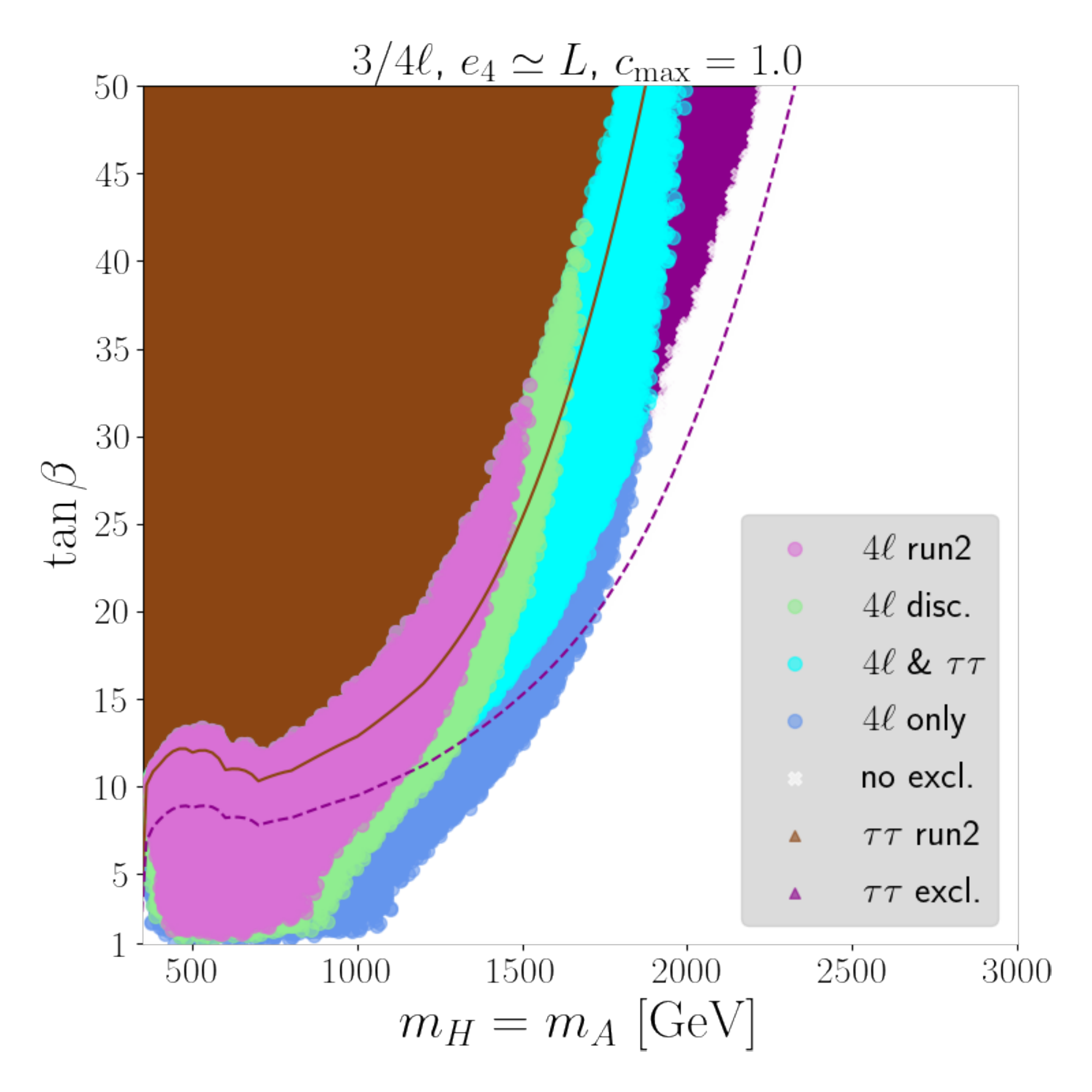}
    \end{minipage}
    \caption{       \label{fig:lim34leps} 
    Sensitivities of the $3/4\ell$ search to the light-E (light-L) scenario with $c_\mathrm{max} = 1$ in the upper (lower) panels. 
In the labels, ``$4\ell$'' is a short for $3/4\ell$ search and the rest of the labels are the same as those in Fig.~\ref{fig:lim2lMET}.
}
\end{figure}

\begin{figure}[p]
    \centering
    \begin{minipage}[c]{0.48\hsize}
    \centering 
    \includegraphics[height=70mm]{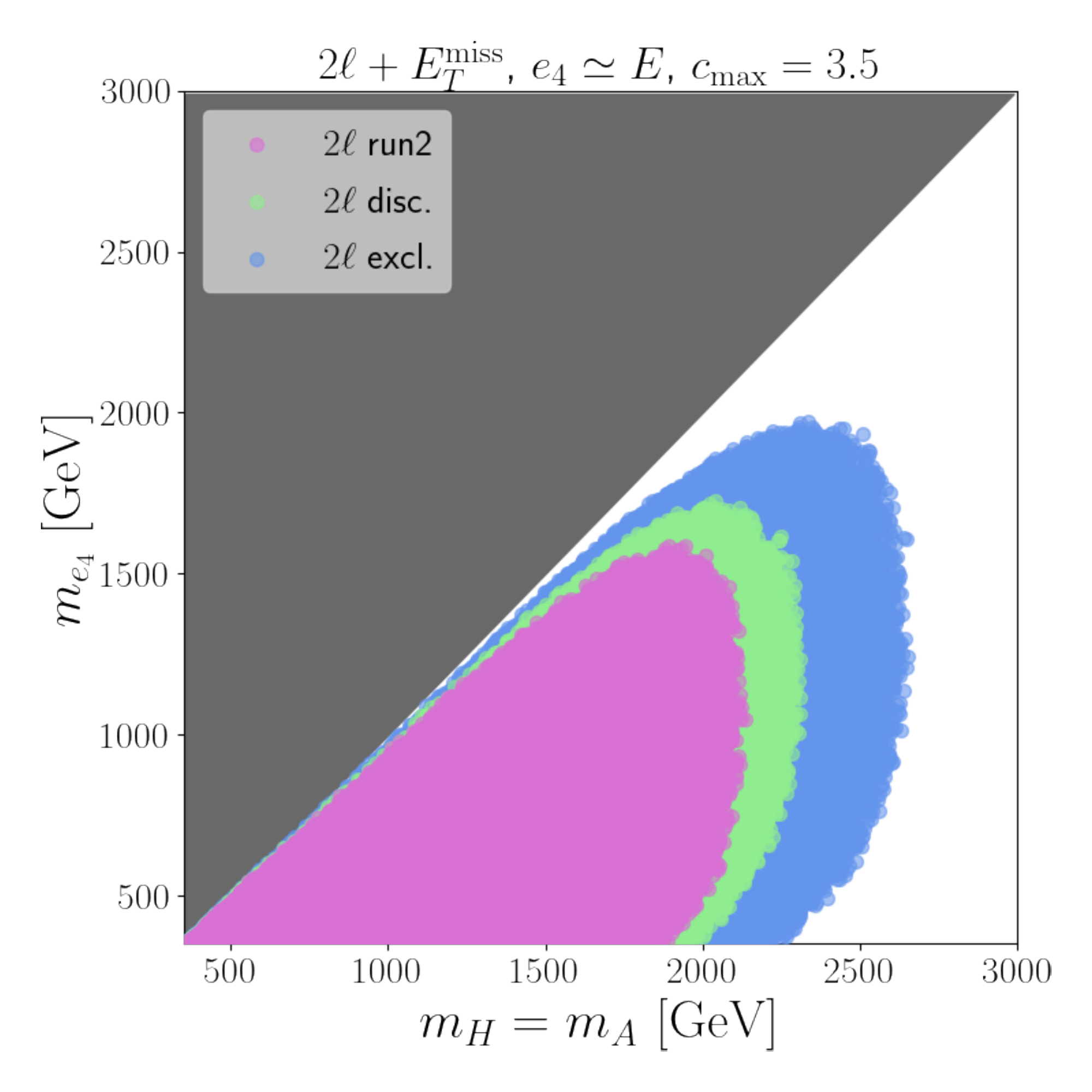}
    \end{minipage}
    \begin{minipage}[c]{0.48\hsize}
    \centering 
    \includegraphics[height=70mm]{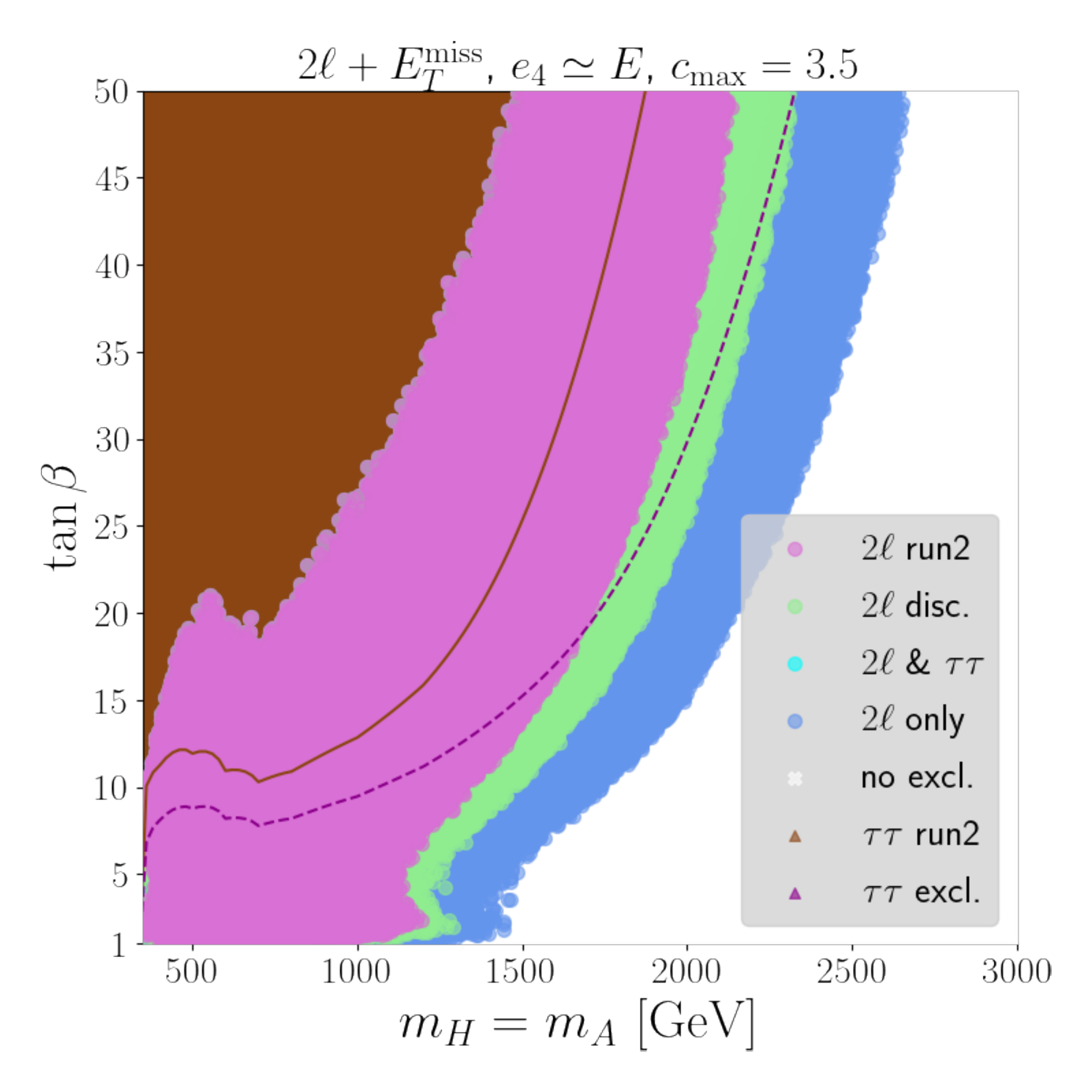}
    \end{minipage}
    \begin{minipage}[c]{0.48\hsize}
    \centering 
    \includegraphics[height=70mm]{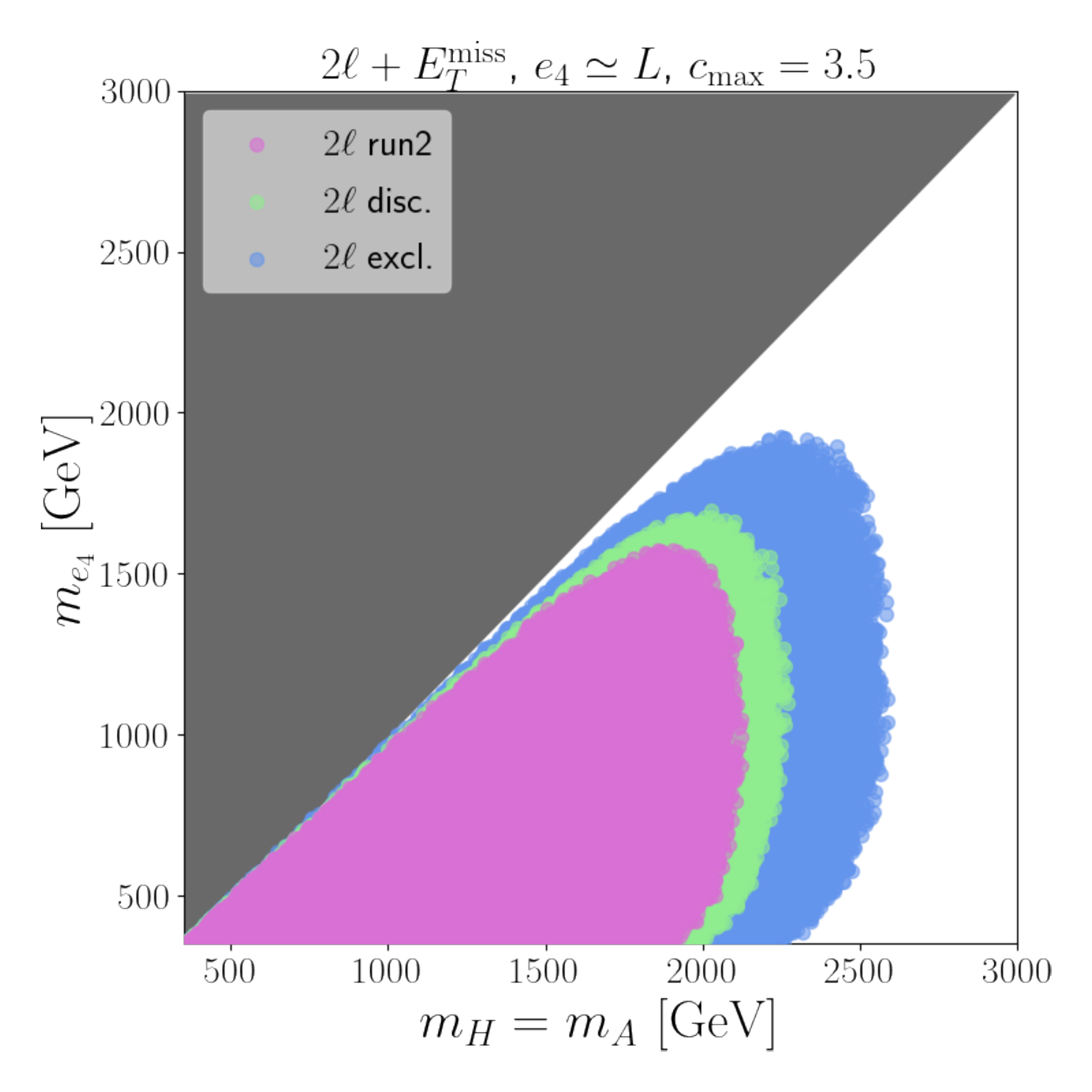}
    \end{minipage}
    \begin{minipage}[c]{0.48\hsize}
    \centering 
    \includegraphics[height=70mm]{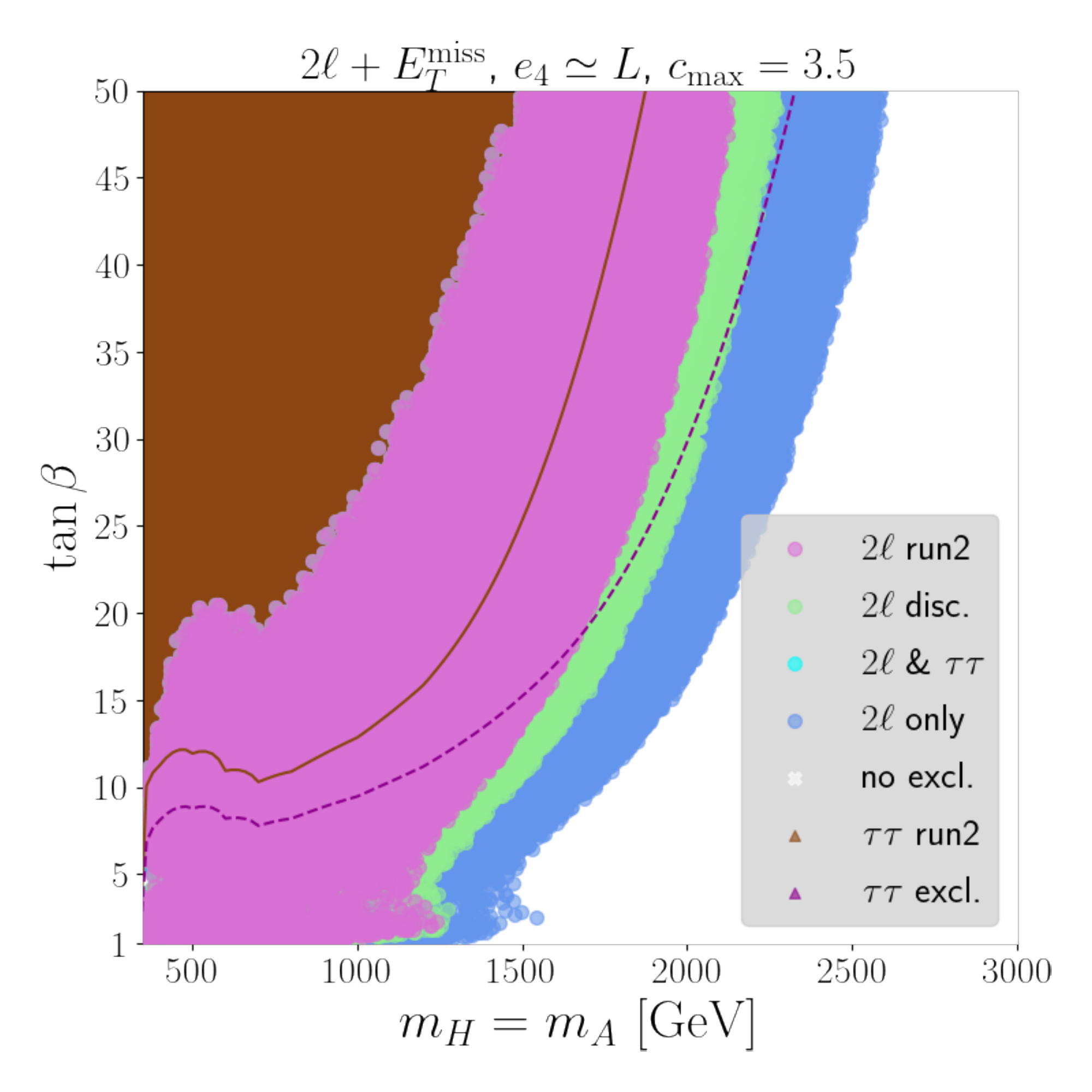}
    \end{minipage}

    \caption{    \label{fig:lim2lMET_cmax3p5}
Sensitivities of the $2\ell+\met$ search 
to the light-E (light-L) scenario with $c_\mathrm{max} = 3.5$ 
in the upper (lower) panels. 
The labels are the same as those in Fig.~\ref{fig:lim2lMET}.
}
\end{figure}

\begin{figure}[p]
    \centering
    \begin{minipage}[c]{0.48\hsize}
    \centering 
    \includegraphics[height=70mm]{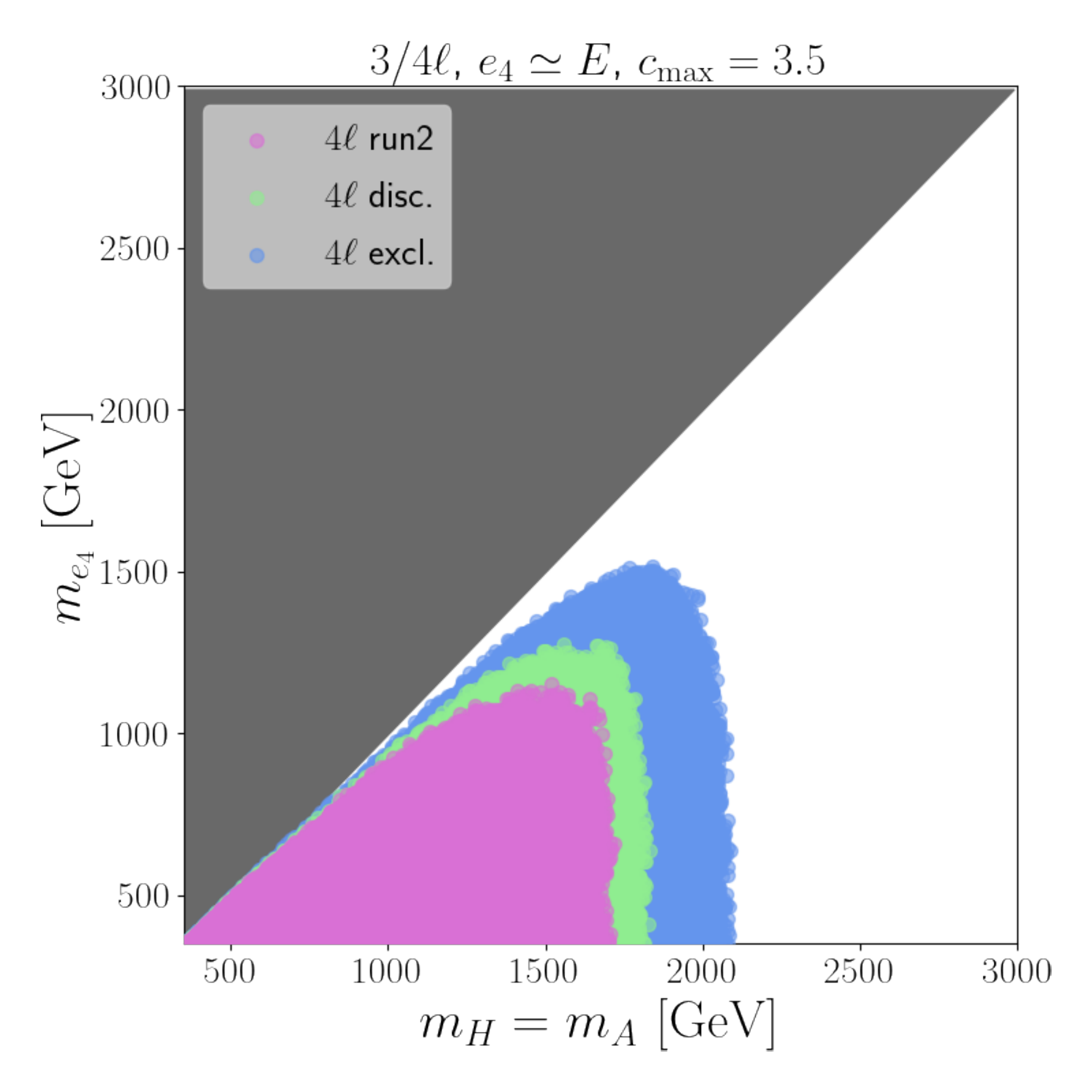}
    \end{minipage}
    \begin{minipage}[c]{0.48\hsize}
    \centering 
    \includegraphics[height=70mm]{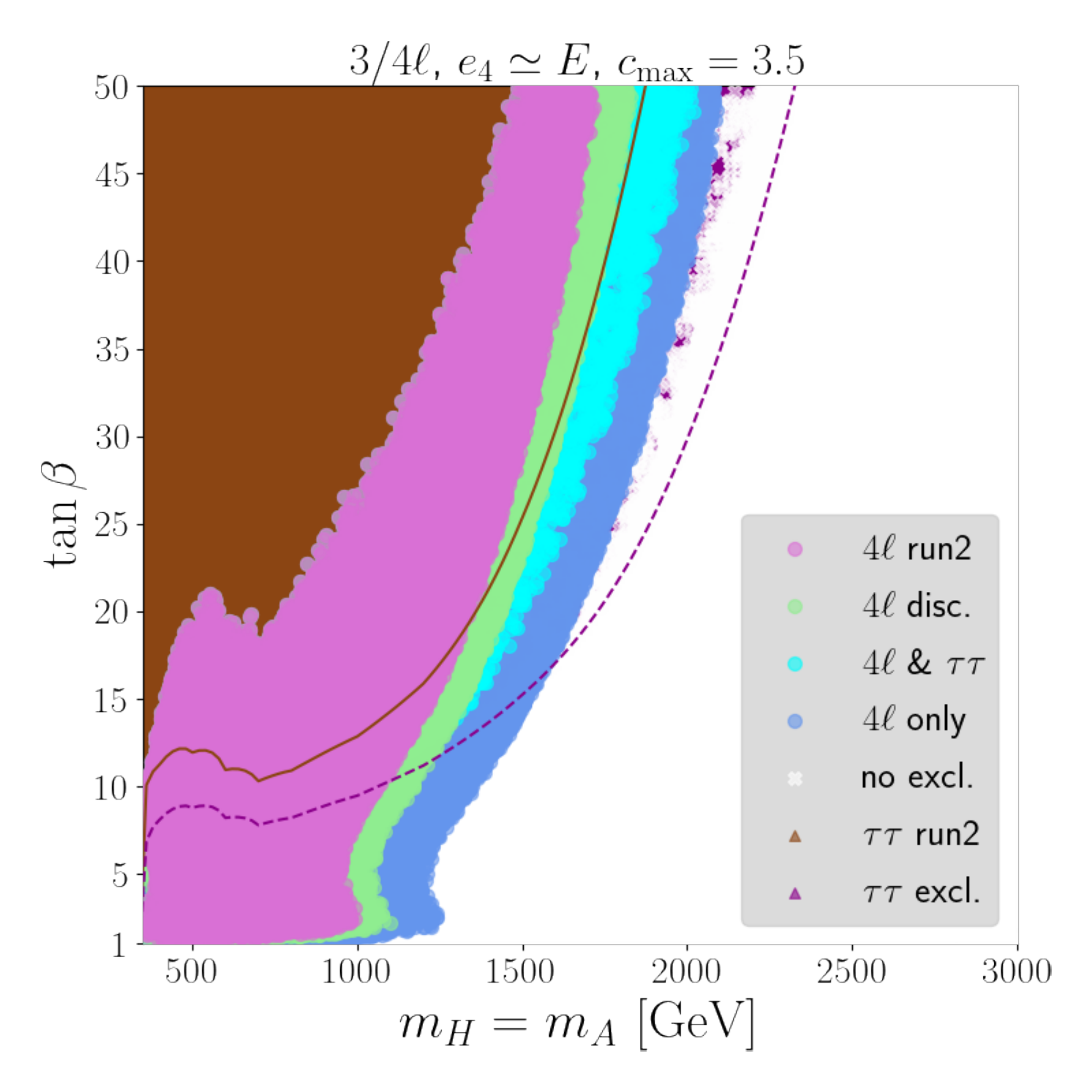}
    \end{minipage}
    \begin{minipage}[c]{0.48\hsize}
    \centering 
    \includegraphics[height=70mm]{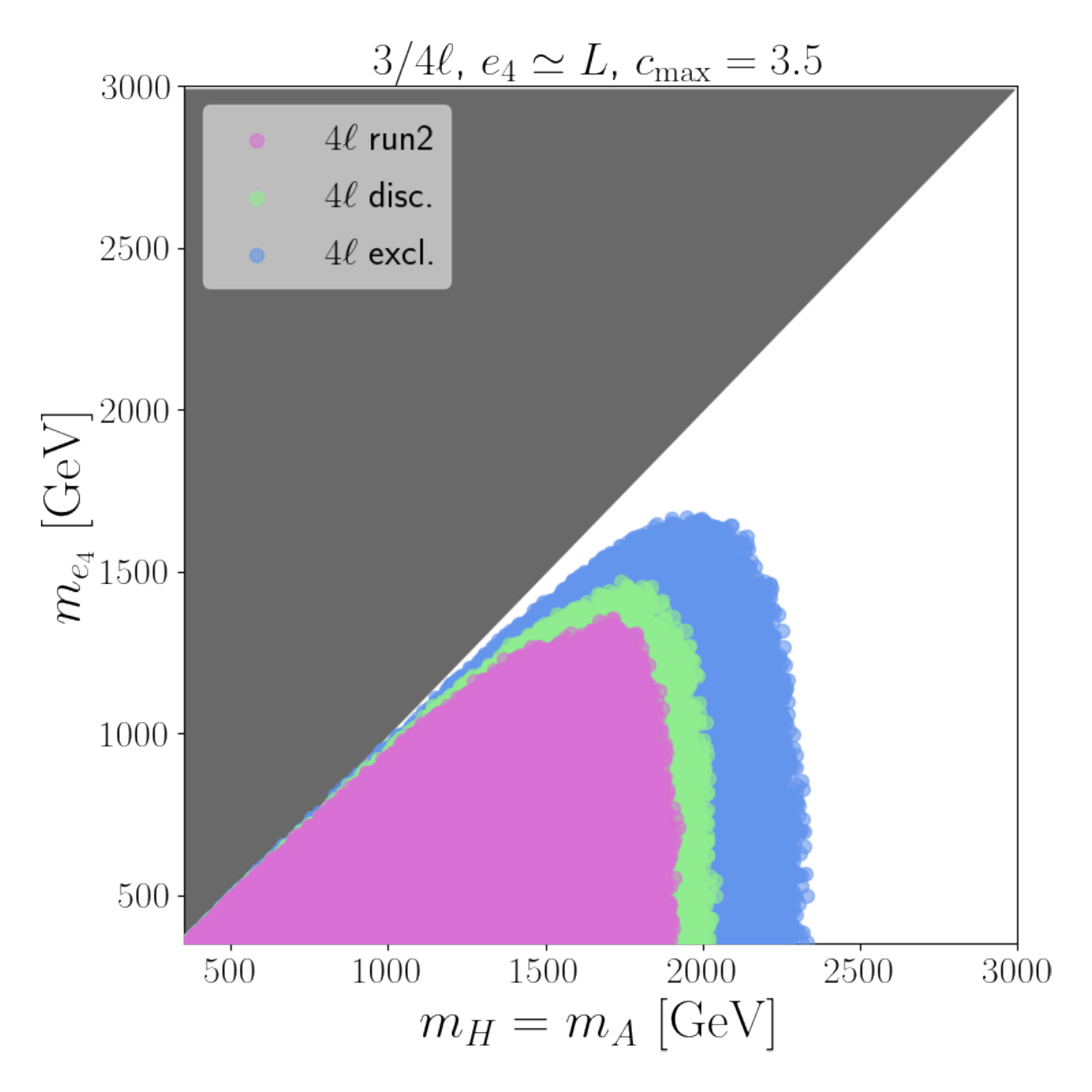}
    \end{minipage}
    \begin{minipage}[c]{0.48\hsize}
    \centering 
    \includegraphics[height=70mm]{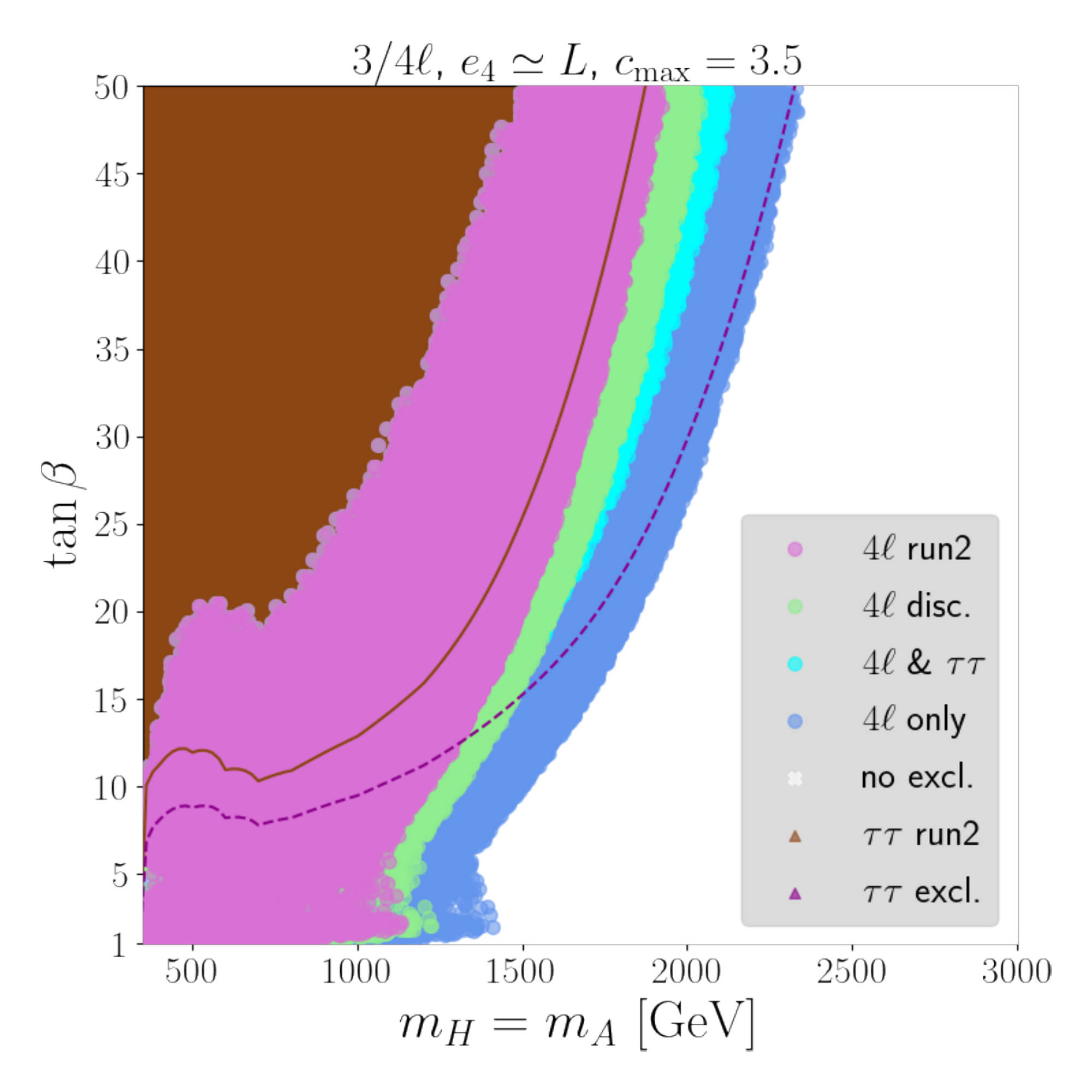}
    \end{minipage}

    \caption{       \label{fig:lim34leps_cmax3p5} 
  Sensitivities of the $3/4\ell$ search to the light-E (light-L) scenario with $c_\mathrm{max} = 3.5$ in the upper (lower) panels. 
  The labels are the same as those in Fig.~\ref{fig:lim34leps}.
}
\end{figure}
\afterpage{\clearpage}

The upper (lower) panels of Fig.~\ref{fig:lim2lMET} show the sensitivities of our reference model in the $2\ell+\met$ analysis for the light-E (light-L) case when $c_\mathrm{max} = 1$.
In the left panels, where the sensitivities are represented in the $m_H - m_{e_4}$ plane, the pink ($2\ell$~run2), green ($2\ell$~disc.) and blue ($2\ell$~excl.) points correspond to the regions which can be covered by our $2\ell+\met$ analysis with current data at 95\% C.L., the corresponding discovery region at the HL-LHC, and the future exclusion region, respectively.
Note that here exclusion regions mean that some choices of parameters would be excluded.
We include the contributions from the three decay modes, EZ, EW and NW, 
as defined in Eq.~\eqref{eq-sifull}. 

The points are plotted in the order given in the legend, i.e. pink points are plotted over green points, and green points over blue ones. 
In the right panels, we classify the blue points ($2\ell$~excl.) by whether they are expected to have sensitivity to the $\Phi \to \tau \tau$ search or not at the HL-LHC. 
The blue ($2\ell$~only) and cyan ($2\ell~\&~\tau\tau$) points are both within the future sensitivity of the $2\ell+\met$ channel but we expect the searches for $\Phi \to \tau \tau$ to also have sensitivity in the cyan region.
The brown ($\tau\tau$~run2) points are excluded by the current $\tau\tau$ search, 
and the purple ($\tau\tau$~excl.) points correspond to the region we expect to be possibly excluded by the future $\tau\tau$ search at the HL-LHC but not by the future $2\ell+\met$ search. 
The white points (no excl.) are not excluded 
by any of the search channels discussed in this paper. 
The brown solid (purple dashed) line corresponds to the nominal exclusion limit  from the current (future) sensitivity of $\Phi \to \tau \tau$ searches {\it without the presence of vectorlike leptons}.
These nominal bounds can be pushed back to the brown or purple scattered dots in the presence of vectorlike leptons depending on the parameter choices.
The (small) purple region in the top-right corner denotes the region within the reach of the $\tau \tau$ search but not of our leptonic cascade process.

We find the current search for the $2\ell + \met$ channel is sensitive to $m_\Phi \leq 1.7$ TeV and $m_{e_4} \leq 1$ TeV when $c_{\rm max} = 1$ and the sensitivity to Higgs masses close to 1 TeV remains for any $\tan\beta > 1$.
This is quite promising even compared to the conventional heavy Higgs search $\Phi \to \tau \tau$.
Moreover the future sensitivities with 3 ab$^{-1}$ extend to $m_\Phi \lesssim 2.2$ TeV and $m_{e_4} \lesssim 1.5$ TeV covering a much wider range of $\tan\beta$ than the conventional searches.

Figure~\ref{fig:lim34leps} shows the corresponding sensitivities for the $3/4\ell$ search.
The labels $4\ell$ indicate that these are limits or sensitivities 
from the $3/4\ell$ search instead of the $2\ell+\met$ search. 
In the light-E case, the limit is much weaker than that 
from the $2\ell+\met$ search 
because the EW decay mode does not contribute to the SRs of the $3/4\ell$ search. 
While in the light-L case, the limit is similar because $\br{e_4}{Z\mu}$ is larger than in the light-E case.
We may be able to test whether the lightest charged vectorlike lepton $e_4$ is almost singlet-like or doublet-like by using both of $2\ell +\met$ and $3/4\ell$ channels in a complementary way.
For example, if we were able to discover the signal at $m_\Phi = 1.7$ TeV and $m_{e_4} = 1$ TeV for $c_{\rm max} = 1$ in both channels, the discovered $e_4$ should be almost doublet-like.

In order to demonstrate the maximal capability of our analysis, we also allow larger Yukawa coupling constants up to $c_\mathrm{max} = 3.5$. The corresponding results are shown in Figs.~\ref{fig:lim2lMET_cmax3p5} and~\ref{fig:lim34leps_cmax3p5}.
The branching fractions can be larger than those in the $c_\mathrm{max}=1.0$ case, and thus the searches can probe heavier mass regions: $m_\Phi \lesssim 2$ TeV and $m_{e_4} \lesssim 1.5$ TeV with current data, and $m_\Phi \lesssim 2.5$ TeV and $m_{e_4} \lesssim 1.8$ TeV at the HL-LHC. 
Note that the presence of vectorlike leptons with large couplings can significantly dilute the typical search for $H\to\tau\tau$ as seen by the absence of cyan points and purple triangles in Fig.~\ref{fig:lim2lMET_cmax3p5} (where they are completely covered by the pink and green points) as compared to Fig.~\ref{fig:lim2lMET}.
We found that the $\tau\tau$ search at the HL-LHC can lose its sensitivity 
up to about the run2 limit, i.e. the brown solid line, 
due to the large Yukawa coupling constant $\la_L$ or $\la_E\sim 3.5$.  
In that case, the leptonic cascade search strategy presented here is necessary to probe the details of the Higgs sector.

\afterpage{\clearpage}

\section{Conclusion} 
\label{sec-cncl}

In this paper, we have studied the potential of leptonic cascade decays of a heavy neutral Higgs boson through vectorlike leptons as a simultaneous probe of an extended Higgs sector and extra matter particles at the LHC.
The fully leptonic final states contribute to multi-lepton signals such as $2\ell + \met$ and $3/4\ell$ which are already under investigation motivated by BSM scenarios such as SUSY and seesaw models.
We have found that by simply recasting the existing searches we can obtain new strong constraints to any BSM scenarios sharing the same event topology and final states as our reference scenario, depicted in Fig.~\ref{fig:Hlcasc}.
Therefore, some of our analysis results are shown in {\it model independent} ways.
The 95\% C.L. model independent upper limits on the total cross sections are found to be $\mathcal O (1-10\,{\rm fb})$ for a heavy neutral Higgs (or a resonant particle producing the same topology) in the mass range between 1 - 3 TeV, using the current run2 data with 139 fb$^{-1}$.
Naively rescaling the size of the data to an integrated luminosity of 3 ab$^{-1}$, assuming the statistical uncertainty dominates, the future sensitivities at the HL-LHC extend to $\mathcal O (0.2 - 1 \,{\rm fb})$. 

The model-independent bounds could be transformed into a useful form, e.g. the total branching fraction of the resonant particle in Fig.~\ref{fig:Hlcasc}, in a wide class of BSM scenarios where the resonant particle production cross sections are the same as (or simply rescaled from) the neutral heavy Higgs production cross sections in 2HDM type-II. 
If the branching fraction of one's interest is 50\%, the heavy Higgs mass up to about 1.3 (2.1) TeV for $m_{e_4} \gtrsim 500$ GeV and $\tan\beta = 10$ (50) can be constrained from the current search results for the $2\ell + \met$ channel, while the coverage slightly reduces for the $3/4\ell$ channel.
The corresponding future coverage at the HL-LHC (for the $2\ell + \met$) extends to 1.8 and 2.8 TeV for $\tan\beta = 10$ and 50, respectively. 

In terms of model-dependent parameters such as $\tan\beta$ and the masses of the new particles, the sensitivities of our leptonic cascade contributing to the $2\ell + \met$ and $3/4\ell$ channels can be better than what are expected in conventional searches for BSM Higgses such as the nominal $H/A \to \tau \tau$ channel.
The current sensitivity covers the region $1\lesssim \tan\beta \lesssim 10$ up to heavy Higgs masses around 1 TeV, beyond the reach of the conventional searches. 
The future sensitivities in this region of $\tan\beta$ extend up to $m_{H/A} \lesssim 2.2$ TeV and $m_{e_4} \lesssim 1.5$ TeV at the HL-LHC.

Although not implemented here, we expect that further investigation of an additional lepton resonance from the decay of $e_4$ would increase the sensitivity of the $3/4\ell$ channel. 
Therefore, we conclude that searches for our leptonic cascade processes can shed light on probing both an extended Higgs sector and extra matter, and are generally more advantageous than conventional heavy Higgs searches in these scenarios.

\medskip

\section*{Acknowledgments} 
The authors thank Kyoungchul Kong and Chan Beom Park for useful discussion.
The work of R.D. was supported in part by the U.S. Department of Energy under Award No. {DE}-SC0010120. The work of J.K. is supported in part by the Institute for Basic Science (IBS-R018-D1) 
and the Grant-in-Aid for Scientific Research from the
Ministry of Education, Science, Sports and Culture (MEXT), Japan No.\ 18K13534. 
TRIUMF receives federal funding via a contribution agreement with the National Research Council of Canada.
S.S. acknowledges support from the National Research Foundation of Korea (NRF-2020R1I1A3072747).

\appendix

\section{Approximated expressions in the reference model} 
\label{appendix-model}

In this appendix, we summarize analytical formulae of the important EWPOs and decay widths 
at leading order in:
\begin{align}
\left| 
\frac{(\la_L,\la_E,\la,\ol{\la},\ka_N,\ka,\ol{\ka})\, v}
{(m_L, m_E, m_N)}\right| \ll 1   \,, 
\label{eq:approx}
\end{align}
where we define $v \equiv \sqrt{v_u^2 + v_d^2} = 174$ GeV with
$v_u = \langle H_u^0 \rangle$ and $v_d = \langle H_d^0 \rangle$.
Here the Lagrangian parameters are given in Eq.~(\ref{eq:lagrangian}).
Further assuming enough splittings between $m_L$ and $m_E$ ($m_N$), the lightest charged (neutral) vectorlike lepton $e_4$ ($\nu_4$) is mostly isodoublet or isosinglet (SM singlet).
Detailed approximate expressions of gauge and Yukawa couplings in the mass eigenstate basis are in Refs.~\cite{Dermisek:2013gta,Dermisek:2015hue}.
Here, we focus on the formulae useful in applying constraints from electroweak precision measurements and several important branching fractions.

\subsection{Electroweak precision measurements}
\label{sec-EWPO}

\begin{table}[t]
\centering 
\begin{tabular}[t]{c|cc} \hline 
Name             & Central                   & Error   \\ \hline\hline 
$G_F$            & $1.1663787\times 10^{-5}$ & $0.060\%$ \\ 
$\br{W}{\mu\nu}$ & $0.22635$                 & $2.4\%$   \\ 
$\ga{Z}{\inv}$   & $0.501464~\GeV$           & $0.50\%$  \\ 
$A_\mu$          & $0.142$                   & $0.015$   \\ 
$A^{\FB}_\mu$    & $0.0169$                  & $0.0013$ \\ 
$R_\mu$          & $20.784$                  & $0.17\%$ \\ 
\hline
$\Delta S$       & $-0.01$                   & $0.10$   \\ 
$\Delta T$       & $0.03$                    & $0.12$   \\ 
$\Delta U$       & $0.02$                    & $0.11$   \\ 
$R_{\gaga}$      & $1.11$                    & $0.095$   \\  \hline 
\end{tabular}
\caption{\label{tab-EWPO}
Summary of the constraints from the electroweak precision measurements~\cite{Zyla:2020zbs} we apply here. 
The errors with $\%$ are relative uncertainties.  
}
\end{table}

The vectorlike leptons in our reference model are constrained by various electroweak precision measurements listed in Table~\ref{tab-EWPO}, which are also discussed in Refs.~\cite{Dermisek:2013gta,Dermisek:2015hue}.
We calculate these observables at tree-level for the first 6 observables and at the one-loop level for the last 4 observables. 

The contributions by the vectorlike leptons to the oblique corrections are almost the same as the formulae in Ref.~\cite{Lavoura:1992np} given for the vectorlike quarks except the definitions of the following functions:
\begin{align}
 \psi^+(y_a, y_\beta) =&\ 
 \frac{2y_a+10y_\beta}{3} + \frac{1}{3} \log\frac{y_a}{y_\beta}+ 
  \frac{y_a-1}{6} f(y_a,y_a) +     \frac{5y_\beta + 1}{6} f(y_\beta,y_\beta), \\ 
 \psi^-(y_a,y_\beta) =&\ 
  - \sqrt{y_a y_\beta}\left(4+\frac{f(y_a,y_a)+f(y_\beta,y_\beta)}{2} \right),    
\end{align}
due to the difference in hypercharges.  
Here, $y_a := m_{\nu_a}^2/m_Z^2$,  
$y_\beta := m_{e_\beta}^2/m_Z^2$ ($a,\beta =1,2,\cdots,5$) 
and the function $f(y_1,y_2)$ is defined in Ref.~\cite{Lavoura:1992np}.~\footnote{
The doublet-like vectorlike lepton can explain the recent anomaly in the $W$ boson  mass~~\cite{CDF:2022hxs} 
if its mass is lighter than 200 GeV
and the mixing with the singlet-like state is sizable~\cite{Kawamura:2022uft}.  
} 
The ratio $R_\gaga := \ga{h}{\gaga}/\ga{h}{\gaga}_\mathrm{SM}$ 
is calculated with the formula shown in Ref.~\cite{Djouadi:2005gi}. 

In our reference model, the contributions of vectorlike leptons to the Fermi constant $G_F$, determined from the muon decay, and $R_\mu := {\Gamma\left(Z\to \mathrm{had}\right)}/{\Gamma\left(Z \to \mu\mu\right)}$ are the most strongly constrained observables in Table~\ref{tab-EWPO}.
Hence, we provide the approximate expressions (when Eq.~(\ref{eq:approx}) is valid) of the leading contributions to $G_F$ and $R_\mu$: 
\begin{align}
\label{eq-roughGF}
 \frac{\delta G_F}{G_F^{\mathrm{SM}}} 
& \sim 
      \frac{v^2/2}{1+t_\beta^2} 
     \left( \frac{\la^2_E}{m_E^2} + t_\beta^2 \frac{\ka_N^2}{m_N^2} \right)
        < 6\times 10^{-4}\,,  \\ 
 \frac{\delta R_\mu}{R_\mu^{\mathrm{SM}}}
  &\sim \frac{4s_W^2}{1-4s_W^2+8s_W^4} \frac{v_d^2 \la_L^2}{m_L^2} 
        + \frac{2(1-2s_W^2)}{1-4s_W^2+8s_W^4} \frac{v_d^2 \la_E^2}{m_E^2} 
 < 1.7\times 10^{-3}\,,     
\end{align}
where $t_\beta = \tan\beta$.
For $t_\beta \gg 1$, the leading contribution to $G_F$ is $\ka_N^2/m_N^2$ 
and the other contributions are suppressed by $t_\beta^2$. 
From Eq.~\eqref{eq-roughGF}, 
the upper bound on $\ka_N$ is estimated as 
\begin{align}
 \ka_N \lesssim \frac{\sqrt{2}m_N}{v s_\beta} \times 6\times 10^{-4} 
         \sim 1\times 10^{-3} \times  
               \left(\frac{m_N/s_\beta}{200~\GeV}\right).  
\end{align}
Therefore, $\ka_N$ should be less than $\order{10^{-3}}$ to be consistent with EWPOs. 
The limits on $\la_L$ and $\la_E$ are weaker by a factor $\tan^2\beta$, 
and hence $\la_{L,E} \sim \order{1}$ is allowed 
for sufficiently large $\tan\beta$ and vectorlike lepton masses.

\subsection{Branching fractions}

\begin{figure}[t]
    \centering
\includegraphics[height=0.9\textwidth]{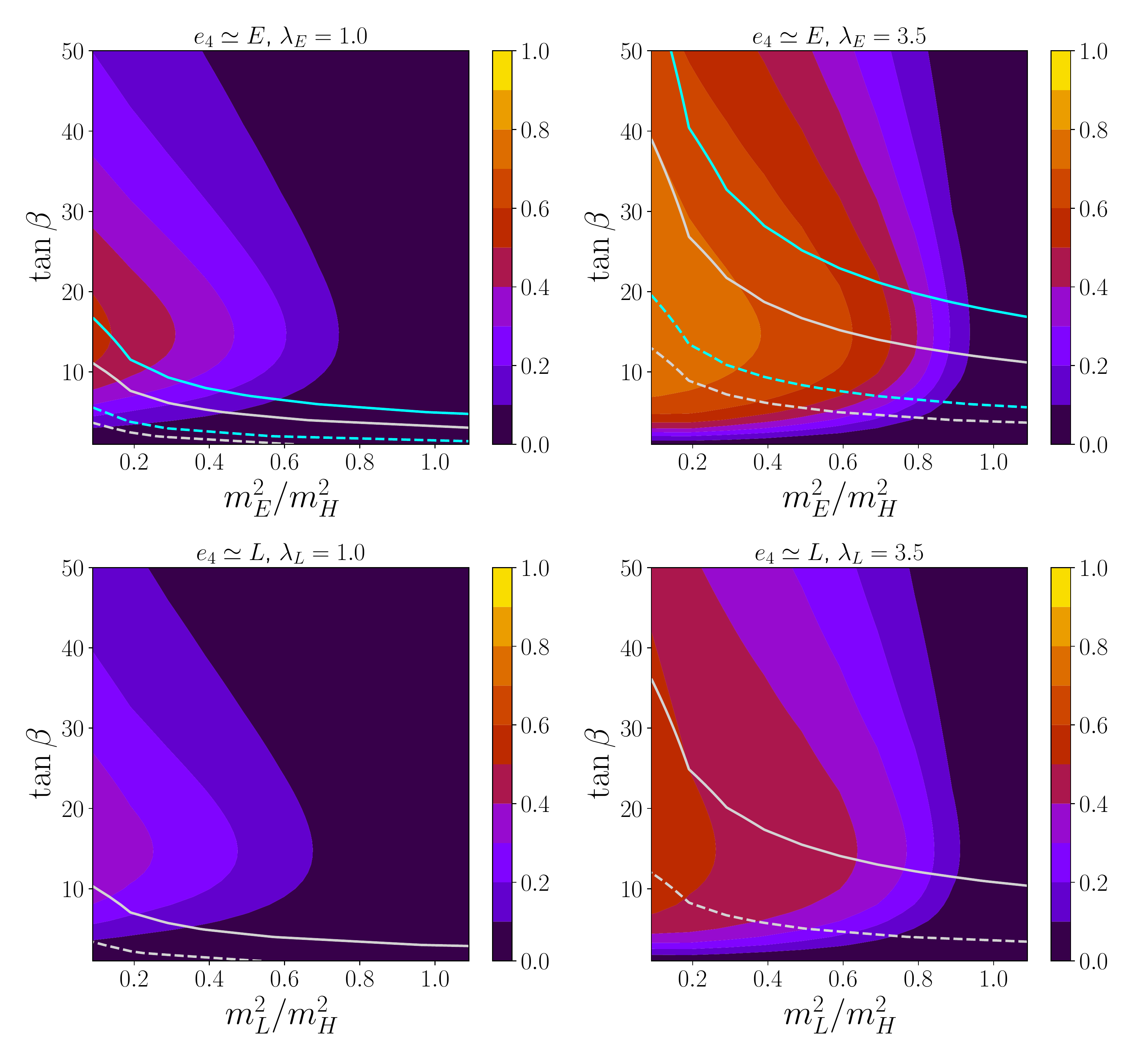} 
    \caption{
Branching fractions of the leptonic cascade decay, 
$H \to \mu e_4 \to V \ell_2 \mu$, 
for $m_H = 1~\TeV$. 
${G_F}$ ($R_\mu$) is more than $2\sigma$ away from the measured value  
below the cyan (gray) solid line. 
The EWPO constraints for $m_H = 3~\TeV$ are shown by the dashed lines. 
} 
    \label{fig:BrLcascade}
\end{figure}

The partial widths for the heavy netural Higgs decays to the charged leptons, $e_a^+ e_b^-$ ($e_a = \mu, e_4, e_5$), are given by
\begin{align}
 &\  \Gamma\left(H\to e_a^+ e_b^- \right) =
  \frac{m_H}{16\pi} \beta\left(\frac{m_{e_a}^2}{m_{H}^2},~\frac{m_{e_b}^2}{m_H^2} \right) 
 \\ \notag 
&\ \times   \left[ 
   \left(\abs{\left[Y^H_\ve\right]_{ab}}^2 
       + \abs{\left[Y^H_\ve\right]_{ba}}^2  \right) 
     \left(1-\frac{m_{e_a}^2}{m_{H}^2}-\frac{m_{e_b}^2}{m_H^2}\right) 
   - \frac{m_{e_a} m_{e_b}}{m_H^2} 
  \mathrm{Re}\left(\left[Y^H_\ve\right]_{ab} \left[Y^H_\ve\right]_{ba} \right) 
   \right],   
\end{align} 
where $\beta(x,y):= \sqrt{1-2(y+x)+(y-x)^2}$. 
Here, $Y^H_\ve$ is the heavy CP-even Higgs Yukawa matrix of the charged leptons in the mass basis.
Those for the CP-odd Higgs boson can be obtained by replacing $H\to A$,
and those for the neutrinos are given by replacing $e\to \nu$ and $\mu\to\nu_\mu$.

When $e_4$ and $\nu_4$ are mostly doublet-like, i.e. $(e_4^-, \nu_4)\sim( L^-, L^0)$, 
the decay widths to a vectorlike lepton and a SM lepton are approximately given by
\begin{align}
\label{eq-HmuL}
 \Gamma\left(H\to \mu^- L^+\right) \sim&\ \frac{m_H \la_L^2 s_\beta^2}{32\pi} 
                                  \left(1-\frac{m_L^2}{m_H^2}\right)^2, \\ 
 \Gamma\left(H\to \nu_\mu L^0\right) \sim&\ \frac{m_H \ka_N^2 c_\beta^2}{32\pi} 
                                  \left(1-\frac{m_L^2}{m_H^2}\right)^2 
                                  \left(\frac{\ol{\ka}v_u}{m_N} 
                    \frac{\ka m_L+ \ol{\ka} m_N}{m_N^2 - m_L^2} v_u \right)^2,
\end{align}
and for mostly singlet-like leptons, i.e. $e_4^- \sim E$, $\nu_4 \sim N$,  
\begin{align}
\label{eq-HmuE}
 \Gamma\left(H\to \mu^- E^+\right) \sim&\ \frac{m_H \la_E^2 s_\beta^2}{32\pi} 
                                  \left(1-\frac{m_E^2}{m_H^2}\right)^2, \\ 
 \Gamma\left(H\to \nu_\mu N\right) \sim&\ \frac{m_H \ka_N^2 c_\beta^2}{32\pi} 
                                  \left(1-\frac{m_N^2}{m_H^2}\right)^2.  
\label{eq-HnuN}
\end{align}
Here the sub-dominant contributions suppressed by 
the new Yukawa couplings times $v^2/m_{\rm VLL}^2$, ${\rm VLL}=E,L,N$, and the SM lepton masses are neglected. 
Note that $H\to \nu_\mu L^0$ appears only at sub-leading order 
and vanishes for $\ka_N=0$. 
Thus, the Higgs boson decays mostly to a charged vectorlike lepton and a SM lepton 
when $\la_L$ ($\la_E$) is large and $e_4^- \simeq L^-~(E)$.

In the numerical analysis, the heavy Higges decay to SM particles,  
$H\to cc$, $bb$, $tt$, $\tau\tau$, $\gamma\gamma$, $gg$, $hh$ and $A\to cc$, $bb$, $tt$, $\tau\tau$, $\gamma\gamma$, $gg$ are calculated using the formulas presented in Ref.~\cite{Djouadi:2005gj}.  
We assume the MSSM for the triple Higgs coupling.  
The decays to gauge bosons, $H\to ZZ,~WW$, and $A\to hZ$ vanish
in the alignment limit.

The partial decay widths of the charged vectorlike lepton $e_4$ is given by 
\begin{align}
 \Gamma\left(e_4\to h \mu \right) 
=&\ \frac{m_{e_4}}{64\pi} \beta( x_h, y_\mu )
\left[ \left(\abs{\left[Y_\ve^h\right]_{\mu e_4}}^2
                 +\abs{\left[Y_\ve^h\right]_{ e_4 \mu}}^2\right) \left(1+y_\mu-x_h\right) 
\right.   \notag \\ 
&\     \left.  \hspace{4.0cm}   + 4 \sqrt{y_\mu}~\mathrm{Re} 
 \left(\left[Y_\ve^h\right]_{ e_4 \mu} \left[Y_\ve^h\right]_{\mu e_1}\right) \right]\,, \\  
 \Gamma\left(e_4\to Z \mu \right) 
=&\ \frac{m_{e_4}}{32\pi x_Z} \beta( x_Z, y_\mu)
 \left[ \left(\abs{\left[g^Z_{e_L}\right]_{\mu e_4}}^2 
            + \abs{\left[g^Z_{e_R}\right]_{\mu e_4} }^2\right)
   \right.   \\ \notag 
&\   \left.    
\hspace{-1.0cm}\times  \left\{ \left( 1-y_\mu \right)^2 + x_Z \left(1+y_\mu\right) - 2 x_Z^2\right\} 
 - 3 x_Z \sqrt{y_\mu}~\mathrm{Re}\left(\left[g^Z_{e_L}\right]_{\mu e_4} 
                                       \left[g^Z_{e_R}\right]_{\mu e_4} \right) 
    \right]\,, \\ 
 \Gamma\left(e_4\to W \nu_\mu \right) 
=&\ \frac{m_{e_4}}{32\pi x_W} 
 \left(\abs{\left[g^W_{L}\right]_{\nu_\mu e_4}}^2 
 + \abs{\left[g^W_{R}\right]_{\nu_\mu e_4}}^2\right) 
 \left(1-x_W\right)^2 \left(1+2x_W\right)\, ,   
\end{align}
where $Y^h_\ve$, $g^Z_{e_{L,R}}$ and $g^W_{{L,R}}$ 
are the coupling matrices to the SM bosons in the mass basis, 
see Refs.~\cite{Dermisek:2013gta,Dermisek:2015hue} for the details. 
The mass squared ratios are defined as 
$y_\mu:= m_\mu^2/m_{e_4}^2$ and $x_B := m_B^2/m_{e_4}^2$ with $B=h,Z,W$. 
Those for the vector-like neutrino $\nu_4$ can be obtained by formally replacing $e\to \nu$ 
and $[g^W_{L,R}]_{\nu_\mu e_4} \to [g^W_{L,R}]_{\nu_4 \mu}$ in the decay to a $W$ boson.

In our analysis, we assume only one of the vectorlike leptons is lighter than the heavy Higgs for simplicity. 
For the ``light-E'' scenario, i.e., $e_4 \sim E$, the partial widths are approximately given by
\begin{align}
 \Gamma(e_4 \to h \mu) \sim&\  
   \frac{c_\beta^2 \la_E^2 m_E}{64\pi}  (1-x_h)^2\,,   \\
 \Gamma(e_4 \to Z \mu) \sim&\  
   \frac{c_\beta^2 \la_E^2 m_E}{64\pi}  (1-x_Z)^2 (1+2x_Z)\,,  \\
 \Gamma(e_4 \to W \nu_\mu) \sim &\  
   \frac{c_\beta^2 \la_E^2 m_E }{32\pi} (1-x_W)^2 (1+2x_W)\,.
\end{align}
For the ``light-L'' scenario, i.e., $e_4^- \sim L^-$, the partial decay widths are approximately given by 
\begin{align}
 \Gamma(e_4 \to h \mu) \sim&\  
   \frac{c_\beta^2 \la_L^2 m_L}{64\pi}  (1-x_h)^2\,,   \\
 \Gamma(e_4 \to Z \mu) \sim&\  
   \frac{c_\beta^2 \la_L^2 m_L}{64\pi}  (1-x_Z)^2 (1+2x_Z)\,,  \\
 \Gamma(e_4 \to W \nu_\mu) \sim &\  
   \frac{c_\beta^2 \la_E^2 m_L }{32\pi} 
       \left( \frac{\ol{\la} m_E + \la m_L}{m_E^2-m_L^2} v_d \right)^2 (1-x_W)^2 (1+2x_W)\,.
\end{align}

\noindent If $\nu_4$ is mostly isodoublet-like, then we have
\begin{align}
 \Gamma(\nu_4 \to h \nu_\mu) \sim&\  
   \frac{s_\beta^2 \ka_N^2 m_L}{64\pi}  (1-x_h)^2 
  \left(\frac{\ol{\ka} v_u}{m_N} + \frac{\ka m_L + \ol{\ka} m_N}{m_N^2-m_L^2} v_u\right)^2\,,
   \\
 \Gamma(\nu_4 \to Z \nu_\mu) \sim&\  
   \frac{s_\beta^2 \ka_N^2 m_L^3}{64\pi m_N^2}  (1-x_Z)^2 (1+2x_Z)
\left(  \frac{\ol{\ka} m_L + \ka m_N}{m_N^2-m_L^2} v_u\right)^2\,,
   \\
 \Gamma(\nu_4 \to W \mu) \sim &\  
   \frac{c_\beta^2 \la_L^2 m_L }{32\pi} (1-x_W)^2 (1+2x_W)\,,  
\end{align}
while if it is mostly singlet-like, then
\begin{align}
 \Gamma(\nu_4 \to h \nu_\mu) \sim&\  
   \frac{s_\beta^2 \ka_N^2 m_N}{64\pi}  (1-x_h)^2\,,  
   \\
 \Gamma(\nu_4 \to Z \nu_\mu) \sim&\  
   \frac{s_\beta^2 \ka_N^2 m_N}{64\pi}  (1-x_Z)^2 (1+2x_Z)\,, 
    \\
 \Gamma(\nu_4 \to W \mu) \sim &\  
   \frac{s_\beta^2 \ka_N^2 m_N }{32\pi} (1-x_W)^2 (1+2x_W) 
   \left[1+ 
 \left(\frac{\la_L m_N v_d}{\ka_N m_L} \frac{\ka m_L+\ol{\ka}m_N}{m_N^2-m_L^2}\right)^2  
  \right]\,.   
\end{align}
In the limit of $\ka_N \ll \la_L, \la_E$, the lepton $\nu_4$ can dominantly decay to a $W$ boson.

Figure~\ref{fig:BrLcascade} shows the values of the branching fractions of the leptonic cascade decay $H\to e_4 \mu \to V \ell_2 \mu$, with $V\ell_2 = Z\mu,~W\nu_\mu$. 
In the upper (lower) panels, the hierarchy $m_E < m_H < m_L, m_N$ ($m_L < m_H < m_E, m_N$) 
is assumed such that $e_4 \simeq E~(L)$. 
The Higgs mass is $1~\TeV$ and the heavier vectorlike lepton masses are set to be $3~\TeV$.
The Yukawa coupling constant $\la_E$ ($\la_L$) is set to $1.0$ or $3.5\sim \sqrt{4\pi}$ on the left and right panels, respectively, 
for the case of $e_4 \simeq E~(L)$.  
The other Yukawa coupling constants are set to $10^{-3}$ for simplicity. 
In addition, we fix $\ka_N = 0$ to suppress the contribution to $G_F$.  
The regions below the cyan (gray) line is $2\sigma$ away from the measured value of $G_F$ ($R_\mu$). $G_F$ provides the stronger constraint for an isosinglet-like $e_4$, while only that of $R_\mu$ provides a constraint for an isodoublet-like $e_4$ due to the assumption $\ka_N=0$.  
When $\la_E~(\la_L) = 1$, the branching fraction can be as large as 50\% (40\%) for a singlet-like (doublet-like) $e_4$. If we allow larger coupling constants, $\br{H}{e_4 \mu} \sim 1$ is possible, 
thus, the total branching fraction can be as large as $\br{e_4}{V\ell}$, 
i.e., 50\% (75\%) for the doublet-like (singlet-like) $e_4$. 
The decay to the neutral component $L^0$ vanishes because of $\ka_N = 0$.

\begin{figure}[t]
    \centering
\includegraphics[height=0.45\textwidth]{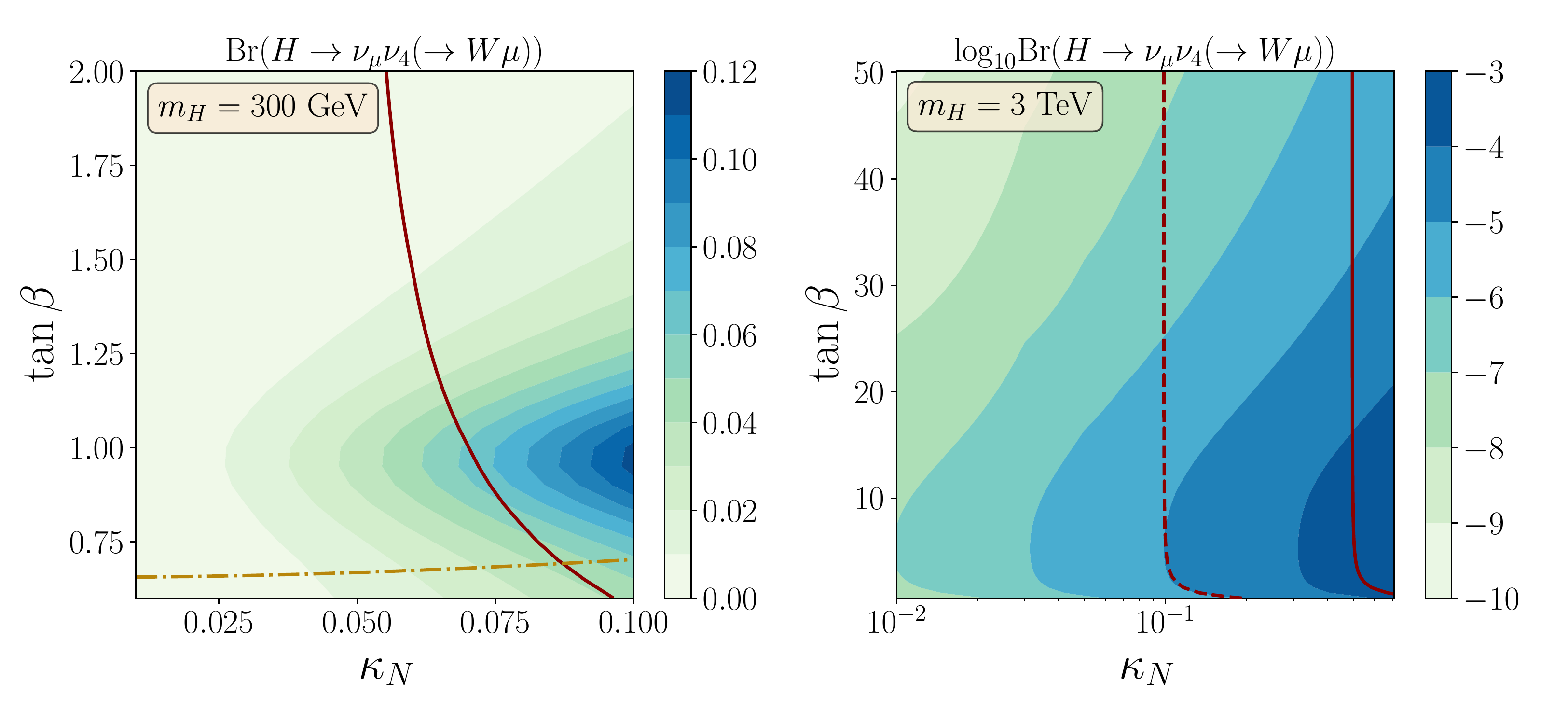} 
    \caption{
Branching fractions of the leptonic cascade decay via the lightest neutral vectorlike lepton: $H\to \nu_\mu \nu_4 \to V \ell_2 \nu_\mu$ for almost SM singlet-like $\nu_4$.  
We set $m_H = 300~(3000)~\GeV$ in the left (right) panel. 
The value of ${G_F}$ is more than $2\sigma$ away from the measured value  
to the right of the red lines for $m_N = 250~(2500)~\GeV$ in the left (right) panel. 
In the left panel, the region above the yellow dash-dot line is excluded by the $H\to\tau\tau$ search. 
The red dashed line in the right panel shows the ${G_F}$ constraint for $m_N = 500~\GeV$. 
} 
    \label{fig:BrNcascade}
\end{figure}

Figure~\ref{fig:BrNcascade} shows the branching fraction $H \to \nu_4 \nu_\mu \to W \mu \nu_\mu$ where $\nu_4$ is the SM singlet-like.
The constraints from the electroweak precision measurements and the $H \to \tau \tau $ searches are also displayed with the brown and yellow lines, respectively.
The value of ${G_F}$ is more than $2\sigma$ away from the measured value  
to the right of the red lines for $m_N = 250~(2500)~\GeV$ in the left (right) panel. 
In the left panel, the region above the yellow dash-dot line is excluded by the $H\to\tau\tau$ search. 
The red dashed line in the right panel shows the ${G_F}$ constraint for $m_N = 500~\GeV$. 
In the left panel, we set $m_H = 300~\GeV < 2m_t$, where the branching fraction can be as large as $10~\%$. 
However, a wide range of parameter space $\tan\beta \gtrsim 0.75$ is already excluded by the recent $H \to \tau \tau$ search.
Below the yellow line, the CP-even (CP-odd) Higgs decay is dominated by $H\to hh$ ($A\to gg$) because we assume a MSSM-like Higgs potential. 
Hence the decay to tau leptons are suppressed. 
This region can be further excluded by the searches for $H\to hh$ depending on the quartic couplings of Higgs potential~\cite{CMS:2017hea,ATLAS:2021ifb}.   
The value of $\br{H}{\nu_4 \nu_\mu}$ is even smaller for larger $\tan\beta$ because the decay widths to bottom quarks and tau leptons are enhanced by $\tan^4\beta$ compared with the decay to the decay mode into $\nu_4$. 
On the right panel, we assume $m_H=3~\TeV$ which is well above the limit from the di-tau decay channel. 
Even if $m_N = 2.5~\TeV$, $\ka_N \lesssim 0.5$ is required to be consistent with EW precision measurements. 
The Higgs width is dominated by decays to SM fermions, 
top quarks for small $\tan\beta$ while bottom quarks for large $\tan\beta$, 
and $\br{H}{\nu_4 \nu_\mu}$ is at most $\order{10^{-3}}$. 
The values of the branching fraction would be similar (or slightly smaller) in the regions $m_H < 3$ TeV where our analysis has sensitivities and hence we conclude the leptonic process $H \to \nu_4 \nu_\mu \to W \mu \nu_\mu$ would not be constrained in most of our parameter space.

\bibliography{ref}

\providecommand{\href}[2]{#2}\begingroup\raggedright\begin{thebibliography}{10}

\bibitem{Djouadi:2005gj}
A.~Djouadi, \emph{{The Anatomy of electro-weak symmetry breaking. II. The Higgs
  bosons in the minimal supersymmetric model}},
  \href{http://dx.doi.org/10.1016/j.physrep.2007.10.005}{\emph{Phys. Rept.}
  {\bf 459} (2008) 1--241}, [\href{http://arxiv.org/abs/hep-ph/0503173}{{\tt
  hep-ph/0503173}}].

\bibitem{Branco:2011iw}
G.~C. Branco, P.~M. Ferreira, L.~Lavoura, M.~N. Rebelo, M.~Sher and J.~P.
  Silva, \emph{{Theory and phenomenology of two-Higgs-doublet models}},
  \href{http://dx.doi.org/10.1016/j.physrep.2012.02.002}{\emph{Phys. Rept.}
  {\bf 516} (2012) 1--102}, [\href{http://arxiv.org/abs/1106.0034}{{\tt
  1106.0034}}].

\bibitem{Dimopoulos:1981zb}
S.~Dimopoulos and H.~Georgi, \emph{{Softly Broken Supersymmetry and SU(5)}},
  \href{http://dx.doi.org/10.1016/0550-3213(81)90522-8}{\emph{Nucl. Phys. B}
  {\bf 193} (1981) 150--162}.

\bibitem{Nilles:1983ge}
H.~P. Nilles, \emph{{Supersymmetry, Supergravity and Particle Physics}},
  \href{http://dx.doi.org/10.1016/0370-1573(84)90008-5}{\emph{Phys. Rept.} {\bf
  110} (1984) 1--162}.

\bibitem{Zhitnitsky:1980tq}
A.~R. Zhitnitsky, \emph{{On Possible Suppression of the Axion Hadron
  Interactions. (In Russian)}}, {\emph{Sov. J. Nucl. Phys.} {\bf 31} (1980)
  260}.

\bibitem{Dine:1981rt}
M.~Dine, W.~Fischler and M.~Srednicki, \emph{{A Simple Solution to the Strong
  CP Problem with a Harmless Axion}},
  \href{http://dx.doi.org/10.1016/0370-2693(81)90590-6}{\emph{Phys. Lett. B}
  {\bf 104} (1981) 199--202}.

\bibitem{Chacko:2005pe}
Z.~Chacko, H.-S. Goh and R.~Harnik, \emph{{The Twin Higgs: Natural electroweak
  breaking from mirror symmetry}},
  \href{http://dx.doi.org/10.1103/PhysRevLett.96.231802}{\emph{Phys. Rev.
  Lett.} {\bf 96} (2006) 231802},
  [\href{http://arxiv.org/abs/hep-ph/0506256}{{\tt hep-ph/0506256}}].

\bibitem{Bar-Shalom:2012vvt}
S.~Bar-Shalom, M.~Geller, S.~Nandi and A.~Soni, \emph{{Two Higgs doublets, a
  4th generation and a 125 GeV Higgs: A review.}},
  \href{http://dx.doi.org/10.1155/2013/672972}{\emph{Adv. High Energy Phys.}
  {\bf 2013} (2013) 672972}, [\href{http://arxiv.org/abs/1208.3195}{{\tt
  1208.3195}}].

\bibitem{Dermisek:2012as}
R.~Dermisek, \emph{{Insensitive Unification of Gauge Couplings}},
  \href{http://dx.doi.org/10.1016/j.physletb.2012.06.037}{\emph{Phys. Lett. B}
  {\bf 713} (2012) 469--472}, [\href{http://arxiv.org/abs/1204.6533}{{\tt
  1204.6533}}].

\bibitem{Dermisek:2012ke}
R.~Dermisek, \emph{{Unification of gauge couplings in the standard model with
  extra vectorlike families}},
  \href{http://dx.doi.org/10.1103/PhysRevD.87.055008}{\emph{Phys. Rev. D} {\bf
  87} (2013) 055008}, [\href{http://arxiv.org/abs/1212.3035}{{\tt 1212.3035}}].

\bibitem{Dermisek:2017ihj}
R.~Dermisek and N.~McGinnis, \emph{{Mass scale of vectorlike matter and
  superpartners from IR fixed point predictions of gauge and top Yukawa
  couplings}}, \href{http://dx.doi.org/10.1103/PhysRevD.97.055009}{\emph{Phys.
  Rev. D} {\bf 97} (2018) 055009}, [\href{http://arxiv.org/abs/1712.03527}{{\tt
  1712.03527}}].

\bibitem{Dermisek:2018hxq}
R.~Derm\'\i{}\v{s}ek and N.~McGinnis, \emph{{Top-bottom-tau Yukawa coupling
  unification in the MSSM plus one vectorlike family and fermion masses as IR
  fixed points}},
  \href{http://dx.doi.org/10.1103/PhysRevD.99.035033}{\emph{Phys. Rev. D} {\bf
  99} (2019) 035033}, [\href{http://arxiv.org/abs/1810.12474}{{\tt
  1810.12474}}].

\bibitem{Dermisek:2018ujw}
R.~Derm\'\i{}\v{s}ek and N.~McGinnis, \emph{{Seven largest couplings of the
  standard model as IR fixed points}},
  \href{http://dx.doi.org/10.1103/PhysRevLett.122.181803}{\emph{Phys. Rev.
  Lett.} {\bf 122} (2019) 181803}, [\href{http://arxiv.org/abs/1812.05240}{{\tt
  1812.05240}}].

\bibitem{Dugan:1984hq}
M.~J. Dugan, H.~Georgi and D.~B. Kaplan, \emph{{Anatomy of a Composite Higgs
  Model}}, \href{http://dx.doi.org/10.1016/0550-3213(85)90221-4}{\emph{Nucl.
  Phys. B} {\bf 254} (1985) 299--326}.

\bibitem{Arkani-Hamed:2001nha}
N.~Arkani-Hamed, A.~G. Cohen and H.~Georgi, \emph{{Electroweak symmetry
  breaking from dimensional deconstruction}},
  \href{http://dx.doi.org/10.1016/S0370-2693(01)00741-9}{\emph{Phys. Lett. B}
  {\bf 513} (2001) 232--240}, [\href{http://arxiv.org/abs/hep-ph/0105239}{{\tt
  hep-ph/0105239}}].

\bibitem{Kim:1979if}
J.~E. Kim, \emph{{Weak Interaction Singlet and Strong CP Invariance}},
  \href{http://dx.doi.org/10.1103/PhysRevLett.43.103}{\emph{Phys. Rev. Lett.}
  {\bf 43} (1979) 103}.

\bibitem{Shifman:1979if}
M.~A. Shifman, A.~I. Vainshtein and V.~I. Zakharov, \emph{{Can Confinement
  Ensure Natural CP Invariance of Strong Interactions?}},
  \href{http://dx.doi.org/10.1016/0550-3213(80)90209-6}{\emph{Nucl. Phys. B}
  {\bf 166} (1980) 493--506}.

\bibitem{Dine:1981gu}
M.~Dine and W.~Fischler, \emph{{A Phenomenological Model of Particle Physics
  Based on Supersymmetry}},
  \href{http://dx.doi.org/10.1016/0370-2693(82)91241-2}{\emph{Phys. Lett. B}
  {\bf 110} (1982) 227--231}.

\bibitem{Nappi:1982hm}
C.~R. Nappi and B.~A. Ovrut, \emph{{Supersymmetric Extension of the SU(3) x
  SU(2) x U(1) Model}},
  \href{http://dx.doi.org/10.1016/0370-2693(82)90418-X}{\emph{Phys. Lett. B}
  {\bf 113} (1982) 175--179}.

\bibitem{Alvarez-Gaume:1981abe}
L.~Alvarez-Gaume, M.~Claudson and M.~B. Wise, \emph{{Low-Energy
  Supersymmetry}},
  \href{http://dx.doi.org/10.1016/0550-3213(82)90138-9}{\emph{Nucl. Phys. B}
  {\bf 207} (1982) 96}.

\bibitem{Dine:1993yw}
M.~Dine and A.~E. Nelson, \emph{{Dynamical supersymmetry breaking at
  low-energies}}, \href{http://dx.doi.org/10.1103/PhysRevD.48.1277}{\emph{Phys.
  Rev. D} {\bf 48} (1993) 1277--1287},
  [\href{http://arxiv.org/abs/hep-ph/9303230}{{\tt hep-ph/9303230}}].

\bibitem{Dine:1994vc}
M.~Dine, A.~E. Nelson and Y.~Shirman, \emph{{Low-energy dynamical supersymmetry
  breaking simplified}},
  \href{http://dx.doi.org/10.1103/PhysRevD.51.1362}{\emph{Phys. Rev. D} {\bf
  51} (1995) 1362--1370}, [\href{http://arxiv.org/abs/hep-ph/9408384}{{\tt
  hep-ph/9408384}}].

\bibitem{Dine:1995ag}
M.~Dine, A.~E. Nelson, Y.~Nir and Y.~Shirman, \emph{{New tools for low-energy
  dynamical supersymmetry breaking}},
  \href{http://dx.doi.org/10.1103/PhysRevD.53.2658}{\emph{Phys. Rev. D} {\bf
  53} (1996) 2658--2669}, [\href{http://arxiv.org/abs/hep-ph/9507378}{{\tt
  hep-ph/9507378}}].

\bibitem{Frank:2020smf}
M.~Frank and I.~Saha, \emph{{Muon anomalous magnetic moment in
  two-Higgs-doublet models with vectorlike leptons}},
  \href{http://dx.doi.org/10.1103/PhysRevD.102.115034}{\emph{Phys. Rev. D} {\bf
  102} (2020) 115034}, [\href{http://arxiv.org/abs/2008.11909}{{\tt
  2008.11909}}].

\bibitem{Chun:2020uzw}
E.~J. Chun and T.~Mondal, \emph{{Explaining $g-2$ anomalies in two Higgs
  doublet model with vector-like leptons}},
  \href{http://dx.doi.org/10.1007/JHEP11(2020)077}{\emph{JHEP} {\bf 11} (2020)
  077}, [\href{http://arxiv.org/abs/2009.08314}{{\tt 2009.08314}}].

\bibitem{Dermisek:2020cod}
R.~Dermisek, K.~Hermanek, N.~McGinnis and N.~McGinnis, \emph{{Highly Enhanced
  Contributions of Heavy Higgs Bosons and New Leptons to Muon
  $g$\ensuremath{-}2 and Prospects at Future Colliders}},
  \href{http://dx.doi.org/10.1103/PhysRevLett.126.191801}{\emph{Phys. Rev.
  Lett.} {\bf 126} (2021) 191801}, [\href{http://arxiv.org/abs/2011.11812}{{\tt
  2011.11812}}].

\bibitem{Dermisek:2021ajd}
R.~Dermisek, K.~Hermanek and N.~McGinnis, \emph{{Muon g-2 in two-Higgs-doublet
  models with vectorlike leptons}},
  \href{http://dx.doi.org/10.1103/PhysRevD.104.055033}{\emph{Phys. Rev. D} {\bf
  104} (2021) 055033}, [\href{http://arxiv.org/abs/2103.05645}{{\tt
  2103.05645}}].

\bibitem{Dermisek:2015oja}
R.~Dermisek, E.~Lunghi and S.~Shin, \emph{{Two Higgs doublet model with
  vectorlike leptons and contributions to $pp\to WW$ and $H\to WW$}},
  \href{http://dx.doi.org/10.1007/JHEP02(2016)119}{\emph{JHEP} {\bf 02} (2016)
  119}, [\href{http://arxiv.org/abs/1509.04292}{{\tt 1509.04292}}].

\bibitem{Dermisek:2015vra}
R.~Dermisek, E.~Lunghi and S.~Shin, \emph{{Contributions of flavor violating
  couplings of a Higgs boson to $pp \to WW$}},
  \href{http://dx.doi.org/10.1007/JHEP08(2015)126}{\emph{JHEP} {\bf 08} (2015)
  126}, [\href{http://arxiv.org/abs/1503.08829}{{\tt 1503.08829}}].

\bibitem{Dermisek:2015hue}
R.~Dermisek, E.~Lunghi and S.~Shin, \emph{{New decay modes of heavy Higgs
  bosons in a two Higgs doublet model with vectorlike leptons}},
  \href{http://dx.doi.org/10.1007/JHEP05(2016)148}{\emph{JHEP} {\bf 05} (2016)
  148}, [\href{http://arxiv.org/abs/1512.07837}{{\tt 1512.07837}}].

\bibitem{Dermisek:2016via}
R.~Dermisek, E.~Lunghi and S.~Shin, \emph{{New constraints and discovery
  potential for Higgs to Higgs cascade decays through vectorlike leptons}},
  \href{http://dx.doi.org/10.1007/JHEP10(2016)081}{\emph{JHEP} {\bf 10} (2016)
  081}, [\href{http://arxiv.org/abs/1608.00662}{{\tt 1608.00662}}].

\bibitem{CidVidal:2018eel}
X.~Cid~Vidal et~al., \emph{{Report from Working Group 3}: {Beyond the Standard
  Model physics at the HL-LHC and HE-LHC}},
  \href{http://dx.doi.org/10.23731/CYRM-2019-007.585}{\emph{CERN Yellow Rep.
  Monogr.} {\bf 7} (2019) 585--865},
  [\href{http://arxiv.org/abs/1812.07831}{{\tt 1812.07831}}].

\bibitem{Dermisek:2019vkc}
R.~Derm\'\i{}\v{s}ek, E.~Lunghi and S.~Shin, \emph{{Hunting for Vectorlike
  Quarks}}, \href{http://dx.doi.org/10.1007/JHEP04(2019)019}{\emph{JHEP} {\bf
  04} (2019) 019}, [\href{http://arxiv.org/abs/1901.03709}{{\tt 1901.03709}}].

\bibitem{Dermisek:2019heo}
R.~Dermisek, E.~Lunghi and S.~Shin, \emph{{Cascade decays of heavy Higgs bosons
  through vectorlike quarks in two Higgs doublet models}},
  \href{http://dx.doi.org/10.1007/JHEP03(2020)029}{\emph{JHEP} {\bf 03} (2020)
  029}, [\href{http://arxiv.org/abs/1907.07188}{{\tt 1907.07188}}].

\bibitem{Dermisek:2020gbr}
R.~Dermisek, E.~Lunghi, N.~McGinnis and S.~Shin, \emph{{Signals with six bottom
  quarks for charged and neutral Higgs bosons}},
  \href{http://dx.doi.org/10.1007/JHEP07(2020)241}{\emph{JHEP} {\bf 07} (2020)
  241}, [\href{http://arxiv.org/abs/2005.07222}{{\tt 2005.07222}}].

\bibitem{Dermisek:2022kyh}
R.~Dermisek, J.~Kawamura, E.~Lunghi, N.~McGinnis and S.~Shin, \emph{{Combined
  signatures of heavy Higgses and vectorlike fermions at the HL-LHC}},  in
  \emph{{2022 Snowmass Summer Study}}, 3, 2022.
\newblock \href{http://arxiv.org/abs/2203.03852}{{\tt 2203.03852}}.

\bibitem{Ellis:2016jkw}
J.~Ellis, \emph{{TikZ-Feynman: Feynman diagrams with TikZ}},
  \href{http://dx.doi.org/10.1016/j.cpc.2016.08.019}{\emph{Comput. Phys.
  Commun.} {\bf 210} (2017) 103--123},
  [\href{http://arxiv.org/abs/1601.05437}{{\tt 1601.05437}}].

\bibitem{Dohse:2018vqo}
M.~Dohse, \emph{{TikZ-FeynHand: Basic User Guide}},
  \href{http://arxiv.org/abs/1802.00689}{{\tt 1802.00689}}.

\bibitem{ATLAS:2019lff}
{\scshape ATLAS} collaboration, G.~Aad et~al., \emph{{Search for electroweak
  production of charginos and sleptons decaying into final states with two
  leptons and missing transverse momentum in $\sqrt{s}=13$ TeV $pp$ collisions
  using the ATLAS detector}},
  \href{http://dx.doi.org/10.1140/epjc/s10052-019-7594-6}{\emph{Eur. Phys. J.
  C} {\bf 80} (2020) 123}, [\href{http://arxiv.org/abs/1908.08215}{{\tt
  1908.08215}}].

\bibitem{ATLAS:2021wob}
{\scshape ATLAS} collaboration, G.~Aad et~al., \emph{{Search for new phenomena
  in three- or four-lepton events in pp collisions at s=13 TeV with the ATLAS
  detector}},
  \href{http://dx.doi.org/10.1016/j.physletb.2021.136832}{\emph{Phys. Lett. B}
  {\bf 824} (2022) 136832}, [\href{http://arxiv.org/abs/2107.00404}{{\tt
  2107.00404}}].

\bibitem{Dermisek:2014qca}
R.~Dermisek, J.~P. Hall, E.~Lunghi and S.~Shin, \emph{{Limits on Vectorlike
  Leptons from Searches for Anomalous Production of Multi-Lepton Events}},
  \href{http://dx.doi.org/10.1007/JHEP12(2014)013}{\emph{JHEP} {\bf 12} (2014)
  013}, [\href{http://arxiv.org/abs/1408.3123}{{\tt 1408.3123}}].

\bibitem{Sahinsoy:2015clx}
{\scshape ATLAS} collaboration, M.~Sahinsoy, \emph{{Searches for vector-like
  quarks with the ATLAS detector at the LHC}},
  \href{http://dx.doi.org/10.22323/1.234.0152}{\emph{PoS} {\bf EPS-HEP2015}
  (2015) 152}.

\bibitem{Allanach:2015gkd}
B.~Allanach, F.~S. Queiroz, A.~Strumia and S.~Sun, \emph{{$Z^\prime$ models for
  the LHCb and $g-2$ muon anomalies}},
  \href{http://dx.doi.org/10.1103/PhysRevD.93.055045}{\emph{Phys. Rev. D} {\bf
  93} (2016) 055045}, [\href{http://arxiv.org/abs/1511.07447}{{\tt
  1511.07447}}].

\bibitem{Kawamura:2019rth}
J.~Kawamura, S.~Raby and A.~Trautner, \emph{{Complete vectorlike fourth family
  and new U(1)' for muon anomalies}},
  \href{http://dx.doi.org/10.1103/PhysRevD.100.055030}{\emph{Phys. Rev. D} {\bf
  100} (2019) 055030}, [\href{http://arxiv.org/abs/1906.11297}{{\tt
  1906.11297}}].

\bibitem{Kawamura:2019hxp}
J.~Kawamura, S.~Raby and A.~Trautner, \emph{{Complete vectorlike fourth family
  with U(1)' : A global analysis}},
  \href{http://dx.doi.org/10.1103/PhysRevD.101.035026}{\emph{Phys. Rev. D} {\bf
  101} (2020) 035026}, [\href{http://arxiv.org/abs/1911.11075}{{\tt
  1911.11075}}].

\bibitem{Kawamura:2021ygg}
J.~Kawamura and S.~Raby, \emph{{Signal of four muons or more from a vector-like
  lepton decaying to a muon-philic Z' boson at the LHC}},
  \href{http://dx.doi.org/10.1103/PhysRevD.104.035007}{\emph{Phys. Rev. D} {\bf
  104} (2021) 035007}, [\href{http://arxiv.org/abs/2104.04461}{{\tt
  2104.04461}}].

\bibitem{Dermisek:2013cxa}
R.~Dermisek, J.~P. Hall, E.~Lunghi and S.~Shin, \emph{{A New Avenue to Charged
  Higgs Discovery in Multi-Higgs Models}},
  \href{http://dx.doi.org/10.1007/JHEP04(2014)140}{\emph{JHEP} {\bf 04} (2014)
  140}, [\href{http://arxiv.org/abs/1311.7208}{{\tt 1311.7208}}].

\bibitem{CMS:2018rmh}
{\scshape CMS} collaboration, A.~M. Sirunyan et~al., \emph{{Search for
  additional neutral MSSM Higgs bosons in the $\tau\tau$ final state in
  proton-proton collisions at $\sqrt{s}=$ 13 TeV}},
  \href{http://dx.doi.org/10.1007/JHEP09(2018)007}{\emph{JHEP} {\bf 09} (2018)
  007}, [\href{http://arxiv.org/abs/1803.06553}{{\tt 1803.06553}}].

\bibitem{ATLAS:2020zms}
{\scshape ATLAS} collaboration, G.~Aad et~al., \emph{{Search for heavy Higgs
  bosons decaying into two tau leptons with the ATLAS detector using $pp$
  collisions at $\sqrt{s}=13$ TeV}},
  \href{http://dx.doi.org/10.1103/PhysRevLett.125.051801}{\emph{Phys. Rev.
  Lett.} {\bf 125} (2020) 051801}, [\href{http://arxiv.org/abs/2002.12223}{{\tt
  2002.12223}}].

\bibitem{ATLAS:2021ayy}
{\scshape ATLAS} collaboration, \emph{{hMSSM summary plots from direct and
  indirect searches}}, .

\bibitem{Harlander:2012pb}
R.~V. Harlander, S.~Liebler and H.~Mantler, \emph{{SusHi: A program for the
  calculation of Higgs production in gluon fusion and bottom-quark annihilation
  in the Standard Model and the MSSM}},
  \href{http://dx.doi.org/10.1016/j.cpc.2013.02.006}{\emph{Comput. Phys.
  Commun.} {\bf 184} (2013) 1605--1617},
  [\href{http://arxiv.org/abs/1212.3249}{{\tt 1212.3249}}].

\bibitem{Harlander:2016hcx}
R.~V. Harlander, S.~Liebler and H.~Mantler, \emph{{SusHi Bento: Beyond NNLO and
  the heavy-top limit}},
  \href{http://dx.doi.org/10.1016/j.cpc.2016.10.015}{\emph{Comput. Phys.
  Commun.} {\bf 212} (2017) 239--257},
  [\href{http://arxiv.org/abs/1605.03190}{{\tt 1605.03190}}].

\bibitem{CMS-NOTE-2011-005}
{\scshape ATLAS, CMS, LHC Higgs Combination Group} collaboration,
  \emph{{Procedure for the LHC Higgs boson search combination in Summer 2011}},
  .

\bibitem{Cowan:2010js}
G.~Cowan, K.~Cranmer, E.~Gross and O.~Vitells, \emph{{Asymptotic formulae for
  likelihood-based tests of new physics}},
  \href{http://dx.doi.org/10.1140/epjc/s10052-011-1554-0}{\emph{Eur. Phys. J.
  C} {\bf 71} (2011) 1554}, [\href{http://arxiv.org/abs/1007.1727}{{\tt
  1007.1727}}].

\bibitem{Lester:2014yga}
C.~G. Lester and B.~Nachman, \emph{{Bisection-based asymmetric M$_{T2}$
  computation: a higher precision calculator than existing symmetric methods}},
  \href{http://dx.doi.org/10.1007/JHEP03(2015)100}{\emph{JHEP} {\bf 03} (2015)
  100}, [\href{http://arxiv.org/abs/1411.4312}{{\tt 1411.4312}}].

\bibitem{Alwall:2014hca}
J.~Alwall, R.~Frederix, S.~Frixione, V.~Hirschi, F.~Maltoni, O.~Mattelaer
  et~al., \emph{{The automated computation of tree-level and next-to-leading
  order differential cross sections, and their matching to parton shower
  simulations}}, \href{http://dx.doi.org/10.1007/JHEP07(2014)079}{\emph{JHEP}
  {\bf 07} (2014) 079}, [\href{http://arxiv.org/abs/1405.0301}{{\tt
  1405.0301}}].

\bibitem{Degrande:2011ua}
C.~Degrande, C.~Duhr, B.~Fuks, D.~Grellscheid, O.~Mattelaer and T.~Reiter,
  \emph{{UFO - The Universal FeynRules Output}},
  \href{http://dx.doi.org/10.1016/j.cpc.2012.01.022}{\emph{Comput. Phys.
  Commun.} {\bf 183} (2012) 1201--1214},
  [\href{http://arxiv.org/abs/1108.2040}{{\tt 1108.2040}}].

\bibitem{Alloul:2013bka}
A.~Alloul, N.~D. Christensen, C.~Degrande, C.~Duhr and B.~Fuks,
  \emph{{FeynRules 2.0 - A complete toolbox for tree-level phenomenology}},
  \href{http://dx.doi.org/10.1016/j.cpc.2014.04.012}{\emph{Comput. Phys.
  Commun.} {\bf 185} (2014) 2250--2300},
  [\href{http://arxiv.org/abs/1310.1921}{{\tt 1310.1921}}].

\bibitem{Christensen:2008py}
N.~D. Christensen and C.~Duhr, \emph{{FeynRules - Feynman rules made easy}},
  \href{http://dx.doi.org/10.1016/j.cpc.2009.02.018}{\emph{Comput. Phys.
  Commun.} {\bf 180} (2009) 1614--1641},
  [\href{http://arxiv.org/abs/0806.4194}{{\tt 0806.4194}}].

\bibitem{Artoisenet:2012st}
P.~Artoisenet, R.~Frederix, O.~Mattelaer and R.~Rietkerk, \emph{{Automatic
  spin-entangled decays of heavy resonances in Monte Carlo simulations}},
  \href{http://dx.doi.org/10.1007/JHEP03(2013)015}{\emph{JHEP} {\bf 03} (2013)
  015}, [\href{http://arxiv.org/abs/1212.3460}{{\tt 1212.3460}}].

\bibitem{Sjostrand:2007gs}
T.~Sjostrand, S.~Mrenna and P.~Z. Skands, \emph{{A Brief Introduction to PYTHIA
  8.1}}, \href{http://dx.doi.org/10.1016/j.cpc.2008.01.036}{\emph{Comput. Phys.
  Commun.} {\bf 178} (2008) 852--867},
  [\href{http://arxiv.org/abs/0710.3820}{{\tt 0710.3820}}].

\bibitem{deFavereau:2013fsa}
{\scshape DELPHES 3} collaboration, J.~de~Favereau, C.~Delaere, P.~Demin,
  A.~Giammanco, V.~Lema\^\i{}tre, A.~Mertens et~al., \emph{{DELPHES 3, A
  modular framework for fast simulation of a generic collider experiment}},
  \href{http://dx.doi.org/10.1007/JHEP02(2014)057}{\emph{JHEP} {\bf 02} (2014)
  057}, [\href{http://arxiv.org/abs/1307.6346}{{\tt 1307.6346}}].

\bibitem{Cacciari:2008gp}
M.~Cacciari, G.~P. Salam and G.~Soyez, \emph{{The anti-$k_t$ jet clustering
  algorithm}},
  \href{http://dx.doi.org/10.1088/1126-6708/2008/04/063}{\emph{JHEP} {\bf 04}
  (2008) 063}, [\href{http://arxiv.org/abs/0802.1189}{{\tt 0802.1189}}].

\bibitem{Cacciari:2011ma}
M.~Cacciari, G.~P. Salam and G.~Soyez, \emph{{FastJet User Manual}},
  \href{http://dx.doi.org/10.1140/epjc/s10052-012-1896-2}{\emph{Eur. Phys. J.
  C} {\bf 72} (2012) 1896}, [\href{http://arxiv.org/abs/1111.6097}{{\tt
  1111.6097}}].

\bibitem{Caravaglios:1998yr}
F.~Caravaglios, M.~L. Mangano, M.~Moretti and R.~Pittau, \emph{{A New approach
  to multijet calculations in hadron collisions}},
  \href{http://dx.doi.org/10.1016/S0550-3213(98)00739-1}{\emph{Nucl. Phys. B}
  {\bf 539} (1999) 215--232}, [\href{http://arxiv.org/abs/hep-ph/9807570}{{\tt
  hep-ph/9807570}}].

\bibitem{Gunion:2002zf}
J.~F. Gunion and H.~E. Haber, \emph{{The CP conserving two Higgs doublet model:
  The Approach to the decoupling limit}},
  \href{http://dx.doi.org/10.1103/PhysRevD.67.075019}{\emph{Phys. Rev. D} {\bf
  67} (2003) 075019}, [\href{http://arxiv.org/abs/hep-ph/0207010}{{\tt
  hep-ph/0207010}}].

\bibitem{Craig:2013hca}
N.~Craig, J.~Galloway and S.~Thomas, \emph{{Searching for Signs of the Second
  Higgs Doublet}},  \href{http://arxiv.org/abs/1305.2424}{{\tt 1305.2424}}.

\bibitem{Carena:2013ooa}
M.~Carena, I.~Low, N.~R. Shah and C.~E.~M. Wagner, \emph{{Impersonating the
  Standard Model Higgs Boson: Alignment without Decoupling}},
  \href{http://dx.doi.org/10.1007/JHEP04(2014)015}{\emph{JHEP} {\bf 04} (2014)
  015}, [\href{http://arxiv.org/abs/1310.2248}{{\tt 1310.2248}}].

\bibitem{Haber:2013mia}
H.~E. Haber, \emph{{The Higgs data and the Decoupling Limit}},  in \emph{{1st
  Toyama International Workshop on Higgs as a Probe of New Physics 2013}}, 12,
  2013.
\newblock \href{http://arxiv.org/abs/1401.0152}{{\tt 1401.0152}}.

\bibitem{Dermisek:2013gta}
R.~Dermisek and A.~Raval, \emph{{Explanation of the Muon g-2 Anomaly with
  Vectorlike Leptons and its Implications for Higgs Decays}},
  \href{http://dx.doi.org/10.1103/PhysRevD.88.013017}{\emph{Phys. Rev. D} {\bf
  88} (2013) 013017}, [\href{http://arxiv.org/abs/1305.3522}{{\tt 1305.3522}}].

\bibitem{Zyla:2020zbs}
{\scshape Particle Data Group} collaboration, P.~A. Zyla et~al., \emph{{Review
  of Particle Physics}},
  \href{http://dx.doi.org/10.1093/ptep/ptaa104}{\emph{PTEP} {\bf 2020} (2020)
  083C01}.

\bibitem{CDF:2022hxs}
{\scshape CDF} collaboration, T.~Aaltonen et~al., \emph{{High-precision
  measurement of the W boson mass with the CDF II detector}},
  \href{http://dx.doi.org/10.1126/science.abk1781}{\emph{Science} {\bf 376}
  (2022) 170--176}.

\bibitem{ATLAS:2020wop}
{\scshape ATLAS} collaboration, G.~Aad et~al., \emph{{Search for type-III
  seesaw heavy leptons in dilepton final states in $pp$ collisions at
  $\sqrt{s}$ = 13 TeV with the ATLAS detector}},
  \href{http://dx.doi.org/10.1140/epjc/s10052-021-08929-9}{\emph{Eur. Phys. J.
  C} {\bf 81} (2021) 218}, [\href{http://arxiv.org/abs/2008.07949}{{\tt
  2008.07949}}].

\bibitem{Lavoura:1992np}
L.~Lavoura and J.~P. Silva, \emph{{The Oblique corrections from vector - like
  singlet and doublet quarks}},
  \href{http://dx.doi.org/10.1103/PhysRevD.47.2046}{\emph{Phys. Rev. D} {\bf
  47} (1993) 2046--2057}.

\bibitem{Kawamura:2022uft}
J.~Kawamura, S.~Okawa and Y.~Omura, \emph{{$W$ boson mass and muon $g-2$ in a
  lepton portal dark matter model}},
  \href{http://arxiv.org/abs/2204.07022}{{\tt 2204.07022}}.

\bibitem{Djouadi:2005gi}
A.~Djouadi, \emph{{The Anatomy of electro-weak symmetry breaking. I: The Higgs
  boson in the standard model}},
  \href{http://dx.doi.org/10.1016/j.physrep.2007.10.004}{\emph{Phys. Rept.}
  {\bf 457} (2008) 1--216}, [\href{http://arxiv.org/abs/hep-ph/0503172}{{\tt
  hep-ph/0503172}}].

\bibitem{CMS:2017hea}
{\scshape CMS} collaboration, A.~M. Sirunyan et~al., \emph{{Search for Higgs
  boson pair production in events with two bottom quarks and two tau leptons in
  proton\textendash{}proton collisions at $\sqrt s$ =13TeV}},
  \href{http://dx.doi.org/10.1016/j.physletb.2018.01.001}{\emph{Phys. Lett. B}
  {\bf 778} (2018) 101--127}, [\href{http://arxiv.org/abs/1707.02909}{{\tt
  1707.02909}}].

\bibitem{ATLAS:2021ifb}
{\scshape ATLAS} collaboration, G.~Aad et~al., \emph{{Search for Higgs boson
  pair production in the two bottom quarks plus two photons final state in $pp$
  collisions at $\sqrt{s}=13$ TeV with the ATLAS detector}},
  \href{http://arxiv.org/abs/2112.11876}{{\tt 2112.11876}}.

\end{thebibliography}\endgroup
\bibliographystyle{JHEP} 

\end{document}